\documentclass{pasa}%

\title[Measuring the Stellar Initial Mass Function]{The Dawes Review 8: Measuring the Stellar Initial Mass Function}
\author[A. M. Hopkins]{A. M. Hopkins$^1$\thanks{ahopkins@aao.gov.au}\\
\affil{$^1$Australian Astronomical Observatory, 105 Delhi Rd, North Ryde, NSW 2113, Australia}}%
\jid{PASA}
\doi{10.1017/pas.\the\year.xxx}
\jyear{\the\year}

\usepackage[authoryear]{natbib}
\bibpunct{(}{)}{;}{a}{}{,}
\usepackage{aas_macros}
\usepackage{hyperref} 
\hypersetup{colorlinks,citecolor=blue,linkcolor=blue,urlcolor=blue}

\def\lapp{\ifmmode\stackrel{<}{_{\sim}}\else$\stackrel{<}{_{\sim}}$\fi}
\def\gapp{\ifmmode\stackrel{>}{_{\sim}}\else$\stackrel{>}{_{\sim}}$\fi}

\begin{document}%
\begin{abstract}
The birth of stars and the formation of galaxies are cornerstones of modern astrophysics. While much is
known about how galaxies globally and their stars individually form and evolve, one fundamental property that affects
both remains elusive. This is problematic because this key property, the birth mass distribution of stars, referred to as
the stellar initial mass function (IMF), is a key tracer of the physics of star formation that underpins almost all
of the unknowns in galaxy and stellar evolution. It is perhaps the greatest source of systematic uncertainty in star and
galaxy evolution. A star's initial mass determines its luminosity, lifetime, and eventual return of material
enriched by nuclear fusion back to the interstellar medium to be incorporated in later generations of stars.
The distribution of stellar masses created in star formation events therefore determines the evolution of galaxies over
cosmic time, and accurately measuring it is crucial. The past decade has seen a growing number and variety of methods
for measuring or inferring the shape of the IMF, along with progressively more detailed simulations, paralleled by refinements
in the way the concept of the IMF is applied or conceptualised on different physical scales. This range of approaches and
evolving definitions of the quantity being measured has in turn led to conflicting conclusions regarding whether or not the
IMF is universal. Here I review and compare the growing wealth of approaches to our understanding of this fundamental
property that defines so much of astrophysics. I summarise the observational measurements from stellar analyses,
extragalactic studies and cosmic constraints, and highlight the importance of considering potential IMF variations,
reinforcing the need for measurements to quantify their scope and uncertainties carefully, in order for this field to progress.
I present a new framework to aid the discussion of the IMF and promote clarity in the further development of
this fundamental field.
\end{abstract}
\begin{keywords}
stars: formation -- stars: luminosity function, mass function -- galaxies: evolution -- galaxies: formation -- galaxies: star formation
\end{keywords}
\maketitle%
{\em The Dawes Reviews are substantial reviews of topical areas in astronomy, published by authors of
international standing at the invitation of the PASA Editorial Board. The reviews recognise William Dawes
(1762-1836), second lieutenant in the Royal Marines and the astronomer on the First Fleet. Dawes was
not only an accomplished astronomer, but spoke five languages, had a keen interest in botany, mineralogy,
engineering, cartography and music, compiled the first Aboriginal-English dictionary, and was an
outspoken opponent of slavery.}

\section{INTRODUCTION}
\label{intro}

\subsection{Context}
My aim with this review is to unify the various observational and simulation approaches for investigating the stellar initial
mass function (IMF), the mass distribution of stars arising from a star formation event. I do this by summarising
work over the past few decades focusing primarily on observational constraints, and presenting a self-consistent
framework to support future research. I address issues of terminology, definition, and scope of results in a way not previously
attempted, with the goal of minimising ambiguity and assessing the degree of consistency or otherwise in published results
regarding the ``universality" of the IMF.

The significance of understanding the IMF was highlighted by \citet{1998ASPC..142....1K} who wrote:
``Accurate knowledge of the form and mass limits of the stellar initial mass function, and its variation
in different star formation environments, is critical to virtually every aspect of star formation, stellar populations,
and galaxy evolution." And: ``Testing the universality of this initial mass function remains as our primary challenge
for the coming decade." Despite this goal being set two decades ago, the question of the universality of the IMF is
still unresolved with a variety of results over the past decade providing evidence in favour of some kind of variation
\citep[e.g.,][]{2010Natur.468..940V,2010ApJ...709.1195T,2011MNRAS.415.1647G}.
\citet{1998ASPC..142....1K} concluded that, while there was no clear physical reason to expect the IMF to be universal,
there was also ``no compelling evidence for large systematic IMF variations in galaxies."
A contrary view was expressed by \citet{1998MNRAS.301..569L} who summarised a broad range of circumstantial evidence in
favour of a stellar IMF with proportionally more high-mass stars at high redshift compared to the low redshift IMF.

The challenge posed in understanding the IMF is highlighted through the range and frequency of review articles
dedicated to it since the 1980s
\citep{1986IAUS..116..451S,1998ASPC..142..201S,1998ASPC..142....1K,1998MNRAS.301..569L,2002Sci...295...82K,2003PASP..115..763C} with a growing number in recent years
\citep{2007ARA&A..45..565M,2009eimw.confE..14E,2010ARA&A..48..339B,2012EAS....57...45J,2013pss5.book..115K,2014prpl.conf...53O,2014PhR...539...49K}, each touching on different but crucial aspects of the
problem. Major conferences, too, have focussed on the IMF, with a celebration of the 50th anniversary of the IMF
concept in 2005, ``The Initial Mass Function 50 Years Later" \citep{2005ASSL..327.....C}, updating work presented in 1998 at
the ``The Stellar Initial Mass Function (38th Herstmonceux Conference)" \citep{1998ASPC..142.....G}. This was followed in
2010 with ``UP2010: Have Observations Revealed a Variable Upper End of the Initial Mass Function?"
exploring evidence for the possibility of IMF variations \citep{2011ASPC..440.....T},
and in 2016 with a Lorentz Centre workshop ``The Universal Problem of the
Non-Universal IMF"\footnote{https://www.lorentzcenter.nl/lc/web/2016/841/info.php3?wsid=841} to share
updates on the status of the work on IMF variations.
Such levels of activity provide further evidence for the significance of the IMF and the
complexity involved in understanding its details.

The field of IMF studies is vast. A search using the SAO/NASA Astrophysics Data System for papers having
abstracts containing ``initial mass function" or ``IMF" yields more than 15000 publications.
No single reviewer could ever hope to comprehensively summarise
such a prodigious volume of work. Fortunately, existing reviews cover a broad range of
different aspects of the field, and provide a solid basis on which to build.

By way of illustration, \citet{2009eimw.confE..14E} summarises and compares the shape of the IMF (its slope and
characteristic mass) as probed through an extensive range of measurements within and external to the Milky Way, and gives
a high level review of the primary physical processes responsible for star formation and the IMF.
\citet{2010ARA&A..48..339B} provides
a comprehensive review into the question of the universality of the IMF, thoroughly summarising work in the
Galaxy and Local Group along with much of the work that was developing at the time to explore novel extragalactic
approaches. Subsequently these fields have evolved quickly, with a lot of attention on the IMF shape in
early type galaxies in particular. \citet{2013pss5.book..115K} present an extensive and detailed review ranging from
defining the IMF through to the various approaches to measuring the IMF in both stellar and extragalactic
regimes, and discuss the implications in the context of the ``integrated galaxy IMF" (IGIMF) formalism of
\citet{2003ApJ...598.1076K}. \citet{2014prpl.conf...53O} present a detailed summary of work measuring the IMF
in Milky Way star clusters and nearby galaxies, along with an overview of extragalactic work, before
providing a highly comprehensive analysis of analytical and numerical theories behind the form and origins of the
IMF. \citet{2014PhR...539...49K} reviews in detail the physical processes and phenomenology of star formation, and the
status of the theoretical framework used in addressing the problem.

This review is intended to complement these and other reviews, referring to the detailed summaries they provide as needed,
without attempting to duplicate the scope of their work. The aim here is not to deliver a comprehensive review
of a vast body of work, but rather to synthesise the key elements from the work to date in order to
develop a self-consistent framework and set of terminology on which to base future work. It is inevitable that there
will be incompleteness in the references covered below, but the hope is that the main elements are addressed,
and that at least representative results are presented.

\subsection{Scope of this review}
\label{scope}
This review builds on earlier work by summarising traditional approaches and the growing range of more
recent techniques used to measure or infer the IMF
with the aim of establishing their strengths and limitations, and identifying the different regimes
in which they are applicable. I explore issues around the nature of the problem itself, in particular the degree
to which the IMF is even a well-posed concept and whether there is an alternative formalism that might
lend itself better to observational measurement.

The strengths and limitations of different methods are highlighted, and comparisons made
between the typical samples to which they are applied, and the corresponding range of physical conditions
probed. Some examples include the approaches typically used in stellar investigations within the
Milky Way and Local Group galaxies, contrasted against those now becoming routine in extragalactic
analyses. The latter include metrics relying on stellar population synthesis (SPS) tools
\citep[e.g.,][]{2008ApJ...675..163H,2010Natur.468..940V,2011MNRAS.415.1647G}, the comparison of
stellar and dynamical mass-to-light ratios \citep[e.g.,][]{2010ApJ...709.1195T},
kinematics of stellar populations to infer mass-to-light ratios \citep[e.g.,][]{2012Natur.484..485C}, and galaxy census
approaches such as the cosmic star formation history (SFH) and the cosmic stellar mass density evolution
\citep[e.g.,][]{2008MNRAS.385..687W,2008MNRAS.391..363W}.

I investigate the potential for linking the results established from this broad range of
approaches, highlighting areas of actual inconsistency and carefully defining areas where
apparent inconsistencies are potentially a result of different physical conditions accessible to different
methodologies. I then identify opportunities for development of the field through new approaches
to measurement of the IMF to provide a self-consistent and uniform foundation for subsequent work.

The review is structured as follows. In \S\,\ref{IMF} I briefly summarise the history of the IMF,
explore issues of nomenclature and propose some conventions to minimise ambiguities in future work.
\S\S\,\ref{stellar}-\ref{sims} present an overview of the wide variety of measurement approaches taken to date
to constrain the IMF. I present a updated approach to the IMF in \S\,\ref{consistency}, followed by a discussion
in \S\,\ref{discussion} of the constraints and implications from the numerous measurements to date, before
concluding in \S\,\ref{conc}. I assume $H_0=70\,$km\,s$^{-1}$\,Mpc$^{-1}$, $\Omega_M=0.3$ and
$\Omega_\Lambda=0.7$ where necessary for converting between redshift and lookback time.

\section{BACKGROUND AND CHALLENGES}
\label{IMF}
\subsection{Overview and history}
Stars and star clusters form when dense gas collapses through gravitational or turbulent processes.
The physical state of the gas (including temperature, pressure, metallicity, and turbulence) determines which
pockets of gas fragment and collapse, and so ultimately the masses of the stars formed. Since the evolution of
a single star is almost entirely determined by its initial mass (although binary effects also play a role),
and the distribution of mass within a bound system defines its kinematics, the evolution of a cluster of coeval stars is
determined almost entirely by its stellar IMF. The evolution of galaxy composed of such clusters depends intimately on
this (potentially varying) IMF in combination with its SFH.

The IMF establishes the fraction of mass sequestered in sub-solar-mass
stars (down to masses as little as $0.1\,M_{\odot}$) with lifetimes much greater than the age of the
Universe, and the high-mass fraction (stars up to $120\,M_{\odot}$ or perhaps more) that rapidly become
supernovae, returning chemically enriched gas to the interstellar medium to support subsequent generations
of star formation. The less numerous higher mass stars dominate the light from a star cluster or a galaxy,
but the more numerous lower mass stars dominate the mass. This results in a need for different tracers to
probe the high and low mass regimes of the IMF. It also means that the mass-to-light ratio is sensitive
to the IMF shape.

The IMF is consequently the fundamental concept linking each of: (1) The process of star formation itself
through the conversion of molecular clouds (enriched to some degree by heavy elements) into a population of stars;
(2) Feedback and chemical enrichment processes arising from the radiative and mechanical energy returned
to the interstellar medium through stellar winds and supernovae from existing stellar populations, that influence
subsequent generations of stars and their metallicity; (3) The measurements used to convert observables,
(such as broadband luminosities or spectral line measurements), to underlying
physical quantities (such as the current rate of star formation and total stellar mass), in order to enable
studies of star formation and galaxy evolution.

The IMF was first measured by \citet{1955ApJ...121..161S} while working at the Australian National University, by measuring
the luminosity distribution of stars in the solar neighbourhood. It was shown to be consistent with a power law
over the mass range $0.4\lapp m/M_{\odot} \lapp 10$. Numerous measurements of the IMF over the subsequent sixty years
\citep[e.g.,][]{1979ApJS...41..513M,1986IAUS..116..451S,1992ApJ...393..373B,1993MNRAS.262..545K}
show that this power law does not extend to the lowest masses, but has a flatter slope below about half a solar mass.
The original power law slope for high mass stars found by Salpeter extends up to about 120 solar masses
\citep[e.g.,][]{1986IAUS..116..451S,1998ASPC..142..201S,2001MNRAS.322..231K,2003PASP..115..763C}, although with
some variation (but also large observational uncertainties)
in the reported high-mass slope and upper mass limit, and some debate about the value for the characteristic
or ``turn over" mass.

While much of the observational work on the IMF in the late 20th century focused on this same method of using
resolved star counts as the most robust and direct approach available, \citet{1983ApJ...272...54K} pioneered an approach
using integrated galaxy light. Many alternatives were also explored, as summarised by \citet{1986IAUS..116..451S}
and \citet{1998ASPC..142....1K}. These include a range of approaches such as ultraviolet (UV) luminosities
of galaxies \citep{1984A&A...140..325D}, indirect approaches related to chemical evolution and abundance ratios
\citep{1976ARA&A..14...43A}, and others like galaxy mass-to-light ratios that are now more routinely used to estimate IMF
properties \citep[e.g.,][]{2010ApJ...709.1195T}.

The IMF was also used as a probe of cosmology and dark matter. For example,
constraints on the IMF and cosmology were inferred from the evolution of galaxy colours \citep{1972ApJ...178..319T},
number counts of galaxies \citep{1990A&A...227..362G}, and the form of the IMF was invoked to explore the
extent to which stellar
remnants \citep[e.g.,][]{1986A&A...162...80D} or substellar objects \citep[e.g.,][]{1981A&A....98..140S} could explain the
``missing matter" in the Solar vicinity \citep{1984ApJ...287..926B}. The cosmological constraints associated with the IMF are
no longer compelling in the age of precision cosmology \citep[e.g.,][]{1998ApJ...507...46S,2016A&A...594A..13P}. Likewise,
as the numbers of substellar objects have been progressively constrained by observations
\citep[e.g.,][]{1993ApJ...414..279T,1993MNRAS.262..545K} and other approaches matured in
ruling out stellar-related contributions to possible baryonic dark matter
\citep{1996ApJ...456L..49G,1996ApJ...467L..65G,2000ApJ...542..281A,2001ApJ...550L.169A},
this aspect of the IMF has also become less important.
With the establishment of the now standard $\Lambda$CDM model, the focus on the IMF now is primarily connected
to the physics of star formation and galaxy evolution.

Part of the challenge in understanding the IMF as currently conceived is that it is a fundamentally statistical concept,
and not directly observable. \citet{2009eimw.confE..14E} notes that when estimating the IMF for star clusters, ``no cluster IMF
has ever been observed throughout the whole stellar mass range." He explains that to probe the upper mass range of the
IMF needs a very massive cluster, which are rare systems, with the nearest being too far away (a few kpc)
to see the low mass stars. Conversely the nearest clusters, required for measuring the low mass end of the IMF, are
all low mass clusters having few high mass stars. He concludes: ``Until we can observe the lowest mass stars in the highest
mass clusters an IMF makes sense only for an ensemble of clusters or stars." It is notable that the science cases for
the next generation of major telescope facilities, {\em James Webb Space Telescope} (JWST), {\em Giant Magellan
Telescope} (GMT), {\em Thirty Metre Telescope} (TMT), {\em European Large Telescope} (ELT), all include the goal of
studying resolved star formation in such high mass Galactic star clusters.
\citet{2013pss5.book..115K} takes this concept a step further, and details why the IMF is not
ever a measurable quantity, by noting that star formation occurs on Myr timescales. This means that
for stellar systems younger than about $1\,$Myr star formation has not ceased and so the IMF is not yet assembled,
while for systems older than about $0.5\,$Myr higher mass stars are lost through stellar evolutionary effects,
while dynamical processes can also cause the loss of lower mass stars. This means that there is no single
time at which the full ensemble of masses is present and measurable within a discrete spatial volume.
There is hence a need to address the issue of the short but finite time of formation, together with the fact that
star clusters do not form in isolation (typically) but within a complex, multiphase interstellar medium that is
also influenced by, and influencing, adjacent sites of star formation.

A possible solution to this issue arises through considering how many independent samples are required,
and over what spatial scale they must be probed, in order to infer the IMF robustly. By sampling a sufficiently
large number of star forming regions it might be expected that each evolutionary stage is captured and the
ensemble can be used to infer the underlying IMF. \citet{2014MNRAS.439.3239K} describe a general formalism, which
they apply to star formation scaling relations in galaxies, that links the timescale of different phases of a process with the
number of independent samples required to capture all temporal phases and the spatial scale on which the processes
are measured. They note that ``[star formation] relations measured in the solar neighbourhood are fundamentally different
from their galactic counterparts" and conclude that ``\ldots when a macroscopic correlation is caused by a time evolution,
then it must break down on small scales because the subsequent phases are resolved."
Considering the temporal dependencies of star formation, and the range of spatial scales over which we are
interested in characterising it, it may be that the formalism and concept of the IMF itself may need to be restructured
\citep{1998ASPC..142..201S}.

Despite these difficulties, a range of the early approaches toward inferring the IMF have been refined
over the past decade, and are now used routinely. These include an update of the \citet{1983ApJ...272...54K} approach
used by \citet{2008ApJ...675..163H} and \citet{2011MNRAS.415.1647G}, use of the Wing-Ford band to infer dwarf-to-giant
ratio \citep[e.g.,][]{2003MNRAS.339L..12C,2010Natur.468..940V,2012MNRAS.426.2994S} following the early work of \citet{1977ApJ...211..527W}, use of kinematics \citep[e.g.,][]{2012Natur.484..485C}, gravitational lensing observations
\citep[e.g.,][]{2010ApJ...709.1195T,2013MNRAS.434.1964S}, chemical abundance constraints
\citep[e.g.,][]{2004MNRAS.347..691P,2007ApJ...658..367K,2017ApJ...840L..11S} and more.
The $2.3\,\mu$m CO index has also been proposed for probing the dwarf-to-giant ratio 
\citep{1994MNRAS.269..655K,2008ApJ...677..276M}. In the same period other
novel approaches have been developed, such as those using cosmic census measurements to place constraints on the IMF
\citep[e.g.,][]{2003ApJ...593..258B,2006ApJ...651..142H,2008MNRAS.385..687W,2008MNRAS.391..363W}.

With this explosion in the range of approaches now being used to measure or infer the IMF there has been
a related growth in the tension between apparently conflicting results. One example is a need for so-called
``top-heavy" IMFs (a relative excess of high mass stars compared to the nominal Salpeter IMF) in regions
of elevated star formation rate \citep[e.g.,][]{2011MNRAS.415.1647G} that contrasts with the so-called
``bottom-heavy" IMFs (a relative excess of low mass stars) inferred in the cores of massive elliptical
galaxies \citep[e.g.,][]{2012ApJ...760...70V}. It is less clear whether such results are actually in conflict or not. The
different approaches measure different things, and the spatial scales probed are different as is the
epoch for the star formation activity. The current review is aimed at assessing the available wealth
of different metrics and their results in a self-consistent fashion, to begin to unify our approach to
understanding the IMF.

With this context in mind it is first necessary to review the terminology used in discussing the IMF and to explore
conventions of nomenclature.

\subsection{IMF definitions and terminology}
\label{nomenclature}
\begin{figure}[ht]
\begin{center}
\includegraphics[width=8.5cm, angle=0]{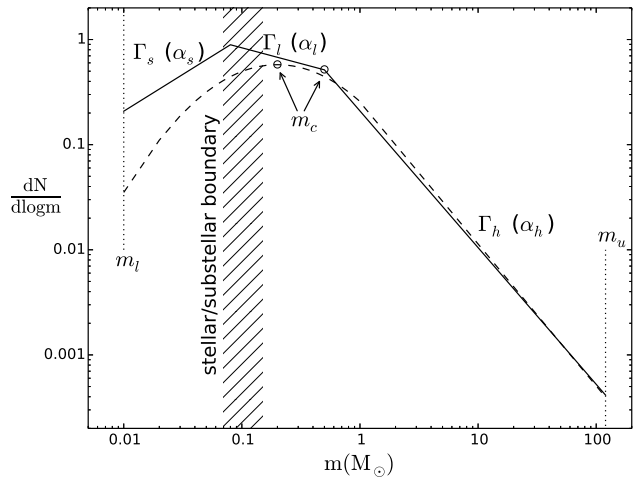}
\caption{An illustration of the key aspects of the IMF as it has been parameterised, either as a piecewise
series of power-law segments \citep[e.g.,][]{2001MNRAS.322..231K} or a log-normal at low masses with a power-law tail
at high masses \citep[e.g.,][]{2003PASP..115..763C}.}\label{form}
\end{center}
\end{figure}

At its most concrete the IMF can be defined as the mass distribution of stars arising from a star formation event. It has
been described and inferred in this sense from observational measurements by innumerable authors over more than 60 years
\citep[e.g.,][]{1955ApJ...121..161S,1979ApJS...41..513M,1983ApJ...272...54K,1986IAUS..116..451S,2001MNRAS.322..231K,2003PASP..115..763C}, who have found that the IMF in many cases follows
a similar form, and established the broad properties of this distribution.
In general, the IMF has a declining power-law shape for masses above about $1\,M_{\odot}$, with a flatter slope at lower
masses down to some minimum mass. Below the stellar/sub-stellar boundary brown dwarfs are often now
included in IMF estimates \citep[e.g.,][]{2013pss5.book..115K}, with a more positive slope below the hydrogen burning
mass limit \citep[although the shape at the lowest masses may be more complex, e.g.,][]{2016MNRAS.461.1734D}.
The observed mass function across the stellar/sub-stellar boundary may be a superposition of two
physically distinct IMFs, inferred from the deficit in models compared to observations of brown dwarfs that form through
direct gravitational collapse in molecular clouds \citep{2015ApJ...800...72T}.
The general shape and key parameters of the IMF are illustrated in Figure~\ref{form}.

Many authors have summarised the range of functional forms used to parameterise the IMF,
with the common choices being piecewise power laws
\citep[e.g.,][their Equations~4 and 5]{2013pss5.book..115K}
or a log-normal form \citep[e.g.,][]{2003ApJ...586L.133C,2005ASSL..327...41C}. Alternative functional forms have been proposed with
varying motivations \citep[e.g.,][]{2005ASSL..327...77D,2011ApJ...726...27P,2013MNRAS.429.1725M} that largely provide the same practical functionality
as the more commonly used forms.

Key parameters are: (1)~the lower mass limit, $m_l$, typically chosen as
$m_l=0.08\,M_{\odot}$ or $m_l=0.1\,M_{\odot}$ (unless
substellar objects are included, in which case $m_l=0.01\,M_{\odot}$ is common); (2)~the upper mass limit,
$m_u$, with typical values of $m_u=100\,M_{\odot}$, $m_u=120\,M_{\odot}$ or $m_u=150\,M_{\odot}$;
(3)~$m_c$, the characteristic
mass, which is the peak in the lognormal form, or the ``turn over" mass where the slope of the power law
representation changes (although as seen in Figure~\ref{form} this isn't necessarily an actual turn over in the relation),
with $m_c$ ranging from about $0.2\,M_{\odot}$ to $1\,M_{\odot}$ depending on the formalism chosen, and
$m_c=0.5\,M_{\odot}$ common in the power law representation;
(4)~the slope parameters for each segment of a piecewise power law relation, or the equivalent
in the lognormal relation defining the width of the relation at low masses, and the power law slope at high masses.
Here and throughout I use $\alpha_s$ for the substellar power law slope, $\alpha_l$ for the low mass slope, and
$\alpha_h$ for the high mass slope (noting $m_c$ for clarity when relevant). This choice avoids a numerical sequence
in which $\alpha_1$ (say) is ambiguous depending on the value of $m_l$, i.e., whether the IMF in question includes
substellar masses or not. Where a single power-law spanning more than one of these segments is assumed,
I use $\alpha$ and define the mass range explicitly.

The same functional form for the mass distribution of stars formed in a single star forming region and called
``the IMF", is also used to describe the average mass distribution of stars formed across a galaxy, a concept
sometimes referred to as the ``integrated galaxy IMF" or IGIMF \citep{2003ApJ...598.1076K}, as well as to the
effective average stellar mass distribution for a population of galaxies, referred to as the ``cosmic IMF" \citep{2008MNRAS.391..363W}.
If the IMF is universal these quantities may be identical\footnote{The IGIMF approach \citep{2003ApJ...598.1076K} presents one
recipe describing how a ``universal" IMF may lead to different IMFs for galaxies and galaxy populations, through
some star clusters being insufficiently populated at the high mass end.} but not otherwise. Such broad application
of the term ``IMF", validated through an underlying presumption of an IMF that is ``universal" until proven otherwise,
may actually be hampering attempts to further understand the properties of the IMF including whether or not it varies.
I return to this point below in \S\,\ref{confusion} and \S\,\ref{consistency}.

Other issues that hamper progress arise from inconsistent conventions describing the IMF. This field
suffers from a wide range of such inconsistencies, and these seem to be growing in number rather than
converging as the breadth of investigations increases. In an effort to stem this flow I explicitly
address these next.

\subsection{Conventions and language usage}
\label{conv}
Ambiguities in the way the IMF is discussed unnecessarily complicate what is already a complex problem.
Different authors adopt different conventions or approaches to the description of the IMF. Different language
is used to describe the same quantity, mass ranges are omitted or assumed implicitly, and ambiguous
terms are introduced. While not a fundamental problem, this definitely leads to confusion and the potential for
misinterpretation, which can be easily avoided if clear conventions and unambiguous language are used.
That this point has been made repeatedly by different authors \citep[e.g.,][]{1998ASPC..142....1K,2010ARA&A..48..339B}, and
still bears repeating is evidence that it deserves attention. Similar issues of convention and usage have been
recognised by the cosmology community \citep{2013PASA...30...52C}, emphasising the importance of striving for clarity.

Here I recount a number of sources of potential confusion and make recommendations for avoiding ambiguity,
while acknowledging that the majority of authors do tend to be diligent.
The bottom line, though, is that because of the many potential sources of confusion in this field
there is an especial need for authors, referees and editors to make extra effort to ensure clarity and consistency.

\subsubsection{Sign conventions}
Different authors have adopted a variety of nomenclature to represent the shape and power
law slope(s) of the IMF, and in particular whether or not a negative sign is given in the power law definition or appears
in the parameter. Opposing sign conventions may even appear within a single publication \citep[e.g.,][]{2009eimw.confE..14E,2009ASSP...10..215T}.
This can lead to unnecessary confusion, especially when discussing the exponent of a power law slope in a
distribution that has opposite signs at the low mass and high mass ends.
It is worth noting that opposing conventions for the use of the negative sign have existed almost as long as work in
this field. \citet{1955ApJ...121..161S} did not use a pronumeral descriptor for the power law slope at all, but
provided the power law value explicitly in his Eq.\,5, a choice followed by \citet{1983ApJ...272...54K}.
\citet{1976ARA&A..14...43A} and \citet{1977ApJ...216..548T} give the negative sign in the equation, a choice subsequently recommended by \citet{1998ASPC..142....1K},
but \citet{1979ApJS...41..513M}, \citet{1986IAUS..116..451S}, and \citet{1994ApJ...435...22K} use the convention that any sign is incorporated into the slope parameter.

I strongly recommend that, to minimise ambiguity, all authors adopt the latter convention
\citep[following, e.g.,][]{1986IAUS..116..451S,1994ApJ...435...22K}:
\begin{eqnarray}
\frac{dN}{dm} & \propto & \left({\frac{m}{M_{\odot}}}\right)^{\alpha} {\rm and} \\
\frac{dN}{d\log m} & \propto & \left({\frac{m}{M_{\odot}}}\right)^{\Gamma}
\end{eqnarray}
where $ \Gamma = \alpha + 1$ and the original Salpeter slope is $\alpha = -2.35$ or $\Gamma=-1.35$.
Contrary to some recent usage \citep[e.g.,][]{2010ARA&A..48..339B,2013pss5.book..115K}, the negative sign is not included in the
relations adopted here, and appears in the quantities $\alpha$ and $\Gamma$ explicitly.
This convention has the advantage that the sign of the parameter and of the power law itself are the same,
not opposing.
It ensures that a sign change from the lowest masses to the highest masses (in a piecewise power law
description, for example) is in the sense that intuition would suggest. It avoids inconsistencies or
clumsy presentation when discussing the value of the power law slope as contrasted with the value of the parameter,
or when inequalities are used to describe slopes flatter or steeper than some nominal parameter value. It
eliminates confusion over the need to swap the sense of asymmetric errors in estimates of the parameter as opposed
to the actual slope.
For internal self-consistency and ease of comparison between published results, I present all IMF slopes discussed
throughout using $\alpha$ as defined above.

\subsubsection{IMF naming conventions}
The use of the phrases ``Salpeter IMF," ``universal IMF," ``typical IMF,"
``normal IMF," or ``standard IMF", often interchangeably, can be confusing because of the
varying assumptions made in relation to the stellar mass range and whether or not a slope change is assumed
at the low mass end. Sometimes what is meant is
the Salpeter power law slope over a given mass range (typically $0.1<(m/M_{\odot})<100$), but also sometimes
extending up to $120-150\,M_{\odot}$, and often including other implicit assumptions. Common omissions that
lead to ambiguity include the mass range being assumed, the existence or degree of a change in
IMF slope at low masses (a ``low mass turn over"), and what the characteristic mass of such a slope
change may be. Clearly in order to avoid such confusion an explicit definition for such terms should be given
when they are introduced, ironically also including the phrase ``the Salpeter IMF" itself, since that terminology
has been used to describe all the above scenarios by different authors.

The use of the phrase ``universal IMF" to mean a Salpeter IMF also leads to or reinforces the unhelpful preconception
of the IMF as a physically universal quantity, and this may act as a stumbling block to further investigation
(see \S\,\ref{confusion} below). I recommend that using the phrase ``universal IMF" as a descriptor of an assumed
IMF in publications be avoided, and that reference to the assumed IMF be given explicitly, to minimise ambiguity
and to limit the impact on preconceptions.

\subsubsection{Parameter ranges}
It is critical to include the stellar mass range over which an IMF is being probed or discussed. This is
necessary to allow comparisons between different work, which may otherwise lead to spurious differences because
of different assumptions about mass ranges, either over the full (assumed) range of the IMF, or over a low or high mass
sub-section. Because of implicit assumptions about the relevant mass range (frequently $0.1<m/M_{\odot}<100$ but
not always, often defined by the choice of SPS code being employed, and commonly related to assuming
a ``Salpeter IMF") it is sometimes omitted, occasionally throughout an entire publication
\citep[e.g., those focused on the ratio of stellar to dynamical mass-to-light ratios, e.g.,][]{2010ApJ...709.1195T,2013MNRAS.434.1964S,2014ApJ...792L..37M}.
Sometimes, while not mentioned explicitly, the mass range may be implied, such as through reference to the IMF chosen
in a population synthesis model \citep[e.g.,][]{2016MNRAS.457..421O}, or through mention of the comparison of total mass-to-light ratios
between different assumed IMFs \citep[e.g.,][]{2013MNRAS.434.1964S}. The specification of the mass range of interest should be
given explicitly to avoid ambiguity.

Language describing mass ranges can quickly become ambiguous if the context is omitted (or described early
and not reiterated). A study of the low mass end ($m\lapp 1\,M_{\odot}$) of the IMF that discusses ``high mass"
stars or the ``high mass end" of the IMF or luminosity function probably means stars above the characteristic mass,
extending up to a solar mass or so. This, though, can easily be misinterpreted by a casual reader to refer to
stars well above $1\,M_{\odot}$ and lead to confusion regarding the truly high mass end of the IMF.
Even using ``high mass" to mean stars with $M\gapp1\,M_{\odot}$ \citep[e.g.,][]{2014prpl.conf...53O} can be misleading.
Clarifying by adding a mass range explicitly avoids such ambiguities.

There is a related ambiguity that may occur when discussing stellar masses given the sometimes significant
change between initial and final masses of high mass stars ($m\gapp 10\,M_{\odot}$) that
undergo rapid mass loss through stellar evolutionary processes. This issue is less prevalent, but
has the potential to be problematic when linking an observed mass function, called the ``present day
mass function" (PDMF), to the IMF, or in star formation simulations.

The IMF has traditionally been estimated by measuring the stellar luminosity function from which the
PDMF can be calculated. For low mass stars with lifetimes longer than a Hubble time, the PDMF is
equivalent to that segment of the IMF, giving the potential for conflating the IMF and the PDMF,
and made especially confusing when mass ranges are omitted from the discussion. It is not uncommon
to see IMF and PDMF used interchangeably in studies of the subsolar IMF. Given the direct
link between the luminosity function and the PDMF, this even leads to the potential for conflating the observed
luminosity function with the IMF in discussions of the two. This is reinforced by the choice of some authors
to publish mass functions with mass decreasing (rather than increasing) to the right in a diagram, to maintain the explicit link
to the underlying luminosity function\footnote{This is a direct consequence of presenting luminosity functions as
a function of magnitude, rather than luminosity, as fainter magnitudes are numerically larger. This tradition arose
from the original choice by Hipparchus over 2000 years ago to label the brightest stars as those of the first magnitude
and counting up for fainter stars. Using modern conventions and physical units where possible should now be preferred.}.

\subsubsection{IMF shape descriptions}
There is ample potential for confusion when describing an IMF slope or shape if language is not chosen carefully.
Any description of a power-law relationship that is expressed variously in linear or logarithmic
units needs to be cautious with words like ``steep/flat (or shallow)", ``increasing/decreasing slope",
``upturn/downturn", or ``turn over". \citet{2010ARA&A..48..339B} notes that an IMF that is ``flat" in logarithmic mass bins will
still be steep if expressed in linear mass bins ($\Gamma = \alpha+1$). Likewise, a ``turn over" apparent in
logarithmic units may not be a ``turn over", merely a change in slope, if illustrated in linear units.
Particularly confusing are descriptions referring to ``increasing/decreasing value of power-law index", given
the explicit ambiguity around whether the negative sign is included in the definition of the index or not.
Using terms such as ``steeper/flatter" or ``more positive/negative slope" instead may be helpful here,
but still need to be worded
carefully, and can be aided by showing the power law value explicitly. Carefully worded clarification around all
such descriptors is necessary to avoid ambiguity, such as being explicit about the binning scheme used,
referring to changes in slope rather than ``turn overs", being clear about the
mass range referred to and whether any ``increase/decrease" is in the higher or lower mass direction, and so on.

With extensive and growing discussion of IMF variations there has been an associated  growth in the
verbal and written shorthand evolving to describe such variations. Commonly seen terms include ``top-heavy",
``bottom-heavy" or ``bottom-light" (but rarely ``top-light" for some reason), ``dwarf-rich", ``Diet-Salpeter",
``heavyweight", and even ``obese" and ``paunchy" \citep{2007MNRAS.379..985F}. This growing range of terminology is often
not well defined and can lead to confusion, such as, for example, interchanging between ``bottom-heavy" and
``dwarf-rich", or the explicit ambiguity between
``bottom-light" and ``top-heavy". \citet{2008MNRAS.385..147D} makes the distinction that ``top-heavy" refers to an IMF that
has a high mass slope less steep than the local Salpeter value, with ``bottom-light" refering to an IMF
with a Salpeter high mass slope but having a deficit of low mass stars. Avoiding such terminology
in favour of simply citing the relevant power-law slope, or range of slopes, for the given mass range,
would eliminate potential ambiguity completely.

\subsubsection{Other issues} New quantities are sometimes labelled using pronumerals that confusingly duplicate existing
conventions. One example is the introduction by \citet{2010ApJ...709.1195T} of an ``IMF mismatch" parameter, called $\alpha$,
to compare mass-to-light ratios (${\rm M/L}=\Upsilon$) inferred through different observational approaches
(gravitational lensing and dynamics as opposed to SPS). This $\alpha$ is not the same as that in common use to
describe an IMF power law slope, although it is directly related, and is consequently an obvious source for potential
ambiguity. Clearly it is impossible to avoid duplication of all variable names, but avoiding common and clearly related
choices is strongly recommended. To avoid this ambiguity while retaining the connection to the originally published
nomenclature, I adopt $\alpha_{mm}=\Upsilon_{\rm LD}/\Upsilon_{\rm SPS}$ for the ``IMF mismatch" parameter
throughout.

There are degeneracies in the way that an IMF can be parameterised. Perhaps the best example are the
very similar shapes defined by the IMFs of \citet{2001MNRAS.322..231K} and \citet{2003PASP..115..763C},
although with completely different parameterisations. There can be more subtle degeneracies between parameters
within a given choice of parameterisation, too, such as that between $\alpha_h$ and $m_u$ or $m_c$
\citep[e.g.,][]{2008ApJ...675..163H,2011MNRAS.415.1647G}. When only the total mass normalisation is
constrained, there is further freedom in specifying the IMF shape, as discussed by \citet{2013MNRAS.432.1862C}.
It is important for authors to acknowledge such degeneracies and to explore the degree to which any inferred
IMF parameters may be influenced.

Misuse of terminology is always a potential source of ambiguity. For example, the
extensive erroneous use of the terms ``bimodal" and ``unimodal" to refer to an IMF shape comprised of
a double or single power-law respectively \citep[e.g.,][]{1996ApJS..106..307V,2013MNRAS.433.3017L,2013MNRAS.432.2632P}. The term ``bimodal" implies
two overlapping distributions with recognisable ``modes" or peaks, such as the model proposed by \citet{1986MNRAS.218..409L}.
Composite power-law relations do not have this characteristic and should be referred to differently.

To conclude this discussion, while these concerns may appear as some combination of
obvious, trivial or nit-picking, the fact that pleas for clarity in presentation have been repeatedly published
by leaders in the field over a span of decades implies a real need for care in this area. Including a statement in
the final paragraph of a paper's introduction, where it has become common to include assumptions regarding the
choice of cosmological parameters, choice of magnitude system, and others, that adds assumptions about
$m_l$, $m_u$, and IMF slope(s) or form, would go a long way to mitigating ambiguities.

Another area which deserves attention,
due as much to its subtle impact as to any overt ambiguity, is the concept of ``universality" of the IMF,
which I address next.

\subsection{The confusion wrought through ``universality"}
\label{confusion}
There are few areas of astrophysics as emotionally charged as the argument over whether the IMF is ``universal,"
that is, the same unchanging distribution regardless of environment and over the entirety of cosmic history.
With conflicting lines of evidence and apparently inconsistent conclusions, emotional attachments
to a particular viewpoint, as opposed to evidence-based conclusions, easily develop and can strongly
influence discussion in person and also in published work. Such an environment by itself makes work in this
area challenging and can limit the depth or scope of investigations and interpretation, independent of any
actual observational limitations.

In the absence of compelling evidence to the contrary the IMF is typically assumed to be universal. Partly
this is an issue of convenience, as it makes interpreting the observations of galaxies easier and allows the direct
calculation of quantities such as stellar mass and star formation rate (SFR) that can then easily be compared
among galaxy populations and over cosmic history. Also, there is
generally an understandable reluctance to invoke the more complex scenario of an IMF that varies if it is
not warranted, and a strong aversion to what is sometimes characterised as giving the theorists and simulators
yet more free parameters to play with. Given this underlying tendency to default to the ``universal" assumption,
there is a preferential inclination for authors to
present results as being consistent with a ``universal" IMF, rather than using measured uncertainties instead to
place limits on the scale of any possible variations for the given mass range, epoch, spatial scale and physical
conditions being probed. This approach hampers efforts to unify IMF studies because of the need to
independently extract the relevant spatial scale and other physical properties, which may not be a trivial process
and serves to provide further opportunity for error. It supports a tendency to acknowlege but then dismiss a host of
observational challenges in inferring the IMF (such as accounting for mass segregation, metallicity effects in the
mass-luminosity relation, dynamical effects, SPS limitations, SFHs and more), by drawing a conclusion
that is consistent with ``universality." This attitude may also lead to a tendency to downplay or dismiss evidence
inconsistent with a ``universal" IMF as arising from observational systematics or model limitations. Such results may
also be relegated to the status of a special case, as with the ``nonstandard IMFs in specific local or extragalactic
environments" noted in the abstract of the review by \citet{2010ARA&A..48..339B}.
A related issue is that the various published IMFs for Milky Way stars can easily
be conflated when arguing that observations are consistent with a ``universal" IMF. As noted by \citet{2010ARA&A..48..339B},
the \citet{1979ApJS...41..513M} and \citet{1986IAUS..116..451S} IMFs are steeper at high masses than the more recently
determined \citet{2001MNRAS.322..231K} or \citet{2003PASP..115..763C} IMFs (for example), and observations consistent
with the former are not necessarily also consistent with the latter.

The assumption of ``universality" has been questioned for about as long as the IMF has been observationally measured
\citep[see discussions in ][]{1986IAUS..116..451S,1998ASPC..142....1K,1998MNRAS.301..569L}, and
arguments for a varying IMF have been put forward since the early 1960s \citep[e.g.,][]{1963ApJ...137..758S}.
Much of the discussion in the 1980s and 1990s touches on the need for an evolving or variable IMF to explain
a variety of puzzles, including some that still remain unresolved. These include the so-called G-dwarf problem
\citep[the deficiency of metal poor stars in the Solar neighbourhood, e.g.,][]{1996AJ....112..948W}, the correlation between
stellar M/L ($\Upsilon_*$) and Mg/H abundance in ellipticals that both increase with galaxy stellar mass
\citep[e.g.,][]{1992ApJ...398...69W,1998MNRAS.301..569L}, the iron abundance in intracluster gas
\citep[e.g.,][]{1995A&A...303..345E,1997ApJ...490L..69W}, and others well summarised in the reviews of
\citet{1986IAUS..116..451S}, \citet{1998ASPC..142....1K} and \citet{1998MNRAS.301..569L}.
Associated with these observational lines of evidence, an extensive number of different models for varying
IMFs have been proposed, both to characterise their impact on different aspects of galaxy evolution, and to
explore different physical mechanisms motivating the IMF variation.
While still maintaining the preference for a ``universal" IMF, there developed some degree of consensus by
the early 2000s that an IMF that was over-represented in high mass stars (through having a low mass cutoff at
several solar masses) at early times, or in high SFR events, could explain many of these
different astrophysical results \citep[e.g.,][]{1998MNRAS.301..569L,2003PASP..115..763C}.
Subsequently, explaining the observed $850\,\mu$m galaxy source counts with semi-analytic
models \citep{2005MNRAS.356.1191B} required invoking such a ``top-heavy" IMF in starbursts, with a flatter high mass
slope ($\alpha = -1$ over $0.15<m/M_{\odot}<125$) compared to quiescent star formation ($\alpha_l=-1.4, m<1\,M_{\odot}$,
and $\alpha_h=-2.5, m>1\,M{\odot}$). This IMF modification reduces the total SFR necessary to produce the observed
$850\,\mu$m flux due to the increased number of high mass stars for a given SFR (see \S\,\ref{sims}).
Such a requirement continues to be developed and refined \citep{2016MNRAS.462.3854L}, although recent observations
may reduce this need somewhat (see \S\,\ref{census}).

There are many systematics involved in estimating an IMF, though, making it challenging to unambiguously conclude
that the IMF is different in different regions. This is highlighted, for example, by \citet{2011ASPC..440...29M}, who
demonstrates that within realistic uncertainties estimates of the slope of the high mass end of the IMF in the
Small Magellanic Cloud (SMC),
Large Magellanic Cloud (LMC) and the Milky Way are all consistent with the Salpeter value, $\alpha=-2.35$.
But appealing to the ``universality" of the IMF based only on the similarity of observed IMFs within nearby regions of
the Milky Way or even within nearby neighbouring galaxies is not justified. The range of physical conditions
being probed in these systems is limited, and does not encompass the extremes seen, for example, in starburst galaxies
or in the early Universe ($z>2$, say). The large observational and systematic uncertainties, too,
place very broad constraints that are equally consistent with the scale of some published claims for IMF variations.
The range of uncertainties for the compilation of measurements shown by \citet{2011ASPC..440...29M}, for example,
means that those results are also consistent with the variations proposed by \citet{2011MNRAS.415.1647G}, with a high
mass IMF slope $-2.5<\alpha_h<-1.8$, seen over a range of almost 2\,dex in SFR surface density, from analysing
a sample of more than 40\,000 galaxies. It would be enlightening to compare local IMF results as a function of
some underlying physical property directly with the extragalactic results, to see whether or not the same trends
hold. This is one example of the limitations on our investigations that arise from an underlying assumption of, or
a tendency to prefer a conclusion for, the ``universality" of the IMF.

More than simply limiting the scope of investigation or interpretion, though, the tendency to default to an
assumed ``universal" IMF has other insidious effects. There is an ever-present danger for investigations to
be internally inconsistent if some elements (such as an SPS model or a numerical or semi-analytic
simulation) make a ``universal" IMF assumption, but the analysis is testing IMF variations.
Any ``variation" identified must be self-consistently present in the underlying models used to infer it.

In addition, because any potential variation of the IMF means that there can be a different effective IMF as a function
of spatial scale, it now becomes important to discriminate between analyses that probe the scale of star clusters,
larger H{\sc ii} complexes or dust and molecular gas clouds, galaxies or even entire galaxy populations. It is
only relevant to compare these directly if the IMF is indeed ``universal," but if not then such comparisons
may easily be misleading. Any comparison must adequately account for any putative variation with the
relevant physical quantity. It also means that measured PDMFs for $m\lapp 0.8\,M_{\odot}$
may not necessarily correspond, as typically assumed, to the IMF \citep{2012EAS....57...45J}.

There are other confusions that arise through the use of the term ``universal." It is easy, for example, to
conflate the concepts of a ``universal" IMF and a ``universal" physical process that gives rise to an IMF
that itself may or may not be ``universal." There are now numerous published models demonstrating how
a common underlying physical process may lead to different IMFs \citep[e.g.,][]{2012MNRAS.423.3601N,2013MNRAS.433..170H}, and
result in IMF variations between galaxies and as a function of time. So a ``universal" physical process
does not necessarily imply a ``universal IMF", and care must be taken to distinguish the two.

Occam's razor is commonly invoked by scientists because there is an elegance to the simplest possible solution,
leading us to prefer not to invoke additional parameters unless clearly warranted by the data.
In the case of the IMF this leads to the well-established assumption that the IMF should be ``universal"
in the absence of compelling evidence otherwise, but I now argue that this approach has been carried too far.
It is clear that the simplest explanations are not always the most accurate or correct, although they may
have the benefit of ease of use (e.g., compare Newtonian and Einsteinian formulations of gravity) and at some level the
definition of ``simplest" is itself a subjective one.
There are some physical motivators for supposing that the IMF is universal, such as the turbulent power spectrum in
molecular clouds apparently having a universal form, which in turn leads to a prediction for a constant high mass
IMF slope \citep[e.g.,][]{2013MNRAS.433..170H}. Even this argument, though, leaves open whether the low mass end
of the IMF may vary. Accordingly, while there may be some physical expectation for some elements of the IMF to be
universal, there is also a large selection of data that question
this picture. As a consequence, I suggest that it is time to turn the basic assumption around. A better assumption
would be the most general scenario, rather than the simplest, that the IMF is not universal. This approach echoes the
sentiment expressed by \citet{1998ASPC..142..201S}, almost twenty years ago! Many of the conclusions by
\citet{1998ASPC..142..201S} are still quite pertinent today, in particular his statement that ``\ldots we are in the rather uncomfortable position of concluding that either the systematic uncertainties are so large that the IMF cannot yet be
estimated, or that there are real and significant variations of the IMF index at all masses above about $1\,M_{\odot}$."

In adopting the default assumption that the IMF may be variable,
we should be aiming to pose research questions that can assess how and the extent to which it varies, what
physical processes are responsible, and couching discussions in language that places constraints on variations
rather than merely asserting that our evidence is consistent with ``universality". Broadly adopting this attitude would lead to
authors presenting the relevant physical scale, mass range, metallicity, SFR, epoch,
and other relevant quantities over which their results hold, making it easier to assess the degree of consistency or not
between different analyses, and improving the community's ability to make progress in this field.

To help with this endeavour it is valuable to develop an ensemble of reference observations that provide a well
defined set of boundary conditions that future measurements can be tested against. It is also critical to summarise
the current state of the constraints on the IMF as a function not only of mass range, but also spatial scale,
epoch and as many relevant additional physical parameters as possible such as metallicity, SFR or SFR surface
density, in order to extend the visual summary introduced by \citet{1998ASPC..142..201S} and referred to by \citet{2002Sci...295...82K} and
\citet{2010ARA&A..48..339B} as the ``alpha plot"\footnote{Note that each of these papers uses a different convention to describe the
IMF slope, and \citet{2010ARA&A..48..339B} retain the terminology of \citet{2002Sci...295...82K} despite showing $\Gamma$ rather than $\alpha$!}. 
Producing such a suite of IMF diagnostics will be invaluable in order to begin the task of quantitatively establishing
whether and the degree to which the IMF may vary.

\section{IMF MEASUREMENT APPROACHES: STELLAR TECHNIQUES}
\label{stellar}
Rather than giving extensive reviews of the many approaches that have been used in inferring IMF measurements,
the intent here and in the following sections is to summarise the main outcomes from the different approaches, identify
a selection of highlights, and to extract the parameter range over which the measurements are valid, in order to begin
the task of unifying our understanding. In reviewing these works I draw primarily on the piecewise power-law
parameterisation, using the $m_l$, $m_c$, $m_u$, $\alpha_l$, $\alpha_h$ notation described above. This is
partly for convenience, as many of the published results use equivalent notation, but also a natural choice
because analyses are often restricted to a mass range where only a single part of the piecewise power law
is being constrained. Also, given the degeneracies in the way the IMF may be parameterised, it is not
necessarily clear that differing measurements for a parameter (a single slope, for example) are inconsistent,
unless the full parameter set is defined and can be compared between two cases.

The physical processes through which stars and star clusters form and chemically enrich their
surroundings, discussed in detail by \citet{2007ARA&A..45..481Z}, \citet{2010ARA&A..48..431P},
\citet{2014prpl.conf..149T}, \citet{2014PhR...539...49K}, and \citet{2014PASA...31...30K}, 
are beyond the scope of the current review, which is aimed instead at exploring the degree to which different
observational probes of the IMF are measuring the same thing. While acknowledging the fundamental
underlying importance of the physics driving star formation, and relying on the results above as needed, I
focus in this and the following four sections below on how we measure and use the IMF in different contexts
to understand star formation and galaxy evolution.

\subsection{Resolved star counts and luminosity functions}
Measuring the IMF directly, even within the Milky Way and nearby galaxies where individual stars can be resolved, is
challenging for several reasons, including: (1) stellar luminosities need to be converted to stellar masses,
requiring information about their ages and metallicities, with more uncertainty at the low-mass end
\citep[e.g.,][]{1993MNRAS.262..545K}; (2) account needs to be taken of the ``missing" stars, those high-mass stars that
have already evolved off the main sequence, using a relation between the stellar mass and main-sequence
lifetime \citep[e.g.,][]{2002AJ....124.2721R,2006ApJ...636..149E}; (3) assumptions need to be made for the fraction
of stars that are unresolved binary systems \citep[e.g.,][]{2010AJ....139.2679B,2012ARA&A..50...65L,2017PASA...34....1D},
with the intrinsic IMF slope being steeper (proportionally fewer higher mass stars) than nominally inferred if this fraction is
underestimated \citep{1986IAUS..116..451S,1991A&A...250..324S}, although \citet{2009MNRAS.393..663W} argue that
this effect is minor for high mass stars, but significant at the low mass end;
(4) the degree to which mass segregation (the effect of high mass stars in a gravitationally bound system moving
toward the centre of a cluster over time) affects the results in stellar clusters or associations
\citep{2007ARA&A..45..481Z,2014prpl.conf..149T,2010ApJ...718..105D}.
The reviews by \citet{2010ARA&A..48..339B}, \citet{2012EAS....57...45J}, \citet{2012ARA&A..50...65L} and \citet{2014prpl.conf...53O} provide a more detailed discussion of these
and related limitations.

Only a relatively small number of stellar systems are accessible to measure directly in this fashion, either within the
Milky Way or in nearby galaxies, with many fewer being the very young systems where high-mass stars are able to
be probed directly. In consequence, much of the work on the IMF in the Milky Way to date has focused on the
low mass end \citep[e.g.,][]{2012EAS....57...45J,2012ARA&A..50...65L}.
The small number of systems available also gives rise to issues of
stochasticity and sampling, which can limit the accuracy when attempting to infer the IMF for individual
star clusters, associations, or dispersed field populations \citep{1999ApJ...515..323E,2014MNRAS.439.3239K}. Apparent
variations between inferred IMFs for different systems may at some level just be a consequence of these observational
limitations, although \citet{2010ApJ...718..105D} argue that all star clusters in the Milky Way, young and old, are consistent
with having a common underlying mass function when dynamical effects are accounted for.
The IGIMF approach \citep{2003ApJ...598.1076K,2013pss5.book..115K} presents an alternative explanation,
where the variations for star clusters are real and depend on, for example, a relationship between the cluster mass
and the highest mass star in the cluster.

Broadly, the IMF shape for field stars in the Milky Way demonstrates a slope somewhat steeper than Salpeter
($\alpha_h \approx -2.7$) at high mass ($m\gapp 0.7\,M_{\odot}$), with a flatter slope ($\alpha_l \approx -0.5$ to
$\alpha_l \approx -1$) at lower masses, as summarised by \citet{2010ARA&A..48..339B} and \citet{2014prpl.conf...53O}.
There are many studies of the local low mass ($m\lapp 1\,M_{\odot}$) IMF, as reviewed by \citet{2003PASP..115..763C}
and \citet{2012EAS....57...45J} for example, but there
are few Galactic studies of the field star IMF in the mass range $1<m/M_{\odot}<10$. These use assumptions about
the Milky Way SFH to infer an IMF with $\alpha_h=-2.65\pm0.2$, as described, for example,
by \citet{2010ARA&A..48..339B}.

A limitation arises from the need to assume a recent SFH in estimating an IMF.
\citet{2006ApJ...636..149E} demonstrate how an assumption of a constant or slowly varying SFH can distort the inferred IMFs
from observed PDMFs, if the true underlying SFH is more stochastic. In particular, an SFH decreasing with time
can be misinterpreted as a steeper IMF if a constant SFH has been assumed. \citet{2006ApJ...636..149E} show that
this explanation can account for apparently steep IMF slopes ($\alpha_h \approx -5\pm0.5$ for
$25<m/M_{\odot}<120$) found for OB associations in the LMC and SMC \citep{1995ApJ...438..188M,2002ApJS..141...81M}.
This demonstrates the need for realistic SFHs to be adopted, and for SFH uncertainties to be incorporated into
uncertainties on the inferred IMF.

There are challenges in constraining the higher mass IMF ($m>1\,M_{\odot}$) for the field star population due
to the short lifetimes of the highest mass stars. These are best studied in OB associations and massive young
clusters \citep[e.g.,][]{2010ARA&A..48..339B}. At $m\gapp 3-10\,M_{\odot}$ \citet{2014prpl.conf...53O} summarise recent
results that suggest such star clusters and associations in the Milky Way have slopes that scatter around the
Salpeter value, $\alpha_h=-2.35$. Mass segregation, the most massive stars tending to be found in a cluster's central
regions, is often invoked as the origin of much of the scatter.
\citet{2015MNRAS.454.3872H} use simulations to argue that the lack of low mass stars observed in some globular
clusters may arise through mass segregation at birth combined with the process of
gas expulsion \citep[see also][for example]{2017MNRAS.467..758Z}.
If mass segregation is primordial, i.e., that the stars form in these
locations, then the IMF must trivially be a variable property, although such segregation is perhaps most easily attributable to
dynamical effects \citep{2007ARA&A..45..481Z,2014prpl.conf..149T}. The existence of mass segregation leads to a
necessity for observations to sample sufficiently large cluster radii in estimating the IMF, in order not to be biased
by the prevalence of high mass centrally located stars.

It can be seen already from this brief and incomplete summary that the broad range of observational challenges in
estimating the IMF for stars in various regions within the Milky Way reinforce a tendency to invoke a ``universal"
IMF. There is an understandable preference to conclude that the observations are not inconsistent with a
``universal" IMF, given the subtleties involved in accounting for the broad range of systematics and
observational limitations.

\subsection{Stellar clusters}
\label{stellarclusters}
\begin{figure}[ht]
\begin{center}
\includegraphics[width=8.5cm, angle=0]{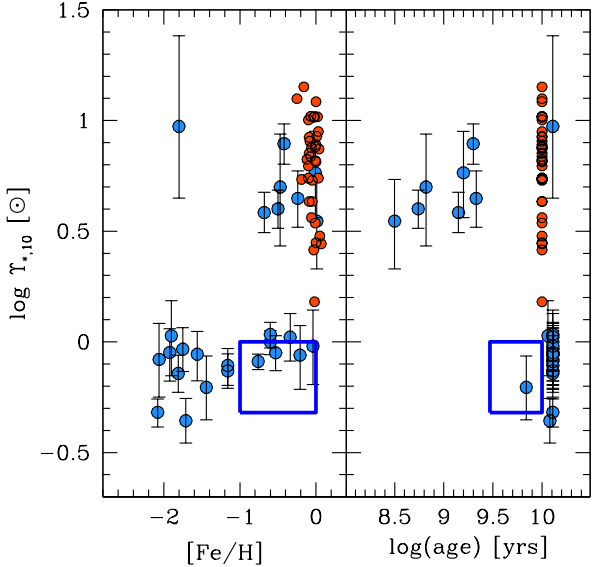}
\caption{The stellar mass-to-light ratio at 10\,Gyr, $\Upsilon_{*,10}$, as a function of metallicity and age, showing two
distinct populations. Clusters (blue data points) with younger ages, or higher metallicities, tend to show higher
mass-to-light ratios, indicative of an IMF similar to Salpeter ($\alpha=-2.35$) over the full mass range. Older, more
metal-poor, clusters have mass-to-light ratios consistent with an IMF having proportionally fewer low mass stars, such as
that of \citet{1993MNRAS.262..545K}. The red data points represent the mass-to-light ratios for early type galaxies,
while the blue box indicates the range of $\Upsilon_{*,10}$ for disk galaxies. See \citet{2014ApJ...796...71Z} for
details. (Figure~9 of ``Evidence for two distinct stellar initial mass functions: Probing for clues to the dichotomy,"
\citet{2014ApJ...796...71Z}, \copyright~AAS. Reproduced with permission.)}\label{Z14fig9}
\end{center}
\end{figure}

In their review of the formation of young massive clusters, \citet{2010ARA&A..48..431P} conclude by noting that
``globular clusters are simply old massive clusters, the logical descendants of young massive clusters
in the early Universe," a view supported by \citet{2015MNRAS.454.1658K}.
\citet{2012ApJ...761...93Z,2013ApJ...770..121Z,2014ApJ...796...71Z} use the stellar mass-to-light ratios in Milky Way and
Local Group stellar and globular clusters to infer, in contrast to the results above, two distinct stellar IMFs in such
systems, which they describe as a ``bimodal" IMF (Figure~\ref{Z14fig9}).

This result was initially established by measuring the stellar mass-to-light ratio in the $V$-band, $\Upsilon_*$,
based on observed velocity dispersions of four key clusters \citep{2012ApJ...761...93Z}, ultimately extended to a sample of
29 clusters among 4 different host galaxies \citep{2014ApJ...796...71Z}. The quantity $\Upsilon_{*,10}$ was introduced,
being the stellar mass-to-light ratio that a cluster would have at the age of 10\,Gyr, based on simple evolutionary models,
in order to more accurately compare between clusters of different ages. After exploring the impact of
stellar binarity on the measured velocity dispersions \citep{2012ApJ...761...93Z},
and the effects of internal dynamical evolution and relaxation driven mass loss \citep{2013ApJ...770..121Z},
they conclude that such effects are not enough to account for the observed differences in
mass-to-light ratio. The resulting conclusion is that the bimodality seen in $\Upsilon_{*,10}$ is evidence for two distinct
IMFs, with young stellar clusters ($\log(t / \rm{yr})\lapp 9.5$) favouring IMFs similar to Salpeter ($\alpha=-2.35$) over
the full mass range \citep[$0.1<m/M_{\odot}<120$][]{2012ApJ...761...93Z},
with older clusters favouring IMFs similar to that of \citet{1993MNRAS.262..545K},
with proportionally fewer low mass stars, and a steeper high mass slope ($\alpha_h=-2.7$). They are careful
to note, though, that neither of these IMFs is a unique solution, given that the constraint is on total mass
arising from the measured mass-to-light ratios.

There are conflicting results regarding the
mass function of globular clusters in the low mass range ($m<1\,M_{\odot}$). Using mass-to-light ratios of 200
globular clusters in M31, \citet{2011AJ....142....8S} find a deficit of low mass stars compared to a Salpeter slope,
with $-1.3<\alpha_l<-0.8$ for $m<1\,M_{\odot}$
\citep[although mass underestimates may change this conclusion, see][]{2015MNRAS.448L..94S}.
\citet{2016ApJ...826...89Z} argue that an excess of high mass stars as a function of metallicity can account for
the lower than expected mass-to-light ratios of metal-rich globular clusters in M31.
\citet{2011ApJ...735L..13V} use stellar absorption features (see \S\,\ref{passive}) measured for four globular clusters in
M31 to infer an IMF consistent with that of \citet{2001MNRAS.322..231K}, with no low mass excess. In contrast, \citet{2014ApJ...780...43G} show that some globular cluster systems (at least the metal-rich population) in elliptical
galaxies have a low mass excess, requiring $-3.0<\alpha_l<-2.3$ over $0.3<m/M_{\odot}<0.8$. 
\citet{2014ApJ...796...71Z} go on to show that the high and low mass-to-light ratios for their stellar clusters are well matched
to those of early and late type galaxies, respectively (Figure~\ref{Z14fig9}), and potentially also consistent with the
variations in IMF proposed for such systems \citep[e.g.,][]{2011MNRAS.415.1647G,2012ApJ...760...70V}.
This suggests an observational approach that can be used for directly linking and comparing studies of stellar and
galactic systems.

Accounting for dynamical evolution is important in understanding how the observed mass function of
a star cluster changes with time, and may be related to its IMF. Some simulations demonstrate that the impact
of tidal fields on star clusters, or that gas expulsion from initially mass segregated globular clusters,
ejects predominantly low mass stars \citep[e.g.,][]{2003MNRAS.340..227B,2015MNRAS.454.3872H}.
Others demonstrate the mechanisms by which star clusters can eject high mass stars
\citep[e.g.,][]{2016A&A...590A.107O,2012A&A...547A..23B}. It is clear that there are many subtleties and
details that need to be considered carefully in inferring an IMF from complex dynamically and astrophysically
evolving systems.

The challenges in measuring star counts and accounting for mass segregation in star clusters when investigating
the high mass end of the IMF can be sidestepped through measuring the hydrogen ionising photon
rate \citep{2010ApJ...719L.158C,2013ApJ...767...51A,2014ApJ...793....4A}. \citet{2014ApJ...793....4A} demonstrate
the presence of high mass stars in young ($t<8\,$Myr) low mass clusters (down to cluster masses of
$M_{\rm cl} \approx 500\,M_{\odot}$) in M83. They conclude that star clusters are populated stochastically, or randomly,
in stellar mass, allowing the existence of high mass stars (up to $120\,M_{\odot}$) even in the lowest mass clusters
($M_{\rm cl} \approx 500\,M_{\odot}$), in direct contrast with the $m_u$-$M_{\rm cl}$ relation of
\citet{2010MNRAS.401..275W} \citep[but see also][]{2014MNRAS.441.3348W}. They conclude that the population of
M83 star clusters they observe have a total H$\alpha$ luminosity to cluster mass ratio consistent with that expected from
a \citet{2001MNRAS.322..231K} IMF with $0.08<m/M_{\odot}<120$.

Also exploring star clusters in a nearby galaxy, \citet{2015ApJ...806..198W} have undertaken a large systematic study of
the colour-magnitude diagrams for 85 young, intermediate mass stellar clusters in M31. Using a framework to infer
the distribution of mass function slopes given a set of noisy measurements \citep{2010ApJ...725.2166H,2014ApJ...795...64F},
they find that the high mass slope of the IMF is best described by $\alpha_h = -2.45^{+0.06}_{-0.03}$, somewhat steeper
than the high mass slope of \citet{2001MNRAS.322..231K} ($\alpha_h=-2.3$) inferred by \citet{2014ApJ...793....4A}. This
is similar to the result of \citet{2004A&A...426..495V} who found $\alpha_h=-2.59\pm0.09$ (for $m\gapp 7\,M_{\odot}$)
using colour-magnitude diagrams for 50 OB associations in the south-western region of M31.

The conclusions of \citet{2014ApJ...793....4A} and \citet{2015ApJ...806..198W} reveal part of the challenge in discussing and understanding the IMF. Both \citet{2014ApJ...793....4A} and \citet{2015ApJ...806..198W} present their results as being
consistent with a ``universal" IMF. The claim is that the IMF for the population of clusters as a whole produces an IMF
consistent with that measured for the Milky Way, although the individual clusters demonstrate observed ``variations".
Any of the individual low mass star clusters of \citet{2014ApJ...793....4A}, for example, that contain a very high
mass star will necessarily demonstrate an IMF skewed to the high mass end, and be different to the IMF of other
clusters in the M83 ensemble, even though their ensemble IMF is consistent with that of \citet{2001MNRAS.322..231K}.
Equally, the M31 cluster IMF slopes found by \citet{2015ApJ...806..198W} show significant
scatter individually (see their Figure~4), while the ensemble is well described by IMFs having high mass
slopes drawn from a normal distribution with a very narrow intrinsic dispersion.

What the results of \citet{2014ApJ...793....4A} demonstrate, but which is not made explicit, is that stochastic or random
sampling of the IMF for a given cluster leads directy to variations in the IMF between clusters. The importance of stochasticity
in the IMF is also highlighted by \citet{2008A&A...484..711B}. The ambiguity here is deeply
buried in the difference between a ``universal" process of star formation compared to a ``universal" mass distribution
produced from any given star formation event, a complication that has led to an entire industry exploring how
the IMF is populated \citep[see, e.g., discussion in][]{2013pss5.book..115K}. The conclusions of \citet{2014ApJ...793....4A}
and \citet{2015ApJ...806..198W} are in support of the former (a ``universal" process), but
not the latter. Given these consistent conclusions, the small but measurable difference in $\alpha_h$ between
M31 \citep{2015ApJ...806..198W} and M83 \citep{2014ApJ...793....4A} is worth noting.
Both cases here also highlight the difference between an IMF inferred for any individual star cluster, and that for a
galaxy taken as a whole, a distinction that will be a recurring theme in this review.

While \citet{2015ApJ...806..198W} also use their technique to show $\alpha_h=-2.15\pm0.1$ for the Milky Way 
and $\alpha_h = -2.3\pm0.1$ for the LMC, they argue that to be robust these values would
need to be calculated using the same homogeneous and principled approach as they applied to M31.
They go on to recommend that their steeper $\alpha_h=-2.45^{+0.06}_{-0.03}$ slope for $m>1\,M_{\odot}$ be used
in the ``universal" IMF shape. When calculating SFRs this IMF slope leads to values 30-50\%
higher than assuming the \citet{2001MNRAS.322..231K} IMF. I return to this point in \S\,\ref{census} below.

Other observations of starburst clusters and super star clusters provide conflicting evidence for the shape of
the IMF. Such systems, containing numerous and unresolved stellar components, have been investigated
using a range of techniques, including spectroscopic observations, in some cases analysed using SPS models,
mass-to-light ratios and dynamics. A deficit of low mass stars has been identified in the super star cluster M82F
within the galaxy M82 \citep{2001MNRAS.326.1027S,2005ApJ...621..278M,2007MNRAS.379.1333B}.
The impact of a deficit of low mass stars on the dynamical evolution of such clusters appears relatively mild
\citep{2014MNRAS.445.2256K}. The Milky Way's nuclear star cluster was analysed by \citet{2013ApJ...764..155L}, who
find an IMF slope of $\alpha_h=-1.7$ (for $1<m/M_{\odot}<150$) and an age of around $3.3\,$Myr.
There is evidence for a Salpeter IMF ($\alpha=-2.35$ for the full mass range
$0.1<m/M_{\odot}<100$) in a massive young cluster in the Antennae \citep{2010ApJ...710.1746G}, which would imply an
excess of low mass stars relative to a \citet{2001MNRAS.322..231K} or \citet{2003PASP..115..763C} IMF, but other
dynamical mass estimates suggest ``standard Kroupa IMFs" (although without detailing
which\footnote{\citet{2008A&A...489.1091M} use the Starburst99 code, which provides an example IMF using the
Kroupa two part power law $\alpha_l=-1.3$ for $0.1<m/M_{\odot}<0.5$ and $\alpha_h=-2.3$ for $0.5<m/M_{\odot}<100$.})
in star clusters in the Antennae and NGC 1487 \citep{2008A&A...489.1091M}.
\citet{2012A&A...547A..23B} use simulations to argue that the true IMF of R136 must have had
an excess of high mass stars, given that their dynamical ejection is efficient, and that the observed mass function
is consistent with that of \citet{2001MNRAS.322..231K}.
The early suggestions of a high value for $m_l$ in starburst nuclei \citep[e.g.,][]{1980ApJ...238...24R} have not been
supported by more recent work,
with \citet{2004ApJ...606..237R} finding evidence using mid-infrared Ne line ratios for either $m_c\approx 40\,M_{\odot}$
(or equivalently a strong steepening in the high mass IMF slope above this value), or that high mass stars
in starbursts are embedded within ultra-compact H{\sc ii} regions, preventing the nebular lines from forming
and escaping, the solution they favour \citep[see also summaries by][]{2005ASSL..329...57E,2007ASPC..362..269E}.
Recent results analysing the 30 Doradus star forming region in the LMC \citep{2018Sci...359...69S} show strong evidence
for an IMF well populated up to $m_u\approx 200\,M_{\odot}$, and with $\alpha_h=-1.90^{+0.26}_{-0.37}$
for $15<m/M_{\odot}<200$. Taken together, such results appear to provide evidence for a relative excess of high
mass stars in some, but not all, starburst clusters.

\subsection{Chemical abundance measurements}
Another common stellar technique used in inferring an IMF relies on the chemical abundances of stars. Since different
elements are produced by stars of different mass, the present day elemental abundances can be used to infer an
indirect measure of the IMF. The summary by \citet{1998ASPC..142...89W} provides an excellent overview of this approach
and its issues and limitations. Broadly, oxygen and the $\alpha$-elements are produced predominantly in
core-collapse (Type II) supernovae, while iron is produced in both core-collapse and thermonuclear (Type Ia)
supernovae. The ratio of [O/Fe] can then be used to probe the high mass star IMF, with a ``plateau" in
this ratio arising from stars pre-enriched only by type II SNe, the quantitative value of which would change
with a change in the high mass IMF slope. A limitation is that the value of this Type II plateau depends on
the theoretical yields assumed for different elements as a function of supernova progenitor mass. \citet{1992AJ....104..144W}
note that varying the IMF slope from $\alpha_h=-2.1$ to $\alpha_h=-3.3$ (for $m>1\,M_{\odot}$) results in [O/Fe] changing
by $\approx 0.4\,$dex, although the difference for the smaller range of $-2.5< \alpha_h < -2.1$ is only
$\Delta{\rm [O/Fe]} \approx 0.1 - 0.15\,$dex depending on the elemental yields assumed (their Tables~1 and 2).
\citet{1998ASPC..142...89W} argues that for the Milky Way stellar halo, thick disk and bulge populations, the measured
abundances are consistent with a Salpeter high mass IMF slope.
A compilation of abundances was used by \citet{2017MNRAS.466.4403N} in introducing a new approach to scaling
abundances with total metallicity, reinforcing the detection of the Type II plateau for the Milky Way, LMC and the
Sculptor Dwarf (shown using [Mg/Fe], their Figure~9).

Combining abundance constraints with
mass-to-light ratio constraints, \citet{1997ApJ...483..228T} find $-2.6<\alpha_h<-2.3$ for $m>1\,M_{\odot}$ for stars
presently in the solar neighbourhood. They also argue for $m_u=50\pm10\,M_{\odot}$,
although this $m_u$ is not necessarily the highest mass of stars formed, but instead is the highest mass of stars that
return chemically enriched material to the interstellar medium. In their analysis stars may exist above this mass, but those that do must end as black holes without ejecting processed material into the interstellar medium. This result has been called
into question by \citet{1998ApJ...501..675G}, though, who demonstrate that using different chemical yield models
relaxes this outcome to a much less stringent constraint of $m_u \approx 30-200\,M_{\odot}$.

The chemical abundances of stars in some dwarf spheroidal (dSph) satellites of the Milky Way show measurable
differences from Milky Way stars. Early work suggested that any abundance differences were still consistent with an IMF
having a Salpeter high mass slope \citep[e.g.,][]{2003AJ....125..707T,2004AJ....128.1177V}.
A high mass truncation of the IMF was discussed as a possible scenario to explain the differences but
a solution arising from the contributions of metal-poor AGB stars was favoured. More recently,
\citet{2011ApJ...736..113T} argue that the deficiency of $\alpha$-elements combined with an enhancement
in $s$-process elements (Ba) found in dSph galaxies provides evidence of a lack of high mass stars
($m\gapp 25\,M_{\odot}$) in these systems, a result in keeping with the idea of a lower
value for $m_u$ in lower SFR environments \citep{2005ApJ...625..754W}.
This result is supported by a different type of
constraint, explored by \citet{2004MNRAS.347..691P}, who find that IMFs typical of the Milky Way
\citep{1983ApJ...272...54K,1998MNRAS.301..569L,2001ApJ...554.1274C}
can explain the mass-to-light ratios seen in Sbc/Sc galaxies but then overestimate the metallicities. They argue
that unless the observed metallicity is underestimated (due to expulsion into the intergalactic medium or through
being locked up in dust grains) the IMF needs to be truncated at high masses. This result was extended
by \citet{2004ApJ...604..579P}, showing that a ``standard solar neighbourhood IMF" \citep{1998ASPC..134..483K}
cannot provide sufficient heavy elements to account for the observed metallicity of galaxies in clusters.
None of these analyses account for galactic winds, which, if included, may help explain the observed chemical signatures
without the need for the lower value of $m_u$, although \citet{2004ApJ...604..579P} note that this would
require ``substantial loss of metals from the solar neighbourhood and from disk galaxies in general."

At the other extreme, a recent analysis of starburst galaxies by \citet{2017MNRAS.470..401R}, using updated chemical
models to track CNO isotopes and accounting for stellar rotation, finds a need for an excess of high mass stars
($\alpha_h=-1.95$, for $m>0.5\,M_{\odot}$) to reproduce observed isotope abundances. This result is consistent
with the conclusions of \citet{2017ApJ...840L..11S} who find a need
for an excess of high mass stars to explain the CO isotopic abundances in the starburst galaxy IRAS 13120-5453.

A related approach links the cosmic microwave background (CMB) to a directly observable stellar chemical signature,
the carbon-enhanced metal-poor (CEMP) stars \citep{2007ApJ...664L..63T}, probing stars in the mass range
$1<m/M_{\odot}<8$. The CMB defines a temperature minimum that may translate to a characteristic fragmentation scale
for star-forming gas \citep{2005MNRAS.359..211L}. The time dependence of the CMB can hence have an impact on the
fraction of carbon-enhanced metal-poor stars as a function of metallicity, and \citet{2007ApJ...664L..63T} find that $m_c$
should increase toward higher redshift. This result is supported by analyses of observed CEMP stars
\citep{2007ApJ...658..367K} explained as arising from binary systems, and as modelled by
binary population synthesis \citep{2013MNRAS.432L..46S}. 
\citet{2007ApJ...664L..63T} go on to point out that such an evolution of the IMF would lead to two clear
systematic errors if it is not accounted for. Early time SFHs for local galaxies derived from colour-magnitude
diagrams assuming a non-evolving IMF would be systematically underestimated, and SFRs from
high-$z$ luminosity tracers, such as UV, would be systematically overestimated. I return to these
points in \S\,\ref{census} below.

While constraints from stellar chemistry in this fashion may turn out to be quite powerful probes of the IMF, in
particular its properties at high redshift, it would be valuable to explore how limitations or systematics in our
understanding of stellar yields and binary star evolution may influence or limit such measurements.
One of the important potential advantages of stellar chemistry, and in particular the most metal-poor stars, is their
potential for probing the first generation of stars, called ``Population III" stars \citep{2010AN....331..474F}, through the
preserved signatures of their supernova chemical yields in subsequent generations of stars. I briefly
discuss this next.

\subsection{Population III stars}
Simulations and physical arguments have demonstrated for many years that Population III stars
are likely to be dominated by high mass objects \citep[e.g.,][]{2002Sci...295...93A,2002ApJ...564...23B}, with typical masses $m>100\,M_{\odot}$
and few or no low mass stars \citep[see review by][and references therein]{2004ARA&A..42...79B}.
Further work has explored the impact of such Population III star properties for the earliest generations of
galaxies \citep{2011ARA&A..49..373B}. The physical processes involved are also summarised well, in the broader
context of structure formation and evolution, by \citet{2006astro.ph..3360L}.
More recently, some simulations now appear to extend the lower mass limit for Population III stars
to lower values \citep[e.g.,][]{2014ApJ...781...60H,2014ApJ...792...32S}. Clearly the IMF for such a
population would be radically different to that in the local Universe and in that sense there is trivially an evolution
in the IMF. This does not address the question that is usually meant regarding the ``universality" of the IMF, though,
which instead is focused on whether the IMF may be different between coeval galaxies, or between
different star forming regions within a galaxy. With that in mind, it is instructive to briefly touch on some
of the results associated with the current measurements probing Population III stars.

Observational probes of the first stars are summarised in the review by \citet{2004ARA&A..42...79B}, who highlight their
reionisation signature, chemical enrichment of subsequent stellar generations (``stellar archaeology"), and gamma-ray bursts
as opportunities then developing. This builds on a significant body of earlier work to explore and explain the lack of
low metallicity stars in the Milky Way and its halo
\citep[e.g.,][]{1981ApJ...248..606B,1985PASP...97..593J,1986A&A...168...81C}
and other novel probes of Population III stars \citep[e.g.,][]{1982Natur.298..711T}.
Subsequently, gamma-ray bursts have been used to constrain the high-$z$ SFH
\citep{2008ApJ...683L...5Y,2009ApJ...705L.104K,2013arXiv1305.1630K}, finding a higher value of the SFR density
for $z>5$ than commonly inferred from deep imaging data \citep[see summary in][]{2014ARA&A..52..415M}, with
implications for reionisation \citep{2010MNRAS.401.2561W}, although
with no direct constraints yet on the Population III IMF. \citet{2015MNRAS.449.3006M} show that none of the gamma-ray
bursts detected at $5\lapp z \lapp 6$ show abundance ratios consistent with an environment dominated by Population III stars.
Opportunities for probing the first stars through stellar archaeology are summarised by \citet{2010AN....331..474F} and in an
extensive review by \citet{2013RvMP...85..809K}. Strong abundance ratio signatures are expected, with
\citet{2010ApJ...724..341H}, for example, demonstrating that increasing $m_l$ from $10-40\,M_{\odot}$ strongly
suppresses the production of elements heavier than aluminium. 

Other observations are now starting to constrain the possible IMF shapes for this earliest
generation of stars. \citet{2015ApJ...808..139S} argue for a ``flat or top-heavy IMF" (lacking stars below $10\,M_{\odot}$)
for Population III stars in a high-redshift ($z=6.6$) Ly$\alpha$ system, although \citet{2017MNRAS.469..448B} dispute this
conclusion based on deeper infrared observations. They argue for a low mass narrow-line active galactic nucleus
or low metallicity starburst to explain the observed infrared colours. \citet{2017MNRAS.468..418F} uses
abundances in extremely metal poor stars to infer a Salpeter IMF slope, $\alpha_h=-2.35^{+0.24}_{-0.29}$, with a
maximum supernova progenitor mass of $m=87^{+13}_{-33}\,M_{\odot}$ and a value of $m_l$ below the minimum mass for
Population III supernovae ($m_l\lapp 9\,M_{\odot}$). \citet{2015MNRAS.447.3892H} demonstrate a technique for constraining
the lower mass limit of Population III stars, finding that they can exclude stars with $m<0.65\,M_{\odot}$ at 95\% confidence.

Future observations hold great promise for constraining the high-$z$ IMF.
Using simulations, \citet{2014MNRAS.442.1640D} show that a few hundred supernovae detections with the
JWST could be sufficient to discriminate between a ``Salpeter and flat mass distribution for high-redshift stars".
\citet{2017A&A...608A..53J} demonstrate the redshift dependent photometric properties of globular clusters and
ultra-compact dwarf galaxies to provide observable IMF diagnostics for anticipated JWST observations.

With this review focused primarily on the consistency of approaches to observational constraints of the IMF
I do not discuss Population III stars further. It is clear that this area will see rapid growth of
a variety of observational constraints in the near future, and I hope that the framework presented here
will be applied to these approaches to aid in understanding this earliest generation of stars.

\subsection{Review}
With the broad range of approaches, techniques and results described above it is worth briefly summarising.
Star clusters sample physical scales on the order of a few pc and a broad range of ages and SFRs, although most
of those observed are high SFR surface density objects. The field star
population in principle probes the full galaxy-wide scale (tens of kpc), and is sensitive to the galaxy-wide SFH.

Although the large uncertainties and systematics involved in these measurements understandably lead to a conclusion
that the IMFs are all similar and broadly consistent with (for example), the IMFs of \citet{2001MNRAS.322..231K} or \citet{2003PASP..115..763C},
it is tantalising to note that the field star and stellar association population have high mass ($m>1\,M_{\odot}$)
slopes on average somewhat steeper ($\alpha_h \approx -2.5$) than the cluster stars
\citep[$\alpha_h \approx -2.2$, see Figure~2 in][]{2010ARA&A..48..339B}. The populations of globular clusters and extragalactic
star clusters, in contrast, show evidence for some differences between IMFs. Probing spatial scales
on the order of pc, a range of IMFs are inferred, with some globular clusters consistent with the IMF
of \citet{1993MNRAS.262..545K} and its steeper high mass slope ($\alpha_h = -2.7$), and others consistent with a Salpeter
slope over the full mass range extending to the lowest masses. It is worth noting the link between the mass-to-light
ratio approaches for globular clusters and galaxy systems as a technique that may lend itself to being applied
self-consistently across a broad range of different physical scales and conditions.

Extragalactic star clusters demonstrate IMFs consistent with \citet{2001MNRAS.322..231K} in M83 ($\alpha_h=-2.3$), or with
high mass slopes somewhat steeper ($\alpha_h=-2.45$) in M31. While this difference is small, the level of
precision of these results begins to suggest that it is not negligible. At least some starburst and super star clusters tend to
show evidence for an excess of high mass stars, either through an increase in $m_c$ or a flatter $\alpha_h$
compared to the IMF of \citet{2001MNRAS.322..231K}.
Approaches using abundance measurements reinforce the slightly steeper high mass slope seen in field stars,
and further suggest the possibility that the high mass limit $m_u$ may need to be somewhat lower than the
typical Milky Way value in nearby dSph galaxies.

Taken as a whole, while the range of uncertainties and systematics makes it easy to assert that there is
no strong evidence measured for IMF variations, adopting the converse approach of placing constraints on
the scale of possible variations opens a different line of argument. Referring below to the \citet{2001MNRAS.322..231K} or
\citet{2003PASP..115..763C} IMF as the ``typical" IMF for the Milky Way, it can be said that:
\begin{itemize}
\item For stars and star clusters within Local Group galaxies, the high mass IMF slope does not vary more than
$\pm0.3$ around $\alpha_h = -2.5$;
\item For extragalactic globular clusters, there are conflicting observations of mass-to-light ratios, which
are parameterised as a constraint on the low mass ($m<1\,M_{\odot}$) slope
ranging across $-3.0 < \alpha_l <-0.8$, from a deficit to an excess compared to the typical Milky Way values;
\item In high SFR regions (starburst or super star clusters) there is an apparent variation with some having
$m_c\gapp 1\,M_{\odot}$ or $\alpha_h>-2.3$, higher than the typical values in the Milky Way;
\item In low SFR regions (dSph galaxies), there may be an apparent variation
with $m_u$ needing to be somewhat lower than the typical value in the Milky Way.
\end{itemize}

It is worth recalling here that there may be degeneracies between $m_u$ and $\alpha_h$, and that the measured
constraints may be able to be reproduced by different parameter combinations. It remains the case, though, that
there does appear to be some difference between the form of the IMF in high and low SFR regions.
These lines of evidence suggest that the IMF is not ``universal", but that there are differences in
different regions, although the details are only qualitatively constrained.

\section{IMF MEASUREMENT APPROACHES: NEUTRAL AND MOLECULAR GAS AND DUST}
\label{molecular}

It is natural when investigating the IMF to turn from already existing stars to the progenitor clouds in which they form,
to probe the earliest stages of the formation process. This is explored most efficiently through measurements
of cold gas and dust in molecular clouds. Due to observational practicalities, much of the work here has focused
on clouds within the Milky Way and nearby galaxies, and is well-summarised in reviews by
\citet{2007ARA&A..45..481Z}, \citet{2014prpl.conf..149T} and \citet{2014prpl.conf...53O}. There is ample evidence
demonstrating that the mass function for the molecular ``cores" (those gas regions of sufficient density to go on to
form stars) has a similar shape to the stellar IMF in the Milky Way
\citep[e.g.,][]{2010A&A...518L.102A,2010A&A...518L.106K,2015A&A...584A..92M,2016MNRAS.459..342M}. Many
core mass funtions (CMFs) show an offset in mass compared to the IMFs of \citet{2001MNRAS.322..231K} or
\citet{2003PASP..115..763C},
with the break in the power law at masses larger by a factor of $\approx 3-4$, and with a similar range of
variations in the high mass slope between different molecular clouds as seen in the various stellar
analyses \citep[see summary by][]{2014prpl.conf...53O}. Despite the range of variations, such results have for
decades been similarly interpreted as consistent with a ``universal" form, and have led naturally to the
idea that the CMF and IMF are linked physically through some star formation efficiency factor.

The connection between pre-stellar CMFs and the IMF is still not clear, although many models have been
proposed to explain it \citep[e.g.,][]{2013MNRAS.433..170H,2015MNRAS.450.4137G,2015ApJ...808...10Z}. One issue is how
a core is defined observationally, and that observational limitations and different threshold levels for
defining a core lead to different results \citep[see discussions by, e.g.,][]{2009eimw.confE..14E,2014prpl.conf...53O}. As noted by
\citet{2014prpl.conf...53O}, ``Different algorithms \ldots even when applied to the same observations, do not always
identify the same cores, and when they do, they sometimes assign widely different masses." Even
if cores can be adequately identified, there is evidence questioning the fragmentation models
leading from cores to stars, and hence linking the IMFs of each \citep{2013MNRAS.432.3534H,2016MNRAS.462.4171B}.

A more robust approach than discrete core identifications is to use the full probability distribution function (PDF)
of observed column densities within a star forming cloud, in order to identify which regions may have sufficient
density to be star forming \citep[e.g.,][]{2014ApJ...795L..25R}.
There is an open question over whether there exists some threshold in column density of molecular hydrogen above
which star formation proceeds efficiently. A universal threshold of $N(H_{2}) \gapp 1.4 \times 10^{22}\,$cm$^{2}$ was
proposed by \citet{2012ApJ...745..190L}, although \citet{2012ApJ...745...69K} argue against the existence of
such a threshold. At least one counterexample, the Galactic centre molecular cloud G$0.253+0.016$, questions the idea of
a universal threshold \citep{2015ApJ...802..125R}.
Dust temperature measurements of this cloud
suggest that star formation may have recently begun, with detection of a cool filament whose hot central region is
undergoing gravitational collapse and fragmentation to form a ``line of protostars" \citep{2016MNRAS.461L..16M}.
Despite this, the central molecular zone of the Milky Way appears to support substantially less star formation
than might be expected from a column density threshold \citep{2013MNRAS.429..987L}.
These results bring the idea of a universal threshold into question, at least for environments with the
extreme high pressures found in the Milky Way central molecular zone, which
may mimic the conditions of star formation at high redshift.

Broadly, the studies of pre-stellar clouds suggest that turbulence and hierarchical fragmentation are dominant processes
in driving the star formation. Turbulence as a dominant contributor to the star formation process
has also been shown to be effective in high pressure environments \citep{2014ApJ...795L..25R}, and may
therefore be significant in starburst nuclei and
high redshift galaxies. High mass stars and clusters can form in filamentary molecular clouds
\citep{2016MNRAS.456.2041C}, although \citet{2017MNRAS.466..340C} note that high mass protoclusters are
very rare in the Galaxy. Young high mass clusters in the Milky Way have been shown to form hierarchically rather
than through monolithic collapse \citep{2015MNRAS.449..715W}, a result seen also in the arms of the grand-design spiral
NGC 1566, where \citet{2017MNRAS.468..509G} demonstrate hierarchical star formation driven by turbulence. 
\citet{2017ApJ...842...25G} show that star cluster formation in eight local galaxies is hierarchical both in space and time,
and that the ages of adjacent clusters are consistent with turbulence driving the star formation.
In contrast, there is evidence for monolithic collapse in the formation of some young Galactic star clusters
\citep[e.g.,][]{2014ApJ...787..158B,2015MNRAS.447..728B,2018ASSL..424..143B}.

Turbulence alone, though, does not seem to be a sufficient mechanism. Using high resolution observations of
molecular gas in M51, \citet{2017ApJ...846...71L} note that observed measures of star formation efficiency are in some
tension with turbulent star-formation models, finding an anticorrelation between the star formation efficiency per free-fall
time with the surface density and line width of molecular gas. \citet{2016ApJ...831...73V}, based on
star forming regions in the Galactic plane, argue that observed relations between SFR and molecular cloud properties
are inconsistent with those seen in extragalactic relations or the model by \citet{2012ApJ...745...69K}. Similarly,
\citet{2016A&A...588A..29H} find low values of star formation efficiency per free-fall time in a sample of Galactic
young stellar objects, noting that the strongest correlations of SFR surface density are with the dense gas
surface density normalized by the free-fall and clump crossing times. They state that models accounting for such
local gas conditions provide a reasonable description of these observations.
\citet{2016ApJ...833..229L} find a rather higher observed
scatter in star formation efficiency for star forming giant molecular clouds in the Milky Way, which they also note
is unable to be explained by constant \citep{2005ApJ...630..250K} or turbulence-related \citep{2011IAUS..270..159H}
star formation. They argue instead for a time-variable rate of star formation noting that ``sporadic small-scale star formation
will tend to produce more massive clusters than will steady small-scale star formation."
By analysing the dense gas in star forming clusters, \citet{2017A&A...606A.123H} argue that both clustered and
non-clustered star forming regions might be naturally explained through the spatial density of dense gas ``sonic fibres"
\citep{2013A&A...554A..55H}.
\citet{2018A&A...610A..77H} extends this approach to propose a unified star formation
scenario that leads naturally to the observed differences between low and high mass clouds, and the origin of clusters.
\citet{2016MNRAS.457.4536W} show that the mass surface density profiles are shallower for gas clouds than for young massive star clusters in the Milky Way. They argue that this implies an evolution requiring mass to continue
to accumulate toward cloud centres in highly star forming clouds after the onset of star formation, in
a ``conveyor-belt" scenario.

It is beyond the scope of this review to explore in depth the range of detailed models of star formation, and
their strengths and limitations. Summaries, however, of some models describing the star formation
process and linking the CMF to the IMF, or that aim to
explain the IMF shape, are presented below in \S\,\ref{sims}. A detailed review of the formation of young high mass
star clusters is given by \citet{2010ARA&A..48..431P}, and \citet{2014prpl.conf..149T} provide
a thorough review of high mass star formation. Interestingly, \citet{2007ARA&A..45..481Z} find strong support for an
IMF upper mass limit of $m_u\approx 150\,M_{\odot}$, and make the case that high mass star formation
proceeds differently from low mass star formation, not just as a scaled up version, but ``partly a mechanism of
its own, primarily owing to the role of stellar mass and radiation pressure in controlling the dynamics." This
conclusion has been questioned by more recent work, summarised by \citet{2014prpl.conf..149T}, who argue
that most observations support a common mechanism for star formation from low to high masses.

A different approach linking star forming gas to the IMF was used by \citet{2008ApJ...682L..13H} to explore the link
between gas consumption and star formation in a cosmic global average sense. They use the Kennicutt-Schmidt
law linking SFR and gas surface densities, following \citet{2005ApJ...630..108H} who convert such a surface
density relation to a volume density relation using the observed redshift distributions of damped Lyman $\alpha$ absorbers.
The IMF dependency arises through the SFR density measurement. Different assumed IMFs will alter the
SFR density calculated from observed luminosity densities, and consequently the corresponding volume
density of gas necessary to sustain such star formation levels. \citet{2008ApJ...682L..13H} infer
that the cosmic mass density of HI at high redshift ($z>1$) implies SFR densities that are not consistent with an IMF
typical of the Milky Way such as \citet{2001MNRAS.322..231K} or
\citet{2003PASP..115..763C}. Instead they require an IMF with a high mass slope flatter than Salpeter ($\alpha_h>-2.35$), \
such as that proposed through the evolving IMF of \citet{2008MNRAS.385..687W}. It would be valuable to
revisit this alternative style of approach in light of more recent work on the relationship between SFR and gas density,
as reviewed for example by \citet{2012ARA&A..50..531K}.

In summary, as with the stellar techniques, the approaches used in measuring the CMF in order to link it to
the IMF are limited by the relatively small samples available within the Milky Way and nearby galaxies, and
the link itself may be unclear \citep{2013MNRAS.432.3534H,2016MNRAS.462.4171B}. As with
the various stellar cluster measurements, a range of CMF high mass slopes is found for different molecular clouds,
with a similar span of uncertainty, and for much the same reason. There are similar levels of variation measured
for the low mass slope, and for the characteristic mass where the CMF slope changes. It is worth reiterating the
argument of \citet{2014MNRAS.439.3239K} regarding the number of independent samples required to capture all phases
and the spatial scale of the processes being measured. It is also worth restating and recommending
the approach of placing constraints on the scale of possible variations rather than defaulting to a ``universal"
conclusion.

There is a further point to be made, picking up on the PDF approach of \citet{2014ApJ...795L..25R}. They note that gas dense
enough to form stars is dense enough to become self-gravitating and undergo runaway collapse. This shows up as a
power-law tail deviation, at the high column-density end, from the otherwise log-normal form of the PDF.
If the gas that will go on to form stars can be identified in this simple and direct way instead, the CMF as an entity,
with all the challenges associated in measuring it, is perhaps not a physically useful quantity.

Such a conclusion reinforces the poorly-posed nature of the definition of such mass
functions. What defines the star forming region of interest over which the CMF or IMF is to be measured? For
gas clouds that have a continuum of densities the challenge in defining boundaries or thresholds (such as with
various clump-finding software tools) is clear \citep[e.g.,][]{2014prpl.conf...53O}, but the PDF approach sidesteps that limitation.
The problem for stars, though, may not be so readily apparent, since for a cluster it would seem straightforward to
focus, for example, on the gravitationally bound stars as a single entity. But even this kind of simple scenario has
been seen to suffer from issues such as mass segregation, dynamical evolution, and so on, leading to systematics
affecting any IMF measurement. This raises the broader concern
of whether the IMF itself is a well-posed concept. If it does not exist as a physical distribution
at any given point in time \citep{2009eimw.confE..14E,2013pss5.book..115K}, and the spatial region over which it is to be measured is
unclear, is there a better entity that can be more well-defined instead? I return to this point in \S\,\ref{consistency} below.

Having raised again the point regarding the spatial scale being probed, I move next to the approaches used in estimating
the IMF for galaxies as a whole.

\section{IMF MEASUREMENT APPROACHES: INTEGRATED GALAXY TECHNIQUES}
\label{galaxy}
\subsection{Background}
Techniques for inferring an IMF for a galaxy, as opposed to a star cluster or stellar population, date back to
early population synthesis work in the 1960s
\citep{1962ApJ...135..715S,1966ApJ...145...36W,1966PASP...78..367S,1971ApJS...22..445S} that claimed an excess of
M dwarfs in the nuclei of nearby galaxies. \citet{1966ApJ...145...36W} also claim a proportionally larger number of giants for
the spiral arms of NGC 224, NGC 3031 (M81), and NGC 5194, ``essentially identical with that for the Solar neighbourhood."
This approach synthesised galaxy spectra by combining stellar spectra to reproduced the observed galaxy colours
and certain spectral features including Mg b and Na D, as well as molecular bands including TiO and CN. These
early analyses were focused largely on understanding the stellar populations and explaining the mass-to-light ratio
for these systems, rather than constraining the IMF explicitly. Subsequently, \citet{1977ApJ...211..527W} showed that the
Wing-Ford band was also a sensitive probe of the dwarf-to-giant ratio, and demonstrated that it could be used to
constrain the IMF for old stellar populations within galaxies. This approach has been revived
\citep[e.g.,][]{2012ApJ...760...70V} and is now used routinely to infer IMF properties for early-type galaxies.

A different kind of approach was used by \citet{1983ApJ...272...54K}, who introduced a diagnostic comparing
the equivalent width of H$\alpha$ (sensitive to the presence of high mass stars) to an optical colour index
(sensitive to low mass stars), as a probe of the IMF. Using this approach, \citet{1994ApJ...435...22K} showed that the
high mass slope of the IMF in nearby spiral galaxies is flatter ($-2.5 \lapp \alpha_h \lapp -2.35$) than the Solar
neighbourhood IMF of \citet{1979ApJS...41..513M} or \citet{1986IAUS..116..451S}
($-3.3 \lapp \alpha_h \lapp -2.7$). They attribute this at least in part to a deficiency of high mass stars
in the small volume of the Galaxy used to construct the local IMF, noting that \citet{1993AJ....106.1471P} find flatter
slopes for the high mass end of the IMF in 30 Dor in the LMC.

Yet another approach is the H$\alpha$-to-UV flux ratio, originally proposed by \citet{1987A&A...185...33B}
as an IMF probe. They used a sample of 31 late type galaxies to find a high mass
($m>1.8\,M_{\odot}$) IMF slope ranging over $-3.1<\alpha_h<-2.3$. They note that the dispersion in slope is
smaller, $\pm0.25$, for the sub-sample of 17 galaxies most similar to the Milky Way. Other approaches, including
UV luminosities and mass-to-light ratio techniques, have been summarised by \citet{1986IAUS..116..451S} and
\citet{1998ASPC..142....1K}. Subsequently many of these approaches have been refined and used extensively. More
recent approaches include the IGIMF approach of \citet{2003ApJ...598.1076K}, which has been explored extensively
over the past decade \citep[e.g.,][]{2017A&A...607A.126Y}.

Different kinds of integrated galaxy techniques have been used for different kinds of galaxies. The
\citet{1983ApJ...272...54K} and \citet{1987A&A...185...33B} diagnostics, and the use of
UV luminosities directly as an IMF probe have focused on currently star
forming galaxies, while the mass-to-light ratio and dwarf-to-giant ratio approaches have focused on
passive galaxies. This is natural given the requirements or limitations of each technique, but it causes
difficulty in directly comparing the results between the approaches. For now I summarise the
various methods based on the galaxy type they are used to probe, and explore the comparisons between them
further below in \S\,\ref{limitations} and \S\,\ref{discussion}.

\subsection{Star forming galaxies}
The IMF for galaxies with active star formation, even at a low level, is able to be probed with diagnostics relying
on emission associated with young high mass stars. The early work by
\citet{1983ApJ...272...54K} and \citet{1994ApJ...435...22K} was extended by \citet{2008ApJ...675..163H} using data
from the Sloan Digital Sky Survey DR4 \citep{2006ApJS..162...38A}. They concluded that the
sample as a whole was best fit by an IMF with high mass ($m>0.5\,M_{\odot}$) slope $\alpha_h=-2.45$ with negligible
random error and systematic error of $\pm0.1$. While noting that a constraint on $\alpha_h$ is degenerate
with one on $m_u$, they also showed that brighter galaxies are better fit
with a slightly flatter slope of $\alpha_h \approx -2.4$, and that fainter galaxies are not well described
by a ``universal" IMF, inferring fewer massive stars leading either to steeper $\alpha_h$ or lower $m_u$.
They point out that stochastic SFHs may mimic or contribute to this kind of signature,
although by testing models for such stochastic SFHs they conclude that to reproduce the measurements for
the observed population of bright galaxies would require an ``implausible coordination of burst times".

\begin{figure*}[ht]
\begin{center}
\includegraphics[width=14cm, angle=0]{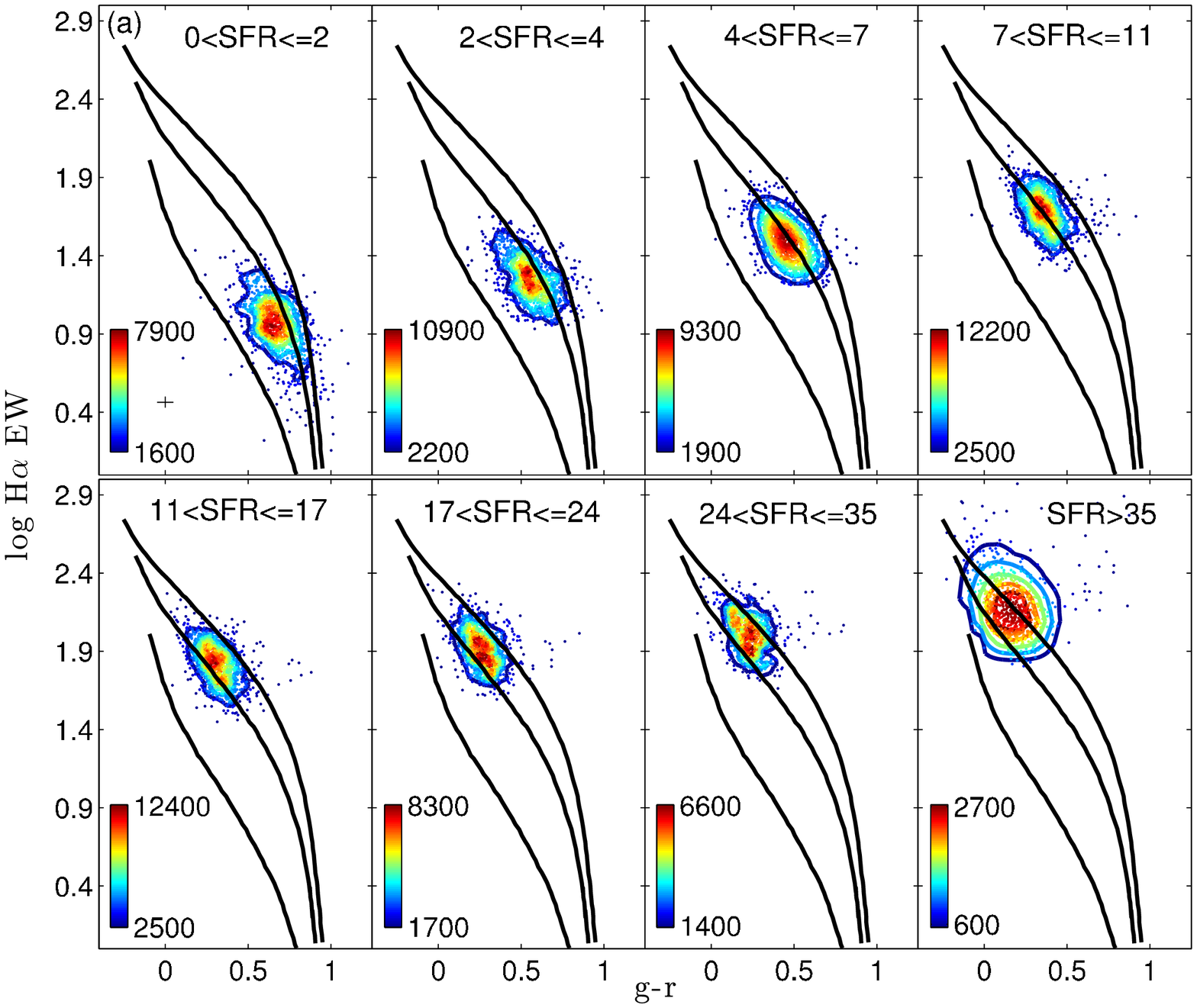}
\caption{Following \citet{1983ApJ...272...54K} and \citet{2008ApJ...675..163H}, this diagnostic shows how IMFs
with different $\alpha_h$ can be discriminated using the equivalent width of H$\alpha$ and an optical colour, in
this case $(g-r)$. The solid tracks are the evolutionary paths followed through a star formation event, showing
(top to bottom) the location expected for $\alpha_h=-2$, $\alpha_h=-2.35$, $\alpha_h=-3$. The data correspond
to galaxies in the highest redshift bin of the three volume limited samples, with $\langle z \rangle = 0.29$, split into
eight bins of SFR. This illustrates the tendency for the higher SFR systems to favour IMFs with flatter high mass
slopes (more positive $\alpha_h$). See \citet{2011MNRAS.415.1647G} for details. Reproduced from Figure~6a of
``Galaxy and Mass Assembly (GAMA): the star formation rate dependence of the stellar initial mass function,"
\citet{2011MNRAS.415.1647G}.}\label{G11fig6a}
\end{center}
\end{figure*}

The same method was used by \citet{2011MNRAS.415.1647G} with data from the Galaxy And Mass Assembly survey
\citep{2011MNRAS.413..971D,2013MNRAS.430.2047H,2015MNRAS.452.2087L}. Using three independent volume-limited
galaxy samples spanning $0.1\lapp z \lapp0.35$, they find a trend between the high mass ($m>0.5\,M_{\odot}$) IMF slope
and SFR (Figure~\ref{G11fig6a}) or SFR surface density ($\Sigma_{\rm SFR}$).
With the large sample available they are able to explore in detail the dependency of $\alpha_h$ on these properties,
which they quantify through the relations $\alpha_h~\approx~0.36 \log ({\rm SFR}/M_{\odot}\,{\rm yr}^{-1})~-~2.6$
or $\alpha_h~\approx~0.3 \log (\Sigma_{\rm SFR}/M_{\odot}\,{\rm yr}^{-1}\,{\rm kpc}^{-2})~-~1.7$.
For the range of SFR or $\Sigma_{\rm SFR}$ they probed, this leads to a variation in the high mass IMF slope
from $\alpha_h \approx -2.5$ for the least active star forming systems (${\rm SFR}\approx 0.1\,M_{\odot}$\,yr$^{-1}$)
to $\alpha_h \approx -1.7$ for the most active (${\rm SFR}\approx 50\,M_{\odot}$\,yr$^{-1}$).
They point out that for the current SFR of the Milky Way, these results would imply a value of $\alpha_h\approx-2.35$,
which is an encouraging consistency check. They also comment that the SFR dependence would imply a flatter
high mass IMF slope ($\alpha_h>-2.35$) in the Milky Way's early history, given that it had an elevated SFR in the past.

\citet{2011MNRAS.415.1647G} note that while the SFR is an IMF-dependent quantity, applying an IMF-dependent SFR
calibration would not qualitatively alter their conclusions, since the variation seen is monotonic, and would only have the
effect of reducing the range of SFR sampled. \citet{2011MNRAS.415.1647G} also explore degeneracies in the variation
of the IMF that would lead to the same observed combination of
H$\alpha$ equivalent width and $g-r$ colour. They show that while the high mass slope can be fixed at
the Salpeter value ($\alpha_h=-2.35$) and the results explained by allowing $m_c$ to increase, it requires
a very high value of $m_c\approx 10\,M_{\odot}$ to account for the highest star forming systems,
reminiscent of the early results for starburst nuclei \citep{1980ApJ...238...24R} that are no longer favoured
\citep{2007ASPC..362..269E}.

\begin{figure}[ht]
\begin{center}
\vspace{3mm}
\includegraphics[width=7cm, angle=0]{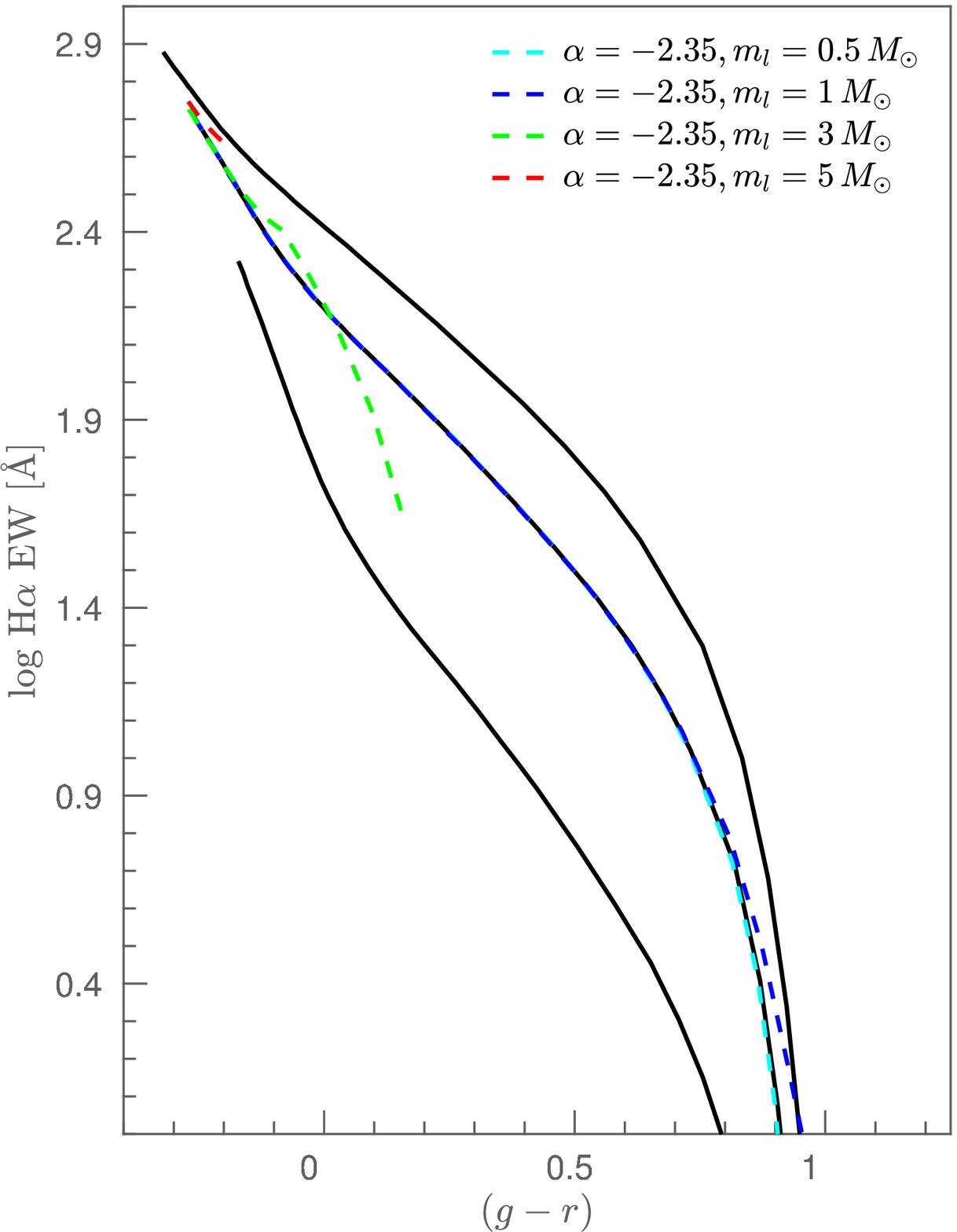}
\caption{The IMF diagnostic used by \citet{2011MNRAS.415.1647G} to identify variations in the slope of the high mass
($m>0.5\,M_{\odot}$) IMF. The black solid lines show the evolutionary tracks expected for galaxies with, from top
to bottom, $\alpha_h=-2,-2.35,-3$. The additional tracks (dashed coloured lines) illustrate the
effect of a fixed high mass slope ($\alpha=-2.35$) but varying $m_l$. The tracks become shorter as $m_l$ increases,
due to the shorter lifetimes of the higher mass stars. Figure courtesy of M. Gunawardhana.}\label{tracks}
\end{center}
\end{figure}

Again to rule out the possibility that stochastic SFHs could be the origin of the observed signature,
\citet{2011MNRAS.415.1647G} extended the analysis of \citet{2008ApJ...675..163H} by demonstrating that both stellar
mass and mass-doubling time for the observed galaxies vary smoothly along the SPS model tracks. They found
no signature corresponding to the significant bursts of star formation that would be seen if stochastic SFHs were the
dominant effect. Subsequently, \citet{2017MNRAS.468.3071N}, too, demonstrated quantitatively that stochastic star
formation histories could not explain their measurements, in a sample at much higher redshift.
\citet{2011MNRAS.415.1647G} further demonstrate
that the result is robust to the implementation of dust correction and choice of population synthesis model.
While there is a degeneracy between a varying high mass $\alpha_h$ and a varying $m_c$, it is possible to
exclude a variation in $m_l$ (Figure~\ref{tracks}). It is clear that a varying low mass cutoff has a very different
signature in this diagnostic than a varying high mass slope or $m_c$. This confirms that the results of
\citet{2011MNRAS.415.1647G} are not attributable to variations in $m_l$ while maintaining a Salpeter value for
the high mass slope. 

These results were extended to $z\approx 2$ by \citet{2017MNRAS.468.3071N}, finding that systems with the highest
H$\alpha$ equivalent widths could only be explained either by rotating extremely metal poor stars with binary
interactions, or IMFs with $\alpha_h>-2$ ($0.5<m/M_{\odot}<120$). They note that in the latter case, no single IMF slope
could reproduce the data, implying a need for a stochastically varying high mass IMF slope. They explore
the trend with SFR derived from H$\alpha$ and from the combined UV and far-infrared (FIR) luminosities, finding
a weak trend with the H$\alpha$ SFR in the same sense as that of \citet{2011MNRAS.415.1647G}, but none with
the SFRs from the UV+FIR (their Figure~21). This arises from the
lowest H$\alpha$ SFR systems having higher SFRs as measured by the UV+FIR. In order to avoid internal inconsistencies,
it would be valuable for all analyses of this kind to present results in terms of IMF-independent quantities (such as
luminosities) or to calculate SFRs or other IMF-dependent properties self-consistently assuming whatever
IMF-dependency on $\alpha_h$ or other parameter is being tested.

At similar redshifts, \citet{2018Natur.558..260Z} measure the $^{13}$C/$^{18}$O abundance ratio,
probed through the rotational transitions of the $^{13}$CO and C$^{18}$O isotopologues, for four starburst galaxies.
These galaxies are gravitationally lensed submillimetre galaxies spanning $2.3<z<3.1$.
They find low ratios compared to chemical evolutionary models for the Milky Way, implying considerably more high
mass stars in such starbursts than in typical spiral galaxies, and with a high mass slope flatter than that
of \citet{2001MNRAS.322..231K}.

At the opposite end of the scale, the \citet{1987A&A...185...33B} approach relying on H$\alpha$-to-UV flux ratios has more
recently been used to focus on potential variations to the IMF for low mass or low SFR galaxies. The results of
\citet{2009ApJ...695..765M}, \citet{2009ApJ...706..599L}, and \citet{2009ApJ...706.1527B} all suggest that for low SFR
(or low luminosity, or low mass) galaxies,
there is evidence of a deficiency of high mass stars, characterised as an IMF slope steeper than Salpeter
($\alpha<-2.35$ for $0.1<m/M_{\odot}<100$). As the H$\alpha$ luminosity (or surface density) decreases within
various samples of nearby galaxies, these authors show that the H$\alpha$ luminosity (or flux) declines
faster than that of the UV. This is in the opposite sense than could be explained through dust obscuration, which affects
the UV proportionally more than H$\alpha$, although different levels of attenuation for the line and continuum
emission may play some role \citep[e.g.,][]{2000ApJ...539..718C}. \citet{2009MNRAS.395..394P} demonstrate how this
result can arise within the IGIMF formalism.

After exploring and ruling out a range of possible scenarios, including dust obscuration, SFH,
stochastic population of the IMF, metallicity, and escape fraction, \citet{2009ApJ...695..765M} conclude that the most likely
scenario is a variation in the high mass end of the IMF, with either $m_u$ ranging over
$30\lapp m_u/M_{\odot} \lapp 120$ or having a varying IMF slope spanning $-3.3 \lapp \alpha \lapp -1.3$
(for $0.1<m/M_{\odot}<100$), in the sense that the lower SFR systems have either lower $m_u$ or steeper $\alpha$.
It is worth noting the use of the single power law IMF slope extending to the lowest masses here,
and questioning whether a slope change below some $m_c$ would affect the results. Since the diagnostic uses flux
ratios of indicators sensitive only to high mass stars, though, it seems unlikely that changing this assumption would
alter the IMF conclusions significantly. 

The results of \citet{2009ApJ...695..765M} are qualitatively similar to those of \citet{2011MNRAS.415.1647G}, with a high
mass IMF slope that gets progressively steeper as $\Sigma_{\rm SFR}$ becomes lower. Quantitatively, though,
the results are somewhat different, with \citet{2009ApJ...695..765M} inferring steeper IMF slopes at a given value of
$\Sigma_{\rm SFR}$ (or equivalently, $\Sigma_{{\rm H}\alpha}$) than \citet{2011MNRAS.415.1647G}, when retaining the
same $m_u$. Specifically, as an example, at $\log(\Sigma_{\rm SFR}/M_{\odot}\,{\rm yr}^{-1}\,{\rm kpc}^{-2})=-2.7$
\citet{2011MNRAS.415.1647G} find $\alpha_h\approx-2.4$ (their Figure~13c). This value of $\Sigma_{\rm SFR}$
corresponds to $\log(\Sigma_{{\rm H}\alpha}/{\rm W}\,{\rm kpc}^{-2})=31.4$, for which \citet{2009ApJ...695..765M} find a
typical value of $\log(({\rm F}_{{\rm H}\alpha}/{\rm f_{FUV}})/{\rm \AA})\approx 0.65$ (their Figure~3), and which leads to
$\alpha_h\approx -3.0$ (their Figure~10b). I return to exploring the details of these quantitative differences below. 

Again in a qualitatively similar fashion, \citet{2009ApJ...706.1527B} find that the IMF slope (for $0.1<m/M_{\odot}<100$)
spans $-2.6\lapp \alpha \lapp -2.3$, with steeper slopes for the lower mass galaxies. Comparing quantitatively to
\citet{2011MNRAS.415.1647G}, these slopes are closer than inferred by \citet{2009ApJ...695..765M} but still rather
steeper for a given stellar mass or sSFR. Based on Figure~10 of \citet{2009ApJ...706.1527B}, $\alpha=-2.5$ is favoured
for galaxies with $9.2<\log(M/M_{\odot})<9.8$, for example. This corresponds broadly to a range of
$-10\lapp \log({\rm sSFR}/{\rm yr}^{-1}) \lapp-9$ \citep[see Figure~4 of][]{2013MNRAS.434..209B} for the GAMA sample in the
redshift range analysed by \citet{2011MNRAS.415.1647G}. For this range of sSFR, \citet{2011MNRAS.415.1647G} find high
mass IMF slopes spanning $-2.4\lapp \alpha_h \lapp -2.1$ (their Figure~13b), somewhat flatter than the
$\alpha=-2.5$ of \citet{2009ApJ...706.1527B}.

\citet{2009ApJ...706.1527B} invoke bursty SFHs to explain their findings in preference to IMF variations,
contradicting the claim by \citet{2009ApJ...695..765M} that the ``gasps" between such bursts would have to be unrealistically
synchronised between galaxies to produce the range of measured flux ratios. \citet{2009ApJ...706.1527B}
note that the star formation in dwarf galaxies can
be dominated by a single H{\sc ii} region, which can lead to larger scatter in the H$\alpha$ to UV flux ratio than
for high mass galaxies, arising from the different lifetimes of the OB stars ($\sim 10^6$\,yr) responsible for the
H$\alpha$ emission and the A stars ($\sim 10^7$\,yr) responsible for the UV emission.

\citet{2009ApJ...706..599L} use a nearby sample of galaxies to probe to extremely low
levels of SFR, down to ${\rm SFR} \approx 10^{-4}\,M_{\odot}\,{\rm yr}^{-1}$, and see very similar effects to
the results of \citet{2009ApJ...695..765M} and \citet{2009ApJ...706.1527B}. They
do not attempt to constrain the IMF directly but instead test (and rule out) a variety of explanations,
while noting that the IGIMF predictions of \citet{2009MNRAS.395..394P} match their observations surprisingly well, before also
implying a preference for SFH variations (bursty or ``flickering") as a possible
explanation \citep[see also][]{2011ASPC..440..179L}. In contrast, \citet{2011ApJ...741L..26F} argue, based on a
stochastic code for synthetic photometry, that these observed H$\alpha$-to-UV flux ratios in low SFR galaxies can be
explained by random sampling from a ``universal" IMF, combined with stellar evolution. A similar result was found
by \citet{2012ApJ...744...44W} who note that such stochastic SFHs can explain the observed
H$\alpha$ to UV flux ratios in low mass galaxies, without invoking IMF variations.

Bearing in mind the likely contributions of stochastic SFHs, there does seem to be some evidence for
the idea that systems with lower luminosities, masses or SFRs favour IMFs lacking in high mass stars relative to a
\citet{2001MNRAS.322..231K} or \citet{2003PASP..115..763C} IMF. This is reinforced by studies of the LMC, SMC,
and other dwarf, low metallicity, or low surface brightness galaxies.
Using UV imaging of the LMC and models jointly constraining IMF slope,
obscuration and age, \citet{1998AJ....116..180P} find $\alpha_h=-2.80\pm0.09$ (for $m\gapp 7\,M_{\odot}$).
\citet{2013ApJ...763..101L} find $\alpha_h=-3.3\pm0.4$ (for $m>20\,M_{\odot}$) in the SMC based on spectra for a
spatially complete census of field OB stars. The dwarf starburst galaxy NGC 4214 was found by
\citet{2007AJ....133..932U} to have $\alpha_h<-2.83\pm0.07$
(for $20<m/M_{\odot}<100$), the upper limit arising due to the presence of unresolved binaries, the neglect
of which acts to flatten the inferred IMF. \citet{2004MNRAS.353..113L} use the high M/L ratios for seven low surface
brightness disk galaxies to estimate $\alpha=-3.85$ (for $0.1<m/M_{\odot}<60$).
\citet{2017IAUS..329..313G} summarise studies of high mass stars in Local Group dwarf galaxies with very low metallicity,
oxygen abundances less than 1/7 of the solar value. Although relying on a sample of only four galaxies, with incomplete
observations of the high mass stellar population, they note that the highest mass stars so far identified
have initial masses of $m \approx 60\,M_{\odot}$.
\citet{2015MNRAS.447..618B} explore the IMF of the dark-matter dominated,
extremely low SFR, blue compact dwarf galaxy NGC 2915, and reinforce the possibility of
high mass IMF slopes rather steeper than Salpeter for such systems. They combine colour magnitude diagrams
for the stellar populations with an assumed recent SFH to find a high mass IMF slope $\alpha_h=-2.85$
(for $m\gapp 4\,M_{\odot}$) and a poorly constrained upper mass limit of $m_u=60\,M_{\odot}$.
Noting the impact of assuming a constant recent SFH \citep{2006ApJ...636..149E}, the IMF may not be quite this extreme,
although it is not inconsistent with the other results highlighted here, in similar environments.
The same approach, assuming a constant star formation rate over the dynamical timescale, was used in inferring the
IMF of the dwarf irregular galaxy DDO 154 by \citet{2018MNRAS.477.5554W}, to derive $\alpha_h=-2.45$ with a
similarly poorly constrained $m_u=16\,M_{\odot}$.

For the population of ultra-compact dwarf galaxies (UCDs), known to have high $V$-band
mass-to-light ratios ($\Upsilon_V$), \citet{2009MNRAS.394.1529D} use population synthesis modelling to infer
$\alpha_h$, after arguing that the non-baryonic dark matter contribution is too low to influence the dynamics of these
systems \citep{2009ApJ...691..946M}. Contrary to the results above for other dwarf galaxy systems, they
show that an IMF with a relative excess of high mass stars, $\alpha_h \approx -1.6$ to $\alpha_h \approx -1.0$
(depending on the assumed age), is required to account for the observed $\Upsilon_V$. This perhaps highlights
star formation rate density as a significant factor in shaping the IMF. In contrast to the local dwarf systems, which are
typically low surface density galaxies, the UCDs are expected to have formed rapidly, with SFRs as high as
perhaps $10-100\,M_{\odot}\,$yr$^{-1}$, and a correspondingly high $\Sigma_{\rm SFR}$ given their small physical sizes.
For a physical scale of 10\,pc this would give $5 \lapp \log(\Sigma_{\rm SFR}/M_{\odot}\,{\rm yr}^{-1}\,{\rm kpc}^{-2})\lapp 6$,
substantially higher than in the population of higher mass galaxies discussed above.
The elevated values of $\alpha_h$ inferred are qualitatively consistent with the results
for higher mass star forming galaxies, although any potential dependence of $\alpha_h$ on
$\Sigma_{\rm SFR}$ is quantitatively inconsistent with the nominal linear relations of \citet{2009ApJ...695..765M},
\citet{2011MNRAS.415.1647G} or \citet{2017MNRAS.468.3071N}.

Using UV spectral lines as a constraint, \citet{2011ASPC..440..309L} combines measurements for a sample of 28 nearby
galaxies (distances less than 250\,Mpc) to construct the average UV spectrum, and concludes that
the high mass ($m>0.5\,M_{\odot}$) IMF slope is constrained to lie between $-2.6\lapp \alpha_h \lapp -2.0$.
This range encompasses most of the variation in $\alpha_h$ inferred from the studies described above,
and consequently does not substantially rule out the proposed variations.

With the exception of the dwarf galaxies and low surface brightness galaxies, which show rather steeper high
mass ($m>0.5\,M_{\odot}$) IMF slopes (although noting the opposite result for UCDs, which are perhaps
more analogous to starburst systems), the range spanned by $\alpha_h$ from these
investigations is not large, with most galaxies having $-2.5\lapp \alpha_h \lapp -1.8$ \citep[e.g.,][]{2011MNRAS.415.1647G},
similar to the range seen in nearby stellar populations and star clusters, and variously attributed to
observational or astrophysical limitations. It is notable that \citet{2011MNRAS.415.1647G} place
the Milky Way on their relationship between $\alpha$ and SFR or $\Sigma_{\rm SFR}$, finding a
Salpeter IMF slope ($\alpha_h=-2.35$) consistent with the bulk of analyses of Galactic stellar populations.

In summary, the observations for star forming galaxies suggest a broadly consistent picture in favour of IMF
variations, with $\alpha_h$ larger (flatter) for higher SFR or $\Sigma_{\rm SFR}$, and smaller (steeper) for
lower SFR or $\Sigma_{\rm SFR}$. This is in line, qualitatively at least, with the results seen above for
starburst or super star clusters.

\subsection{Passive galaxies}
\label{passive}
The descriptor ``passive" is intended here to refer to the degree of star formation, or rather its absence at any
significant level, and is independent of the existence or not of an active galactic nucleus (AGN).
The methods of \citet{1983ApJ...272...54K} and \citet{1987A&A...185...33B} are not typically able to be applied to
passive galaxies, as they have little or no H$\alpha$ and UV emission arising from high mass young stars.
Where such emission is present it is likely to arise from, and be dominated by, an AGN rather than star formation.
Also, since the stars being probed in these passive galaxies are only those low mass objects remaining from
star formation episodes much earlier, any IMF measured in such systems is in a sense a ``relic" IMF,
analogous to the PDMF in Milky Way stellar systems, and in many cases this has led naturally to a focus
on the low mass end of the IMF.

A novel approach was explored by \citet{2008ApJ...674...29V}, noting the opposing effect of the IMF on
luminosity and colour evolution identified by \citet{1980FCPh....5..287T}, such that an IMF with proportionally
more high mass stars would lead to stronger luminosity evolution and weaker color evolution.
He compared the rate of luminosity and colour evolution for high mass
elliptical galaxies in clusters spanning $0\le z \le 0.83$, and found a need for an IMF with
$\alpha=-0.7^{+0.7}_{-0.4}$ at masses around $m\approx 1\,M_{\odot}$, much flatter than a \citet{2001MNRAS.322..231K}
or \citet{2003PASP..115..763C} IMF at these masses, to explain the observations. Casting this as an estimate of
$m_c=1.9^{+9.3}_{-1.2}\,M_{\odot}$ at $z=3.7^{+2.3}_{-0.8}$ (the estimated formation redshift of stars at
this stellar mass) for a Chabrier-like IMF, and comparing with the
lower inferred values of $m_c$ at lower redshift, he goes on to explore the impact
of an evolving $m_c$. He notes the effect of reducing the apparent discrepancy between the cosmic SFH and
stellar mass density, confirming similar results \citep{2006ApJ...651..142H,2007MNRAS.379..985F,2008MNRAS.385..687W}, and which I discuss in detail below in \S\,\ref{census}. The result of \citet{2008ApJ...674...29V} was questioned by
\citet{2012ApJ...760...70V} who show instead that
when comparing galaxies at a given velocity dispersion, rather than stellar mass, the observed luminosity and
colour evolution is consistent with the standard Salpeter slope.

More recent approaches have focused on using spectral signatures sensitive to the dwarf-to-giant ratio,
stellar mass-to-light ratios and kinematics to explore the inferred IMF.

The dwarf-to-giant ratio approach evolved from early work using Na {\sc i} D lines as a tracer of dwarf star populations
\citep{1962ApJ...135..715S} which concluded that the most luminous and high mass elliptical galaxies showed evidence
for proportionally more dwarf stars than found in lower mass galaxies. Additional spectral features (Mg {\sc i}, Ca {\sc i},
Ca {\sc ii}, TiO, CN, CH, CaH, MgH) sensitive to different stellar populations
\citep{1966PASP...78..367S,1966ApJ...145...36W,1971ApJS...22..445S} enabled improvement of the ability to characterise
the stellar populations present in integrated galaxy spectra. \citet{1977ApJ...211..527W} incorporated measurements
of the Wing-Ford molecular band at 9910\,\AA\ to argue against the earlier results that favoured an excess of dwarf stars
in such galaxies, ruling out IMFs with $\alpha \le -3$, and finding results supporting an IMF with $\alpha \approx -2$.

\begin{figure*}[ht]
\begin{center}
\includegraphics[width=16cm, angle=0]{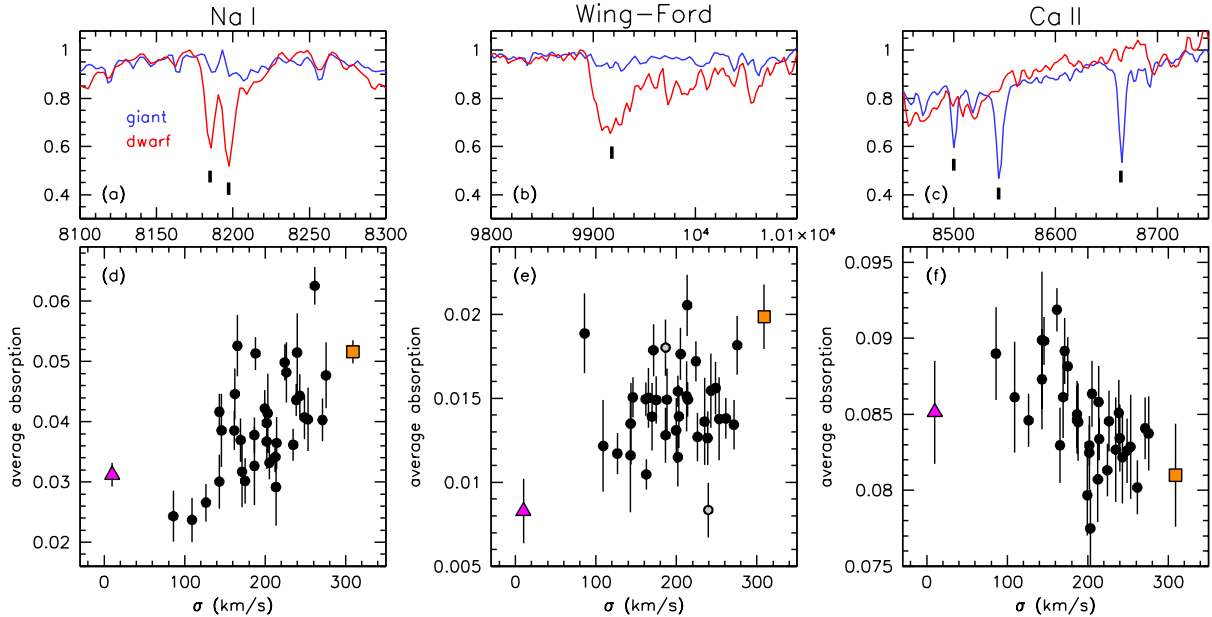}
\caption{Three spectral regions showing features sensitive to the presence or absense of low mass stars
(upper panels), and the trend in the absorption strength of those feature seen with velocity dispersion of the
galaxies (lower panels). This demonstrates that galaxies with higher velocity dispersion, and hence higher stellar mass,
have a tendency to favour an excess of dwarf, or low mass, stars. See \citet{2012ApJ...760...70V} for details.
(Figure~10 of ``The stellar initial mass function in early-type galaxies from absorption line spectroscopy. I.
Data and empirical trends," \citet{2012ApJ...760...70V}, \copyright~AAS. Reproduced with permission.)}\label{vCD12fig10}
\end{center}
\end{figure*}

This general technique has been developed significantly
through recent work \citep[e.g.,][]{2010Natur.468..940V,2012ApJ...760...70V}
and is illustrated by Figure~\ref{vCD12fig10}, which has been reproduced from \citet{2012ApJ...760...70V}.
A key result was that of \citet{2010Natur.468..940V}, who apply the high resolution SPS models
of \citet{2009ApJ...699..486C} to estimate the proportion of low mass dwarf stars ($m<0.3\,M_{\odot}$)
necessary to explain the observed absorption in
the Na{\sc i}$\lambda\lambda$\,8183,8185\AA\ doublet and the Wing-Ford FeH band around $\lambda$\,9916\AA\
for eight nearby elliptical galaxies in the Virgo and Coma
clusters. They infer $\alpha_l \approx -3$ for the low mass range
($0.1<m/M_{\odot}<0.3$), significantly steeper than the Salpeter slope. As these stars would have formed at
high redshift ($z= 2 - 5$), they argue against high redshift IMFs with a deficit of low mass stars such as the
truncated IMF of \citet{2005MNRAS.356.1191B}. Recall that for the Milky Way $-1.4 \lapp \alpha_l \lapp -0.6$
from the compilation of \citet{2010ARA&A..48..339B}.
\citet{2012ApJ...760...70V} and \citet{2012ApJ...760...71C} extended this result to the bulge of M31 and 34 galaxies
from the SAURON integral field spectroscopy sample \citep{2001MNRAS.326...23B}, adding the
Ca{\sc ii} $\lambda\lambda$\,8498,8542,8662\AA\ triplet to
the diagnostics, and concluding that the low mass $m<0.3\,M_{\odot}$ IMF slope varies systematically with galaxy
velocity dispersion and $\alpha$-enhancement, with steeper slopes in more massive and high-abundance galaxies
(Figure~\ref{vCD12fig10}).

Using similar approaches, both \citet{2012ApJ...753L..32S} and \citet{2013MNRAS.429L..15F} infer IMF slopes for the
same mass range steeper than
Salpeter ($\alpha_l \approx -3$) for high velocity dispersion systems ($\sigma\gapp 300$\,km\,s$^{-1}$), and
\citet{2013MNRAS.433.3017L} also find IMF slopes steeper than Salpeter for systems with $\sigma\gapp 220\,$km\,s$^{-1}$.
\citet{2014ApJ...780L...1Z} find a strong correlation between UV colour and $\Upsilon_*$ for the same sample of galaxies
analysed by \citet{2012ApJ...760...70V}, and conclude that this correlation is attributable to varying populations
of extreme horizontal branch stars arising from the IMF variations.

For lower stellar mass systems, \citet{2012MNRAS.426.2994S} analysed 92 red-sequence galaxies in the Coma cluster,
including more galaxies with lower velocity dispersions than \citet{2012ApJ...760...70V} and \citet{2012ApJ...760...71C}.
They found no clear dependence of IMF slope on velocity dispersion with a Salpeter slope adequate for
$100\lapp \sigma/{\rm km\,s}^{-1} \lapp 250$ (although they do not rule out steeper IMF slopes at higher velocity dispersions),
but they do see a dependence on $\alpha$-enhancement (Mg/Fe ratio), concluding that the IMF variation arises as a result
of star formation mode, with rapid bursts leading to proportionally more low-mass stars ($m<0.3\,M_{\odot}$).

In parallel with these analyses, gravitational lensing and dynamical constraints were being explored as a tool for inferring
mass-to-light ratios ($\Upsilon$) and placing associated constraints on IMF slopes. This approach was refined by
\citet{2010ApJ...709.1195T} who infer $\Upsilon$ for 56 galaxies from the SLACS survey
\citep{2006ApJ...638..703B}. They introduce an IMF mismatch parameter, also denoted $\alpha$, which is defined
to be the ratio of $\Upsilon$ inferred independently from the lensing and dynamical analysis
compared to that from SPS modelling. For clarity, I refer to the IMF mismatch parameter as
$\alpha_{mm}=\Upsilon_{\rm LD}/\Upsilon_{\rm SPS}$ throughout.

\citet{2010ApJ...709.1195T} find that, with the assumption of an NFW dark matter profile \citep{1996ApJ...462..563N},
such galaxies tend to favour IMFs such as Salpeter that provide a higher $\Upsilon_*$ than those of
\citet{2003PASP..115..763C}. While stellar mass range is not discussed, the \citet{2003MNRAS.344.1000B} models
used span a mass range of $0.1\le m/M_{\odot}\le100$, and from the scale of the IMF mismatch parameter it
is reasonable to infer that the Salpeter IMF slope ($\alpha=-2.35$) is applied over that full range for their
Salpeter SPS mass estimates (confirmed by T.\ Treu, personal communication).
\citet{2010ApJ...709.1195T} also note that their sample is limited to relatively high velocity dispersion systems, and show
that, while for galaxies with $\sigma\approx 200$\,km\,s$^{-1}$ a \citet{2003PASP..115..763C} IMF provides consistent
$\Upsilon_*$ with the lensing and dynamical estimate, ``heavier" IMFs (i.e., with a greater total mass normalisation) are
required for higher velocity dispersion systems. They conclude that either the IMF varies toward one with a Salpeter-like
mass normalisation for the most massive galaxies, or that dark matter profiles are not universal and the inner slope
is systematically steeper than NFW for the most massive galaxies, or possibly a combination of both. This result
is still not as extreme as the $\alpha_l \approx -3$ found spectroscopically by \citet{2010Natur.468..940V}.

Similarly, \citet{2013MNRAS.434L..31L} use a combination of dynamical models and the SPS and IMF approach of
\citet{1996ApJS..106..307V,2012MNRAS.424..157V} to infer an IMF with a steep high mass slope,
($\alpha_h=-4.2\pm0.1$ for $m>0.6\,M_{\odot}$ constraining $\alpha_l=-1.3$
for $0.1<m/M_{\odot}<0.6$) in a low redshift ($z=0.116$) high mass ($\sigma = 360\,$km\,s$^{-1}$)
early type galaxy with a putative extremely high mass nuclear black hole. In apparent contrast to
these results, though, \citet{2013MNRAS.434.1964S} used gravitational lensing mass estimates to demonstrate that
$\Upsilon_*$ for a high mass ($\sigma\approx 330\,$km\,s$^{-1}$) low-redshift ($z=0.035$) giant elliptical is consistent
with a \citet{2001MNRAS.322..231K} IMF. This result was subsequently reinforced by \citet{2015MNRAS.449.3441S} who
analysed three high mass ($\sigma > 300\,$km\,s$^{-1}$) low redshift ($z\lapp 0.05$) galaxies, showing that
the inferred IMF mass normalisation is consistent with that of \citet{2001MNRAS.322..231K}, and excluding a Salpeter IMF
($\alpha=-2.35$ over $0.1<m/M_{\odot}<100$) at the $3.5\,\sigma$ level.

These results, though, are not inconsistent with the scatter seen by \citet{2010ApJ...709.1195T} in their IMF mismatch
parameter. It is worth reiterating that \citet{2010ApJ...709.1195T} use the observed small range of scatter on
$\alpha_{mm}$ to argue that ``the absolute normalization of the IMF is uniform to better than 25\%", a result that echoes
the relatively small range of potential variations found for the star forming galaxy population, although with a different sense
in the variation itself for the passive galaxies (i.e., higher mass passive galaxies favouring proportionally more low mass
stars, but higher mass star forming galaxies favouring proportionally more high mass stars).

There is a subtlety around gas recycling in SPS that affects the M/L ratio comparison technique. Whether the gas recycled
through stellar evolution is retained or not in the SPS inferred masses has a direct impact on the comparison to masses
inferred from lensing and dynamics. \citet{2010ApJ...709.1195T} explore
the two extreme cases, where all gas lost through stellar evolution is removed, and where it is retained
(the ``zero age" SPS mass). They note that for a \citet{2003PASP..115..763C} IMF the ``zero age" masses
tend to be overestimates compared to the lensing masses, implying that at least some fraction of the gas associated
with recycling is expelled and does not contribute to the baryonic mass in the region probed by the lensing
and dynamical constraints. In this and subsequent work, it is typically just the mass in stars and stellar remnants
derived from the SPS that is compared with the lensing and dynamical mass estimates, which
for early type galaxies with very low gas fractions is likely to be a reasonable assumption. It must be noted,
though, that such mass estimates may be lower limits if some gas component still contributes non-negligibly
to the baryonic mass, and needs to be accounted for in the uncertainties on inferred IMF constraints.

\begin{figure}[ht]
\begin{center}
\includegraphics[width=8.5cm, angle=0]{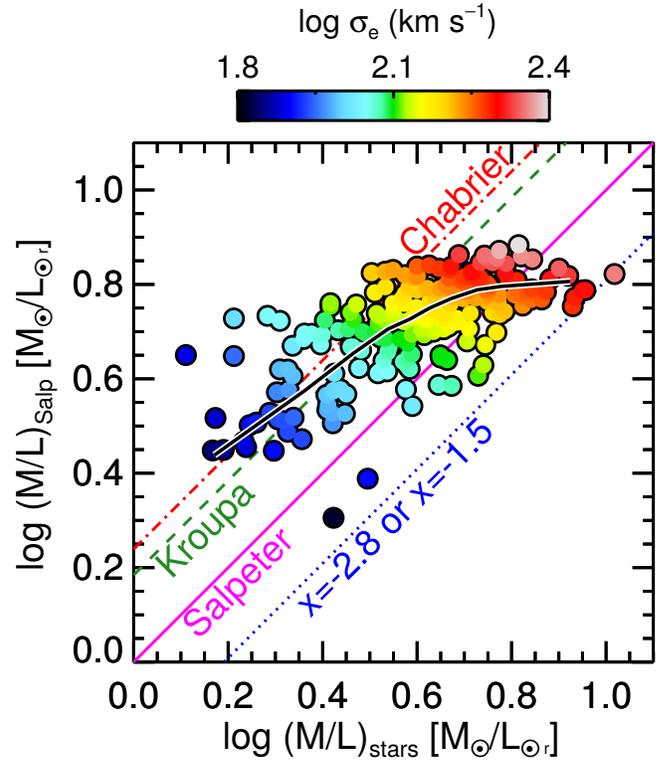}
\caption{The mass-to-light ratio for the stellar component of ATLAS$^{3{\rm D}}$ galaxies estimated using
dynamical models, $(M/L)_{\rm stars}$, compared to that estimated from spectral fitting using SPS models assuming
a fixed Salpeter IMF, $(M/L)_{\rm Salp}$. This demonstrates the trend for the high mass-to-light, or high velocity dispersion,
galaxies in this sample to favour IMFs with an excess of mass compared to the IMFs of \citet{2003PASP..115..763C} or
\citet{2001MNRAS.322..231K}, approaching and exceeding that from a Salpeter slope over the full mass range
(an excess of low mass stars). See \citet{2013MNRAS.432.1862C} for details.
Reproduced from Figure~11 of ``The ATLAS$^{3{\rm D}}$ project -- XX. Mass-size and mass-$\sigma$ distributions
of early-type galaxies: bulge fraction drives kinematics, mass-to-light ratio, molecular gas fraction and stellar initial mass
function," \citet{2013MNRAS.432.1862C}.}\label{C13fig11}
\end{center}
\end{figure}

Another significant development around the same time was the use of stellar kinematics and dynamical models to constrain
the IMF \citep{2012Natur.484..485C,2013MNRAS.432.1862C} illustrated in Figure~\ref{C13fig11}.
They used the ATLAS3D sample of 260 early-type galaxies
\citep{2011MNRAS.413..813C} combining the measured stellar kinematics with detailed axisymmetric
dynamical models, to derive accurate stellar masses and mass-to-light ratios for the population. By comparing the
mass-to-light ratio measured dynamically in this way to that inferred from the photometry using SPS models that
assume a Salpeter IMF over the full mass range ($0.1<m/M_{\odot}<100$), they are able
to show a systematic variation in the inferred IMF normalisation. This variation ranges from a mass
normalisation consistent with \citet{2003PASP..115..763C} or \citet{2001MNRAS.322..231K} at
$\Upsilon_* \approx 2\,M_{\odot}/L_{\odot}$ to one consistent
with a Salpeter slope spanning $0.1\le m/M_{\odot}\le100$ at $\Upsilon_* \approx 6\,M_{\odot}/L_{\odot}$
(Figure~\ref{C13fig11}). This may extend to even higher mass normalisations at the most extreme measured values of
$\Upsilon_* \approx 10\,M_{\odot}/L_{\odot}$, with an IMF characterised equally by $\alpha=-2.8$ (dominated
by low mass stars) or $\alpha=-1.5$ (dominated by high mass stars). This result has been questioned
by \citet{2015MNRAS.449.4091C}, though, who note that these trends could also be produced if the kinematic mass
estimates had Gaussian errors of the order of 30\%.

There are significant degeneracies possible in IMF shape when only the mass normalisation is constrained.
While this is highlighted for the very high mass normalisations by \citet{2013MNRAS.432.1862C}, that of a Salpeter IMF slope
($\alpha=-2.35$) over the full mass range can also be reproduced by an IMF with a Milky Way style low mass slope, and
an excess of high mass stars. By way of illustration, this is achieved (including only the mass in stars and stellar remnants)
by an IMF with $\alpha_l=-1.5$ ($0.1\le m/M_{\odot} \le 0.5$) and $\alpha_h=-1.70$ ($0.5\le m/M_{\odot} \le 100$), or
an IMF with the \citet{2001MNRAS.322..231K} low mass slope $\alpha_l=-1.3$ ($0.1\le m/M_{\odot} \le 0.5$) and $\alpha_h=-1.64$
($0.5\le m/M_{\odot} \le 100$). Different combinations can equally be used to match the zero-age mass normalisation.
The point is that a value of $\Upsilon_*$ consistent with a Salpeter IMF over the full mass range doesn't necessarily imply an
excess of low mass stars. To break this degeneracy, \citet{2013ApJ...776L..26C} quantitatively compared the scale of the IMF
mass normalisation derived using the dwarf-to-giant approach with that from dynamical mass constraints, and find that
they are consistent, inferring that the explanation lies in an excess of low mass stars ($m\lapp 1\,M_{\odot}$).

There have also been a range of enhancements and refinements that combine lensing, dynamical and SPS
constraints. These results largely confirm the need for a low mass IMF slope similar to
the Salpeter value for high mass early type galaxies. \citet{2015MNRAS.452L..21S}, for example, find a low mass slope of
$\alpha_l=-2.37\pm0.12$ and $m_l = 0.131^{+0.023}_{-0.026}\,M_{\odot}$ at a reference point corresponding
to $\sigma=250\,$km\,s$^{-1}$ for 9 early type galaxies, confirming earlier results by \citet{2013MNRAS.436..253B}.
In apparent contrast \citet{2016MNRAS.463.3220L}, using 27 early type galaxies from CALIFA \citep{2014A&A...569A...1W}
and the IMF parameterisation approach of \citet{1996ApJS..106..307V}, rule out a single power-law
IMF and conclude that a double power-law with a varying high-mass slope is required to explain the dynamical
and stellar M/L ratios. This result may not be inconsistent with most of the results above, to the degree that it is only
the mass normalisation that is being constrained through this approach, but it would be inconsistent with
the conclusions of \citet{2013ApJ...776L..26C}. Testing the consistency of IMF measurements between
the SPS and the lensing and dynamical approaches is clearly important, and some work in this area has already begun
\citep{2017ApJ...845..157N}. Using three strongly lensed passive galaxies, \citet{2017ApJ...845..157N} find consistent
IMF estimates between the two techniques for one galaxy, but require a variable low-mass cutoff or a nonparametric
form of the IMF to reconcile the approaches for the remaining two systems. Clearly, extending such analyses to
larger samples is desirable.

More recently these approaches have been extended to identify radial gradients in the IMF shape \citep{2017ApJ...841...68V}.
They characterise their results using a variant of the $\alpha_{mm}$ IMF mismatch parameter which, instead of
comparing dynamical to SPS M/L ratios, compares the $\Upsilon_*$ inferred from a given SPS fit to a canonical Milky Way
IMF such as \citet{2001MNRAS.322..231K} or \citet{2003PASP..115..763C} as a convenient shorthand for encapsulating the
relative mass normalisation. Using $\alpha_{mm}=\Upsilon_{\rm SPS}/\Upsilon_{\rm MW}$ in this way, they find IMFs with
mass normalisations heavier than \citet{2001MNRAS.322..231K} or \citet{2003PASP..115..763C}
($\alpha_{mm}\approx 2.5$) in the central regions for six early type galaxies, but approaching the Milky Way
value as radius increases, with $\alpha_{mm} \approx 1.1$ at $R>0.4\,R_e$. The SPS IMF constraint approach has also
been refined by \citet{2017ApJ...837..166C}, who developed a non-parametric approach to constraining the
$m<1\,M_{\odot}$ shape. Applying this to the centre of NGC 1407, they find an IMF slope consistent with $\alpha_l=-2.7$.
Such radial trends may provide an explanation for the differences found in the IMF properties between the
spectroscopic and the dynamical approaches. It is also a tantalising link to the postulated ``two-phase" formation scenario
for early type galaxies \citep[e.g.,][]{2017A&A...607A.128G}, which proposes that the cores of elliptical galaxies formed
quickly and quenched rapidly, (the high redshift ``red nuggets"), in contrast to their outer regions.
\citet{2015MNRAS.451.1081M} find a ``bottom-heavy" IMF ($1.5 \lapp \alpha_{mm} \lapp 2$) out to $1.5\,R_e$ in
NGC 1277, and argue that this is an example of the kind of ``core" that would evolve through dry merging to the
characteristic masses and sizes of $z\approx 0$ elliptical systems.

\citet{2015ApJ...806L..31M} argue that metallicity, rather than velocity dispersion, is the driver of IMF variations in early
type galaxies, building on their earlier work \citep{2015MNRAS.447.1033M,2015MNRAS.451.1081M} using the SPS and
IMF parameterisation approach of \citet{1996ApJS..106..307V}. Using five key spectral features sensitive to metallicity,
age and the IMF, they add a sample of 24 galaxies from CALIFA to the earlier work, jointly constraining the metallicity and
high mass ($m>0.6\,M_{\odot}$) IMF slope, with the low mass ($m<0.6\,M_{\odot}$) slope fixed at the
\citet{2001MNRAS.322..231K} value ($\alpha_l = -1.3$). They find a metallicity dependence on the high mass
slope expressed as $\alpha_h=-3.2(\pm0.1)-3.1(\pm0.5){\rm [M/H]}$, consistent with a \citet{2001MNRAS.322..231K}
high mass slope ($\alpha_h=-2.3$) for metallicities ${\rm [M/H]} \approx -0.3$, and steeper for higher metallicity. They also
fit for a single power law, finding that the data could also be explained by an IMF with slope
$\alpha=-2.5(\pm0.05)-2.1(\pm0.2){\rm [M/H]}$ over the full mass range ($0.1 \le m/M_{\odot} \le 100$).
They go on to demonstrate that the relationship observed by other authors between IMF slope and stellar velocity
dispersion naturally arises, qualitatively at least, through a combination of the mass-metallicity relation
\citep[e.g.,][]{2004ApJ...613..898T,2013MNRAS.434..451L} and their derived IMF slope relation with metallicity.
\citet{2015ApJ...806L..31M} further assert that the evolution in metallicity during galaxy formation
implies an evolving IMF, drawing on arguments for IMFs dominated by high mass stars in low metallicity environments
\citep{2012MNRAS.422.2246M} transitioning to IMFs with a relative excess of low mass stars as the metallicity rapidly
increases. They also point to similar arguments by \citet{2010MNRAS.402..173A} in favour of an evolving IMF for early
type galaxies, based on chemical evolution in semi-analytic models. This conclusion contrasts with the
analysis of 212 ATLAS3D early type galaxies by \citet{2014ApJ...792L..37M} who
conclude that there are no strong trends between any of the stellar population derived parameters
(age, metallicity, [$\alpha$/Fe]) and the IMF mass normalisation.

More significantly, perhaps, \citet{2016MNRAS.456L.104M} show that, in order for passive galaxies to have
both enhanced [Mg/Fe] ratios and an IMF overabundant in low mass stars relative to the Salpeter slope ($\alpha_l < -2.35$),
they must have had extremely short star formation episodes that imply exceptionally high SFRs at high redshift,
SFR $\approx 10^5\,M_{\odot}\,$yr$^{-1}$, which have not been observed. They present two possible scenarios to
resolve this issue. The first invokes an IMF overabundant in both low and high mass stars. The second argues for
a time varying IMF, initially overabundant in high mass stars to account for the chemical signature, evolving to one
overabundant in low mass stars at later times.

The results described above have generally used relatively low redshift galaxy samples. Exploring the higher-redshift
Universe ($0.9<z<1.5$) \citet{2015ApJ...798L...4M} use the TiO$_2$ IMF-sensitive spectral feature and the
\citet{1996ApJS..106..307V} SPS and IMF parameterisation, finding $\alpha_h = -4.2\pm0.2$ (for $m>0.6\,M_{\odot}$)
for the most massive galaxies (with stellar masses $M_* > 10^{11}\,M_{\odot}$), and slightly less steep
($\alpha_h=-3.7^{+0.4}_{-0.3}$) at lower stellar mass ($2\times10^{10}<M_*/M_{\odot}<10^{11}$). 
With estimated ages of $1.7\pm0.3\,$Gyr, this population would have formed at redshifts of $1.5\lapp z \lapp 3$.
These IMF slopes are similar to those found in the low redshift population for the highest metallicity galaxies by
\citet{2015ApJ...806L..31M}, which have similar formation epochs ($z \approx 2$) based on the ages inferred in that analysis.

Using a different approach again, \citet{2012ApJ...747...72D} measure the low mass X-ray binary (LMXB)
population in globular clusters and UCDs within 11 elliptical galaxies of the Virgo Cluster. They conclude that these are
10 times more frequent in UCDs than expected from a \citet{2001MNRAS.322..231K} IMF, implying an excess of
high mass stars. When the LMXB number is compared against an optical or infrared galaxy luminosity, this provides a
direct constraint of the high mass end of the IMF (the progenitors of the neutron stars and black holes detected as LMXBs)
compared to the low mass stellar population. In contrast, \citet{2014ApJ...784..162P} find no evidence for IMF variations
in a sample of 8 early type galaxies, based on their LMXB population. Their results are based on a small sample, and are
consistent with the range of scatter seen by \citet{2010ApJ...709.1195T}. In either case, this demonstrates
an important complementary approach to
constraining the IMF. The presence of a low mass companion in an LMXB also implies a degree of binarity that is often
neglected in common SPS models. There is clearly scope to expand this kind of analysis to larger samples and through
incorporating such SPS models \citep[e.g.,][]{2012MNRAS.422..794E,2017PASA...34...58E}.

Other complementary approaches that have been recently explored for estimating integrated galaxy IMFs include the
demonstration by \citet{2014MNRAS.437..994R} of how the plateau in the [$\alpha$/Fe] ratio for a galaxy is sensitive
to the galaxy IMF, and can be used as a test of whether IMFs vary between galaxies. \citet{2014MNRAS.437.1950B} present
a hierarchical modelling approach that can be used to derive upper limits on departures from universality in the IMF.
\citet{2013MNRAS.432.2632P} introduce a pixel-based SPS fitting approach, similar to the ``pixel-z" technique
\citep{2003AJ....126.2330C,2008ApJ...677..970W,2009ApJ...701..994W}
that fits pixel colours to infer stellar population ages, obscurations and SFHs. The approach of
\citet{2013MNRAS.432.2632P} allows the IMF slope at the low mass end to be a free parameter in the library of models
generated and fit to the observed pixel colours, in principle allowing the IMF to be inferred within spatially resolved galaxy
images. \citet{2013ApJ...771...29G} use resolved star counts of two nearby ultra-faint dwarf galaxies to determine
$\alpha_l=-1.2^{+0.5}_{-0.4}$ (Hercules) and $\alpha_l=-1.3\pm0.8$ (Leo IV) in the mass range
$0.52\le m/M_{\odot} \le 0.77$, and argue that, in combination with resolved star counts from the Milky Way, SMC and
Ursa Minor, this suggests a trend to flatter low mass IMF slopes (or increasing $m_c$) for systems with lower velocity
dispersion and metallicity, qualitatively consistent with the broad results seen for early type galaxies summarised above
\citep[e.g.,][]{2012ApJ...760...70V}.

As with the star forming galaxies, the range of analyses of passive galaxies reveal a broadly consistent
picture, finding an increased abundance of low mass stars for higher velocity dispersion, or higher metallicity, galaxies,
and potentially with this excess located preferentially in the galaxy cores. These IMF variations for passive galaxies
do seem to contradict the sense of the variation found for star forming galaxies, a point that I return to below.
It would also be interesting to explore the extent to which the recent generation of integral field spectroscopic surveys
may be able to bridge the two, applying techniques so far used only for star forming galaxies but instead
looking at passive systems with some residual star formation. With spatially resolved spectra, any AGN
contributions may be identified and excluded to isolate star formation signatures in otherwise
passive galaxies \citep[e.g.,][]{2017MNRAS.470.3395H,2018MNRAS.tmp..130M}, allowing the
\citet{1983ApJ...272...54K} and related approaches to be applied.

\subsection{Summary and limitations}
\label{limitations}
It is clear that analyses of the star forming and passive galaxy populations have a rather
different focus in terms of the IMF, of necessity, due to the available observational metrics.
Work on the star forming galaxy population has focused on the high mass end of the IMF, and its
potential dependence on luminosity, SFR, or SFR (surface) density, loosely
characterised as star formation intensity. Analysis of the passive galaxy population, instead, has
focused on the low mass end of the IMF. The stars being probed in these systems are only the low
mass stars remaining from star formation episodes many Gyr earlier, analogous to the PDMF in Milky Way
stellar systems. In qualitative terms the results can be summarised as:
\begin{itemize}
\item There is evidence for $\alpha_h$ ($m\gapp 0.5\,M_{\odot}$) variation in star forming galaxies,
with flatter (more positive) values in stronger star forming galaxies (increasing SFR, sSFR, or
luminosity), and vice-versa.
\item There is evidence for $\alpha_l$ ($m\lapp 1\,M_{\odot}$) variation in passive galaxies, (or equivalently,
steeper $\alpha_h$ when $\alpha_l=-1.3$ for $m<0.6\,M_{\odot}$ is constrained)
with steeper (more negative) values in (at least the centres of) more high mass passive galaxies
(higher $\sigma$ or metallicity).
\end{itemize}
There is a range of uncertainty, quantitative difference, and scatter around these general
conclusions, but the broad picture seems to be well established given the observational approaches used.

This broad picture leads to an apparent tension, though, because at first glance the two qualitative results seem
inconsistent. The stellar population in the centres of passive galaxies, that formed at the peak of cosmic star
formation \citep[$z\approx 2$, e.g.,][]{2004ApJ...615..209H,2006ApJ...651..142H}, have a relative excess of low mass stars (that may be
characterised broadly
as $\alpha<-2.35$), while the star forming galaxies seem to imply that high star formation is associated with a
relative excess of high mass stars (similarly characterised broadly by $\alpha>-2.35$). Are these results actually
inconsistent or not? If the IMF is not universal, they are not necessarily inconsistent, as it may be the case that different
physical conditions prevailed in the progenitors of high mass elliptical galaxy nuclei, than seen now in high
SFR galaxies at low redshift. The question needs to be answered by exploring the physical properties dominating
the origin of a particular stellar mass distribution, and the physical conditions prevailing in the different
systems at the time of star formation. This is reviewed in \S\,\ref{sims} below.

Before delving into those issues, it is worth questioning the robustness of the various observational approaches,
and assessing the degree to which they may be systematically biased.

In each case, SPS models play a crucial role in the way an IMF slope is inferred. There are some fundamental
limitations in even the most sophisticated modern SPS models, each of which focuses on one particular area
of strength but without necessarily incorporating other facets, or coarsely modeling them in the interests of
computational efficiency. Two main limitations are the neglect of stellar rotation
\citep[e.g.,][]{2011A&A...530A.115B,2011A&A...530A.116B,2012ApJ...751...67L,2014ApJS..212...14L} and stellar multiplicity
\citep[e.g.,][]{2012MNRAS.422..794E,2017PASA...34...58E}.
These effects are not negligible, with perhaps up to 30\% of high mass main sequence stars produced through
binary interaction \citep{2012Sci...337..444S,2014ApJ...782....7D}. Binarity and stellar rotation may reinforce each other in
some observable properties, as the modeling of binaries can provide effects similar to the inclusion of stellar rotation for
single stars \citep{2011ASPC..440..309L}. In particular, extreme rotation can lead to an increase in the H$\alpha$ equivalent
width by up to an order of magnitude compared to the assumption of no rotation, for the same SFH,
in a window between $10^{6.5}<t/{\rm yr}<10^7$ \citep{2014ApJS..212...14L}. This may lead to a need to reduce inferred
SFRs (for example) in some analyses, and may well have an impact on inferred IMFs. Binaries add an extra dimension too,
as binary mergers and mass transfer can lead to the presence of stars of higher mass than those in the initial population
\citep{2012MNRAS.422..794E,2012MNRAS.426.1416B}. For example, using a binary population synthesis approach
\citet{2013MNRAS.432L..46S} make the case for the Milky Way having an IMF dominated by high mass stars at early times,
subsequently evolving to the presently observed IMF, based on observations of CEMP stars. They note that in chemical
evolution models the current Milky Way IMF would overpredict Type 1.5 SNe. While observational constraints on the
degree of rotation and binarity (or multiplicity) fraction are challenging, it is clear that it is necessary to address these effects
in refining any estimated IMFs. 

It is also worth noting that the SPS models used to analyse star forming galaxies are typically different
from those used to analyse passive galaxies, again for the obvious reason that different codes focus on
producing different diagnostics. It would be highly desirable to be able to cross-compare results between
the two galaxy populations using a common SPS tool to eliminate any possibility that inconsistent conclusions
regarding the IMF are related to SPS model systematics.

SPS models commonly used in the analysis of star forming galaxies include PEGASE \citep{1997A&A...326..950F},
STARBURST99 \citep{1999ApJS..123....3L}, and GALAXEV, \citep{2003MNRAS.344.1000B}, as well as those incorporating
a greater contribution from TP-AGB stars \citep{2005MNRAS.362..799M}. The conclusions regarding IMF properties in star
forming galaxies are generally consistent between these different models, and authors often check that their results are not
strongly dependent on the SPS model used \citep[e.g.,][]{2009ApJ...695..765M,2011MNRAS.415.1647G}.
\citet{2017MNRAS.468.3071N} compared various diagnostics between the PEGASE, STARBURST99 and
BPASS \citep{2016MNRAS.456..485S} SPS tools, to confirm that their results were not sensitive to the choice of
SPS model.

In the analysis of passive galaxies, the two dominant SPS tools are the FSPS code of \citet{2009ApJ...699..486C} and its
recent variants \citep{2017ApJ...837..166C}, and that of \citet{1999ApJ...513..224V} and its recent variants
\citep{2010MNRAS.404.1639V,2012MNRAS.424..157V} based on the MILES/MIUSCAT empirical stellar spectral libraries.
The higher resolution of these libraries is important for constraining the key absorption features sensitive
to the low mass stellar populations. Analyses using the former tend to be cast in terms of single power-law IMF slopes,
while those of the latter explore both single and double power-law forms. Broadly, both approaches tend
to conclude that high mass (or high metallicity) early type galaxies have a need for IMFs with a relative excess of
low mass stars, although this is achieved with different IMF forms depending on the SPS tool used for the analysis.
Either a single power-law IMF with a slope steeper than Salpeter ($\alpha<-2.35$) or a double power-law with a steep
high mass ($m>0.6\,M_{\odot}$) slope, both leading to an excess of stars for $m<1\,M_{\odot}$ compared to
Milky Way type IMFs of \citet{2001MNRAS.322..231K} or \citet{2003ApJ...586L.133C}. \citet{2015ApJ...803...87S} compare these two SPS models explicitly, and conclude that, while quantitatively different, the result implying an excess of low
mass stars for early type galaxies is qualitatively consistent between the two. \citet{2013ARA&A..51..393C} reviews SPS
techniques and discusses what can reliably be measured. In this comprehensive review,
issues of assumed metallicity and abundances, dust obscuration, SFHs, and stellar evolution libraries are addressed.
In particular, \citet{2013ARA&A..51..393C} describes the impact on inferred IMFs arising from limitations in
SPS models in his \S\,7. He focuses primarily on the results arising from the passive galaxy analyses, concluding
that modest IMF variations are supported, as inferred through $\Upsilon_*$ variations of a factor of $2-3$.
He briefly argues that the case for ``top-heavy" IMFs is not compelling, and does not explore
the impact of SPS assumptions in those analyses. 

Other potential issues affecting the use of the dwarf-to-giant ratio sensitive features are rare stellar populations
and parameter degeneracies. \citet{2014MNRAS.442L...5M} argue, for example, that barium stars and extrinsic S stars can explain
the effects seen in the Wing-Ford band and the Na {\sc i} D absorption features, without needing to invoke
varying IMFs. \citet{2015MNRAS.453.4431T} highlight strong degeneracies between the inferred IMF slope and the value of
$m_l$, the extent of AGB populations and variations in elemental abundances. They note that ``increasing evidence
shows that single-burst, single-composition stellar populations oversimplify the underlying stellar systems" and conclude
that it is very difficult to disentangle a steepening of the IMF slope from a decreased contribution of the AGB population
in young metal-rich galaxies. They note that this degeneracy can be addressed using sufficiently
high precision photometry and spectroscopy for old (10\,Gyr) metal-rich populations.
Similarly, \citet{2015MNRAS.454L..71S} used composite $J$-band spectra compiled from over 100 galaxies to show
that it is not possible to jointly constrain the Na abundance and the IMF slope in the most massive galaxies.
They conclude by cautioning against over-reliance on Na lines in such studies.
 
Another limitation was highlighted by \citet{2014MNRAS.443L..69S}, who analysed the dynamical and SPS IMF constraints
for 34 galaxies in common between those used by \citet{2012ApJ...760...71C} and \citet{2013MNRAS.432.1862C}. He found
that while the general results of each are consistent, there is no correlation of the IMF inferred on a galaxy-by-galaxy level
between the two approaches. He argues that the results of \citet{2012ApJ...760...71C} are explained by a trend with Mg/Fe
rather than $\sigma$, \citep[perhaps qualitatively in line with][]{2015ApJ...806L..31M}, but that
\citet{2013MNRAS.432.1862C} finds no such relation. He concludes that a range of confounding factors (dark matter
contributions or abundance patterns) have not been disentangled from the IMF effects in one or both of the methods.
Interestingly, \citet{2016MNRAS.457..421O} show for M87 that introducing radial stellar anisotropy has a strong impact on
the derived M/L, and that this leads to an inferred IMF consistent with \citet{2003PASP..115..763C}, although neglecting the
anisotropy would lead to a value of $\Upsilon_*$ that implies a Salpeter-like IMF over $0.1<m/M_{\odot}<100$.

It is also the case that many samples analysed to date are, in most cases, limited in number to a few tens or
in some cases hundreds of galaxies. Although observationally challenging, there is clearly scope for exploring
large volume-limited or mass-limited samples, as done by \citet{2011MNRAS.415.1647G} for example, in order
to account for systematics and selection effects when interpreting any putative physical dependencies for potential
IMF variations.

\subsection{Linking galaxies to their constituents}
\label{linking}
The link between the IMF for stars or star clusters and that inferred for galaxies has been explored extensively
in the framework of the IGIMF \citep{2003ApJ...598.1076K,2005ApJ...625..754W,2013pss5.book..115K}.
\citet{2010ARA&A..48..339B} nicely summarise the link between the IGIMF and the relation between $m_u$ and cluster
mass \citep{2010MNRAS.401..275W}, noting that the evidence for this relation is mixed
\citep[but see, e.g.,][]{2016A&A...588A..40R,2017ApJ...834...94S}.
It is worth a brief diversion here to consider a direct comparison between potential variations
in the IMF for star clusters and that for galaxies, to test whether there are physical dependencies that may
be consistent. From the summary of \citet{2010ARA&A..48..339B}, their Figure~2 has a suggestion that star clusters tend
to have a somewhat flatter high mass IMF slope ($m>1\,M_{\odot}$) than found in associations or the field. Can this be
characterised as a function of SFR or velocity dispersion, perhaps, to compare against the results from
integrated galaxy measurements?

It is not straightforward, it turns out, to compare star forming clusters directly with galaxies. The young massive star
forming cluster Westerlund 1 in the Milky Way, for example, has $\alpha_h=-2.3$ for stars in the mass range
$3.4<m/M_{\odot}<27$ \citep{2008A&A...478..137B}. To compare this with the star forming galaxies of
\citet{2011MNRAS.415.1647G}, say, we need to estimate one or more of its SFR, sSFR or SFR surface density
($\Sigma_*$). \citet{2008A&A...478..137B} quote a total initial stellar
mass for Westerlund 1 of $m\approx 52000\,M_{\odot}$, an age of $t\approx 3.6\,$Myr, and they measured the IMF in
annuli extending out to $r=3.3\,$pc. We can make use of these estimates by way of illustration, although the similarity
of values for the other starburst clusters in the Milky Way \citep{2008ASPC..387..369B} lead to the same conclusions. Using
the values above, ${\rm SFR}=0.014\,M_{\odot}$yr$^{-1}$, and $\Sigma_*\approx 400\,M_{\odot}$\,yr$^{-1}$\,kpc$^{-2}$
for Westerlund 1. It is apparent that the sSFR (defined as the SFR divided by the total stellar mass, for a galaxy) is not really
a meaningful quantity here, since the total stellar mass is the same as the stellar mass formed in the
star formation event, and would simply equal the inverse of the age for the cluster. This is a
hint that such comparisons are not easily made, quickly supported by the fact that the SFR surface density
is two orders of magnitude higher than seen in the ensemble of galaxies sampled by \citet{2011MNRAS.415.1647G},
where $-2.7\lapp \log(\Sigma_*/M_{\odot}\,{\rm yr}^{-1}\,{\rm kpc}^{-2}) \lapp -0.8$. Clearly the average $\Sigma_*$ for
a galaxy is reduced by the many regions that have no ongoing star formation. Alternatively, we could artificially define a larger
area encompassing Westerlund 1 that extends out to the boundary with the closest neighbouring star forming system in order
to define the value of $\Sigma_*$, but that opens a host of related questions for how to define such a region.
That leaves the direct measure of SFR itself. Here again the comparisons are not straightforward, since now
the quantity for Westerlund 1 is two orders of magnitude lower than the range of SFR seen in the galaxies
($0\lapp \log({\rm SFR/M_{\odot}\,{\rm yr}^{-1}})\lapp 1.7$), unsurprisingly when comparing a single star cluster to a
whole galaxy. If instead the stellar velocity dispersion is used as the linking factor, star clusters again
have $\sigma$ substantially lower than those of galaxies. The super star clusters in M82, for example,
have $10\lapp \sigma/{\rm km\,s}^{-1} \lapp 35$ \citep{2007ApJ...663..844M}, compared to the range of
$100\lapp \sigma/{\rm km\,s}^{-1} \lapp 350$ for the early type galaxies in the analyses described above.
Said another way, approaches that parameterise the IMF as a function of SFR or $\sigma$ only make sense
in the case of an entire galaxy where such parameters themselves are well-defined. Any relation between the IMF
and these parameters must actually reflect an underlying physical dependence on a truly local quantity such as the
gas density, ionisation background, or volume density of SFR, for example.

One approach that may have potential in linking the two is that of \citet{2014ApJ...796...71Z}, using the stellar M/L ratio,
$\Upsilon_*$, scaled to a common $10\,$Gyr age, and which they denote $\Upsilon_{*,10}$. As described in
\S\,\ref{stellarclusters}
above, \citet{2014ApJ...796...71Z} directly compares $\Upsilon_{*,10}$ for early type galaxies and disk galaxies, showing
a general match with the values they identify for the two populations of star clusters. They argue that the
different values of  $\Upsilon_{*,10}$ for the two populations reflect variations in the underlying IMF, but that
there is no characteristic physical property identified yet that maps to these IMF differences, having ruled out
velocity dispersion, surface brightness, half-light radius, metallicity, age, half-mass relaxation time, 
central luminosity and mass densities, escape speed, binding energy, and more. Despite this lack of a clear
physical origin, the use of $\Upsilon_{*,10}$ as a metric to compare galaxies with stellar clusters deserves
further attention.

If the potential physical dependencies of the IMF for a star cluster cannot be directly compared to those for a galaxy,
this reintroduces the concern about the IMF itself being a poorly-posed concept, to which I return in
\S\S\,\ref{consistency} and \ref{discussion} below.

Continuing to increase in scale, I turn next to the constraints on the
IMF that have been explored through galaxy populations, rather than individual systems.

\section{IMF MEASUREMENT APPROACHES: COSMIC CENSUS TECHNIQUES}
\label{census}
Since the seminal results in the 1990s by \citet{1996ApJ...460L...1L} and \citet{1996MNRAS.283.1388M} the luminosity
density and associated SFH of the Universe has been measured in progressively greater detail.
This has been complemented by growing numbers of measurements of the stellar mass density (SMD) of the Universe
from large-scale galaxy surveys, following \citet{2001MNRAS.326..255C}. These two major cosmic census methods,
summarised in the review by \citet{2014ARA&A..52..415M}, provide fundamental boundary conditions on an IMF for
the galaxy population as a whole. Other census-style probes sensitive to the IMF include the extragalactic
background light (another luminosity density metric), and the core-collapse supernova rate. Using constraints
such as these, that effectively sample the entire galaxy population at a given epoch or series of epochs, provides
a direct test of whether the IMF can be ``universal". A ``universal" IMF must be able to reconcile the measurement
of all such census metrics. With the increasing fidelity of and focus on the SFH and SMD, it is perhaps not
surprising that this combination has been used to explore implications for the IMF. Such cosmic census
approaches have an advantage over the analysis of individual galaxies in that the assumption of a relatively smooth
SFH is more likely to be reasonable for the ensemble of a large galaxy population, and less of a source of
systematic uncertainty.

The different sensitivity to an underlying IMF present in the cosmic SFH and the SMD in the universe allows
these properties to be used in combination to infer a constraint on the IMF, averaged, in a sense, over the
ensemble of galaxies sampled. The constraint arises because the SFH measurements
are based on luminosity densities (such as the UV or far-infrared) sensitive to high mass
stars ($m> 5-10\,M_{\odot}$), while the mass density measurements are based on luminosity densities
(such as the optical and near-infrared) sensitive to the low mass stellar population. 
Different IMF shapes will affect these stellar populations, and their
associated luminosity densities, differently. Even as early as \citet{2001MNRAS.326..255C} there was a recognised tension
between the SFH and SMD, with that study noting that the two could only be reconciled for a
``universal" IMF with the assumption of surprisingly little dust obscuration affecting the overall SFH.

The approach was explored explicitly by \citet{2003ApJ...593..258B} who inferred constraints on the high mass
($m>0.5\,M_{\odot}$) slope of an assumed ``universal" IMF. They find a slope of $\alpha_h=-2.15$, slightly more positive
than, but consistent with, that of the Salpeter slope, using the joint constraint of the cosmic SFH
and the $z\approx0.1$ luminosity density. Combining the SFH and the electron antineutrino upper limit arising from the
core-collapse supernova background, \citet{2006ApJ...651..142H} place bounds on the high mass ($m>0.5\,M_{\odot}$)
slope of a ``universal" IMF ($-2.35<\alpha_h<-2.15$), and note \citep{2008ApJ...682.1486H} that for consistency with the
SMD an IMF slope of $\alpha_h=-2.15$ \citep[from][]{2003ApJ...593..258B} is favoured over the Salpeter slope of
$\alpha_h=-2.35$. \citet{2007MNRAS.379..985F} also noted a need for an excess of high mass stars, proposing a
``paunchy" IMF, with an excess of stars in the mass range $1.5<m<4\,M_{\odot}$, ($\alpha=-1$ for
$0.1<m/M_{\odot}<0.5$; $\alpha=-1.7$ for $0.5<m/M_{\odot}<0.4$; and $\alpha=-2.6$ for $4<m/M_{\odot}<100$) to
reconcile joint measurements of the extragalactic background radiation density and the stellar mass density
(or $K$-band luminosity density).

Building on these results, \citet{2008MNRAS.385..687W} showed that the SFH and SMD are inconsistent with a
universal, unevolving IMF. \citet{2008MNRAS.391..363W} quantified a requirement for an IMF with a high mass slope of
$\alpha_h=-2.15$ at low redshift that evolves to a high mass slope with a more positive index
still ($\alpha_h>-2.15$) at $z\gapp 1$. This result has subsequently been questioned \citep{2009ApJ...692..778R},
with the key issues being the extent of obscured star formation at high redshift ($z\gapp 2$) and systematics in the
estimates of the SMD.

The core collapse supernova rate density can also be used as a tracer of the cosmic SFH
\citep[e.g.,][]{2012ApJ...757...70D}. Recent results \citep{2015ApJ...813...93S} suggest that the observed
rates are consistent with the SFR densities derived from dust-corrected UV emission, and inconsistent with
the higher SFH that has been used to infer IMF evolution. These results, which rely on $1.6\,\mu$m
imaging for samples out to $z=2.5$, may still suffer from incompleteness, however, due to extreme obscuration in high star
formation regions. This has been demonstrated through the $2.15\,\mu$m detection of heavily obscured supernovae
in the nuclei of nearby ($z<0.027$) luminous infrared galaxies \citep{2018MNRAS.473.5641K}.

The extensive review of the cosmic SFH by \citet{2014ARA&A..52..415M} argues that the discrepancy
between the SFH and the SMD is not significant enough to require an evolving IMF.
They use a Salpeter IMF over the full mass range ($\alpha=-2.35$ for $0.1<m/M_{\odot}<100$) and a selective
compilation of observations focusing on far-infrared and UV measurements, and argue that the discrepancy
between the SFH and SMD is not as large as previously assserted. They note that their observed 0.2\,dex (60\%)
overestimate between the SMD implied from the SFH and direct SMD measurements can be reduced to 0.1\,dex
with a \citet{2003PASP..115..763C} or \citet{2001MNRAS.322..231K} IMF
\citep[as subsequently demonstrated, for example, by][]{2017A&A...605A..70D},
and argue that this residual discrepancy is not sufficient evidence for variations in the IMF.
The more recent analysis of 570\,000 galaxies by \citet{2018MNRAS.475.2891D} reaches a similar conclusion.
In both cases, though, the highest redshifts are probed through galaxies that are rest-frame UV selected,
and are not sensitive to heavily obscured systems.

The analysis of \citet{2014ARA&A..52..415M} omits high redshift ($z\approx 2.3$) H$\alpha$ measurements
\citep[e.g.,][]{2013MNRAS.428.1128S}, which are somewhat higher than those inferred from the compilation of UV
measurements, perhaps by as much as $\approx 0.1$\,dex at $z\approx 2.3$. Combined with the observation that the
fitted functional form of \citet{2014ARA&A..52..415M} tracks
closer to the lower envelope of their data compilation for $1\lapp z \lapp 3$ than the median, another offset of
about 0.1\,dex, there appears to remain scope for discussion of the consistency between the SFH and SMD.
Subsequent updates include new high redshift ($z>4$) SMD measurements \citep{2015A&A...575A..96G} that also
renew the tension between the SMH and SMD. This was explored in more detail by \citet{2016ApJ...820..114Y} who
highlight in particular a significant mismatch
between the observed SFH and that inferred from the SMD in the range $0.5\lapp z \lapp 6$. It seems that
there are still degrees of inconsistency between the SFH and SMD that remain to be resolved.
This includes the steeper IMF slope ($\alpha_h=-2.45^{+0.06}_{-0.03}$) found in M31 by \citet{2015ApJ...806..198W},
which would be inconsistent with the required $\alpha_h=-2.35$ implied by \citet{2014ARA&A..52..415M}
\citep[but see][]{2016A&A...590A.107O}.
There is also the evidence from Milky Way CEMP stars \citep{2007ApJ...664L..63T} that seems to require an evolution in the
IMF toward a larger proportion of high mass stars at higher redshift (increasing $m_c$),
which would be absent in the ``universal" IMF scenario of \citet{2014ARA&A..52..415M}.

Before the SFH and SMD can be used as an IMF constraint, their robustness must be established. At the lower
redshift end, $z\lapp2-3$, the SFH and SMD are well constrained to the level of $30-50\%$. At higher
redshifts, especially $z\gapp 4$, there has been a growing tension over the past decade in the form
of the SFH evolution. The differences arise depending on whether the SFH is measured using photometric dropout
samples \citep[e.g.,][]{2015ApJ...803...34B}, probes that may be sensitive to low mass galaxies
\citep[such as gamma ray bursts, GRBs, e.g.,][]{2013arXiv1305.1630K} or heavily obscured systems,
\citep[using far-infrared data, e.g.,][]{2013MNRAS.432...23G,2016MNRAS.461.1100R}. Recent work has highlighted issues
with these latter measurements, with \citet{2017MNRAS.471.4155K} arguing that they are overestimated because the
inferred luminosity functions overpredict the observed $850\,\mu$m source counts. \citet{2017MNRAS.471.4155K} show
that results from SCUBA-2 and ALMA are consistent with those inferred from the UV-selected photometric dropout samples.
In contrast, recent results using deep radio observations \citep{2017A&A...602A...5N} find SFR densities at $z>2$
consistent with those of \citet{2013MNRAS.432...23G} and \citet{2016MNRAS.461.1100R}.
These are higher than inferred by \citet{2013ApJ...770...57B}, who updated the SFH compilation of
\citet{2006ApJ...651..142H} based on the UV selected samples at such redshifts that had appeared in the meantime,
inferring a lower SFH fit beyond $z>3$. \citet{2017A&A...602A...5N} conclude that there is substantial dust-obscured
star formation at these high redshifts, finding marginal consistency with the dust-corrected SFH of \citet{2015ApJ...803...34B}.

It is clear from the comparison between the UV and radio luminosity functions by \citet{2017A&A...602A...5N} at
$\langle z\rangle = 3.7$ and $\langle z\rangle = 4.8$ that there is a significant difference at the high luminosity (SFR) end,
with the deep radio data picking up high SFR systems not seen in the UV luminosity functions of
\citet{2015ApJ...803...34B}. This may be a consequence of the much larger survey area
probed by the radio surveys ($\sim 2$\,deg$^2$) than the UV surveys ($\sim 0.3$\,deg$^2$). It is telling that in the
comparison by \citet{2015ApJ...803...34B} with their earlier work in much smaller ($\sim 50$\,arcmin$^2$) survey regions
(their Figure~10), they find that the larger survey area ($\sim 1000$\,arcmin$^2$) reveals uniformly higher bright ends
for the UV luminosity functions at $z>5$, implying larger numbers of higher luminosity systems. It is perhaps not
unreasonable to expect that trend to continue when much larger regions are sampled. An alternative is
significant obscuration, optically thick at UV wavelengths, preventing the high luminosity systems from
being detected at all, and unable to be accounted for when making obscuration corrections to the observed
high redshift UV detected population. Of course, both effects may play a role here.

\citet{2015ApJ...799...32B} used the updated SFH normalisation from \citet{2013ApJ...770...57B} to scale down the
GRB inferred SFH of \citet{2013arXiv1305.1630K} at $z>4$, making them more consistent with their inferred SFH fit.
In a recent review, though, \citet{2016SSRv..202..181C} show that metallicity constraints at $z>2$ from damped
Ly$\alpha$ systems are consistent with the rather more elevated SFH inferred by \citet{2013arXiv1305.1630K}, and
consistent with that of \citet{2013MNRAS.432...23G} and \citet{2016MNRAS.461.1100R}
than the lower SFH of \citet{2015ApJ...799...32B}. It is noteworthy in this discussion that the radio luminosity function
results of \citet{2017A&A...602A...5N} are also consistent with the GRB inferred SFR densities
\citep{2007ApJ...671..272C,2008ApJ...683L...5Y,2009ApJ...705L.104K,2013arXiv1305.1630K}
at these high redshifts, reinforcing the expectation of a steepening low luminosity tail to the high redshift
galaxy luminosity function \citep{2013arXiv1305.1630K,2015ApJ...803...34B}.
Such a steep tail, implying the existence of a low mass population of star forming galaxies at $z>4$, was
argued for by \citet{2014MNRAS.439.1326W}, who show a need for a 10\% duty cycle for star formation based on
observed sSFR at such high redshifts.

To return to the IMF constraints imposed by cosmic census approaches, the metal mass density of the universe
is another worth considering \citep[e.g.,][]{2003MNRAS.341..589D,2005ApJ...630..108H,2006ApJ...651..142H}, as
well as average metallicities of galaxy populations \citep[e.g.,][]{2013MNRAS.430.2622D,2016SSRv..202..181C}. The
limited use to date of such constraints reflects in part the challenge in observationally constructing large samples of
such measurements at high redshift. There would seem to be significant power achievable through a joint cosmic
census constraint combining the local luminosity density \citep{2003ApJ...593..258B}, the SFH/SMD
\citep[e.g.,][]{2006ApJ...651..142H}, the extragalactic background light \citep{2007MNRAS.379..985F}, and the
metal mass density.

Taken together, these results suggest that there is further scope for refining our understanding of the high redshift
end of the SFH and SMD, and the joint constraint they impose on the underlying IMF. In that light, I strongly endorse
objectivity in the selection of observational datasets for future comparisons. There has been a clear tendency in the
community to favour one form of observational constraint over another when comparing new work against old, or models
against data, to present new results in the best light. The wealth of published measurements makes it easy to
overlook or omit data that is inconsistent or introduces tension with the new results, rather than
objectively comparing against the full range of observations, with a critical consideration of their limitations.
With the now significant numbers of published measurements for the SFH and SMD, there is scope for
a critical and thorough review to assess the reliability of each, in order that all published measurements are not simply
each given equal weight in future compilations, and that future work does not have the scope to be selective in the
published measurements against which they compare. Old results that have been superseded should be discarded,
and careful consideration given to the origins of any tension in apparently conflicting results, rather than choosing to favour
one over another. There is a valuable opportunity now to establish a new ``gold standard" of SFH and SMD results for
comprehensive future use.

It is worth considering that the observational constraints summarised by \citet{2014ARA&A..52..415M} set, in a sense, an
absolute bound for a ``universal" IMF. Taking the most robust measurements possible, they still find (and dismiss) a
mild tension between the SFH and SMD. The observations not considered by \citet{2014ARA&A..52..415M}, though, are
those which imply higher values for the SFH, and hence exacerbate the SFH/SMD tension. If any weight is given at all to
these other observations, the high redshift SFH tends to move upward and the tension with the SMD is increased.
In that sense, either the IMF is universal, similar to \citet{2001MNRAS.322..231K} and \citet{2003PASP..115..763C}, and
any higher SFH estimates must be overestimated, or there is evidence for an IMF that varies, with $\alpha_h$ increasing
as redshift increases.

The SFH/SMD constraint is, however, inconsistent with the very precise high mass IMF slope
of M31 from \citet{2015ApJ...806..198W}, with $\alpha_h=-2.45^{+0.06}_{-0.03}$ ($m>1\,M_{\odot}$), and the steeper
slopes found for the LMC, SMC and other dwarf galaxies \citep[e.g.,][]{1998AJ....116..180P,2007AJ....133..932U,2013ApJ...763..101L,2015MNRAS.447..618B}. This steep a
high mass slope, if it were ``universal," would exacerbate the tension between the SFH and the SMD significantly,
as the inferred SFH would need
to be at least 30-50\% higher. It seems reasonable to conclude on this point alone, then, that the IMF is not universal.
It also bears reiterating here that most of the SFH/SMD tension is in the mid-range of redshifts,
$1\lapp z \lapp 4$ \citep[e.g.,][]{2016ApJ...820..114Y}, since there is too little time at the highest redshifts ($z>6$) for
appreciable stellar mass to form, compared to that assembled subsequently \citep[e.g.,][]{2013MNRAS.430.2622D}.

Observations at such high redshifts also begin to probe the epoch of reionisation ($z\gapp 6$). The reionisation of the
universe now seems able to be well-explained by star formation in $z>6$ galaxies
\citep[e.g.,][]{2015MNRAS.450.3032M,2015ApJ...811..140B}. The contributions from Population III stars and the implications
for their IMF are now also being explored \citep[e.g.,][]{2017ApJ...834...49S}. The IMF in such high redshift galaxies is of
critical interest. In an earlier analysis using the UV and $V$-band luminosity densities at $z\approx 6$
and a constraint from the epoch of reionisation, \citet{2008ApJ...680...32C} rule out a Salpeter-like IMF ($\alpha=-2.3$ for
$0.1<m/M_{\odot}<200$) for $z>6$ as not producing enough ionising photons per baryon. Depending on the
details of the reionisation history, \citet{2008ApJ...680...32C} argues that the high redshift ($z>6$) IMF must have a
flatter slope, favouring $\alpha=-1.65$ over $0.1<m/M_{\odot}<200$.
It is tantalising that such a conclusion is in the same sense as would be required from an evolving IMF from
the SFH/SMD constraint, and there is clearly scope for a unified approach to link these observational
constraints on the IMF.

I digress now to take step back and consider some logical inconsistencies in the argument for a ``universal" IMF:
\begin{itemize}
\item If the IMF is universal, it cannot have a Salpeter slope over the full mass range ($\alpha=-2.35$ for
$0.1<m/M_{\odot}<100$), for at least two reasons: It is observed to have a flatter slope at low masses
in the Milky Way \citep[e.g.,][]{2001MNRAS.322..231K,2003ApJ...586L.133C}, and it violates the joint SFH/SMD constraint
\citep[the SMD predicted from the SFH is too high, e.g.,][]{2014ARA&A..52..415M}.
\item If the IMF is universal, it cannot be consistent with the Milky Way at low mass ($\alpha_l \approx -1.3$)
and have a slope steeper than Salpeter at high masses ($\alpha_h<-2.35$) without
violating the joint SFH/SMD constraint \citep[the SMD predicted from the SFH is too high, e.g.,][]{2006ApJ...651..142H}.
\item If the IMF is universal, it cannot be consistent with the Milky Way at low mass ($\alpha_l \approx -1.3$)
and have a slope flatter than Salpeter at high masses ($\alpha_h>-2.35$) because
it is observed to have a Salpeter slope in the Milky Way \citep[e.g.,][]{2001MNRAS.322..231K}.
\item If the IMF is universal, it cannot have a Salpeter high mass slope ($\alpha_h=-2.35$ for $m>1\,M_{\odot}$)
given the high precision steeper slopes found for external galaxies, such as M31
\citep[$\alpha_h=-2.45^{+0.06}_{-0.03}$,][]{2015ApJ...806..198W}, NGC 4214 \citep[$\alpha_h<-2.83\pm0.07$,][]{2007AJ....133..932U},
NGC 2915 \citep[$\alpha_h=-2.85$,][]{2015MNRAS.447..618B}, the LMC
\citep[$\alpha_h=-2.80\pm0.09$,][]{1998AJ....116..180P} and the SMC \citep[$\alpha_h=-3.30\pm0.4$,][]{2013ApJ...763..101L}.
\end{itemize}
Since a universal IMF cannot have a high mass slope that is steeper, flatter or equal to the Salpeter value,
the logical conclusion, then, is that the IMF is not ``universal." The limitations in this argument will be clear, and it
is obviously not a formal proof, but the conclusion that a growing wealth of evidence points against a
``universal" IMF is inescapable.

If the IMF is not universal, then authors must be wary of inconsistent usage of assumed IMFs. As a
naive example, galaxy SFRs may be calculated assuming a nominal IMF, but then compared
against SPS outputs assuming a variety of input IMFs in order to establish which (erroneously)
better matches the data. Such analyses must be careful to ensure self-consistency of IMF assumptions throughout.
This is true of cosmic census analyses as well.

If the IMF is not universal, then there are clearly many observational implications, that can be tested to further
explore the extent of any IMF variation. For example, \citet{2015MNRAS.448L..82F} show that no single IMF with a fixed high mass
($m>0.5\,M_{\odot}$) slope ($\alpha_h=-2.3$) and a low mass slope ranging from $-2.8 \le \alpha_l \le -1.8$
can reproduce the observational constraints from the stellar populations of massive early type galaxies, together with
their observed metallicities. They conclude that an evolving IMF \citep{2013MNRAS.435.2274W} is required to explain the
joint constraint. Some implications of a varying IMF were explored by \citet{2016MNRAS.462.2832C}, who show
the impact of assuming the metallicity-dependent IMF found by \citet{2015ApJ...806L..31M} on the SFR of galaxies, the stellar mass
function, mass-metallicity relation and reionisation. The results range from significant to minimal, depending on how the
dwarf-to-giant ratio of the IMF is implemented, but define a clear set of observational constraints that can be used to
begin ruling out particular IMF forms. The substantial variations in physical distributions seen for some of these
comparisons, many already inconsistent with observation, highlight the significant existing scope to begin a focused
program of quantifying any potential variation in the IMF.

As a thought experiment, consider whether the metallicity-dependent or $\sigma$-dependent ``bottom heavy" IMF
for spheroids ($\alpha_l\lapp -2.35$, $m<1\,M{\odot}$) and the SFR-dependent ``top heavy" IMF for disk galaxies
($\alpha_h\gapp -2.35$, $m>0.5\,M{\odot}$)
might both be consistent with the sense of a putative evolving IMF from the SFH/SMD constraint. We can use
the two-phase model for the evolution of galaxies proposed by \citet{2013MNRAS.430.2622D}. In this model, systems that will become
spheroids dominate the SFH earlier (with a peak around $3\lapp z \lapp 5$)
than those that become disks (with a peak around $z \approx 1$). If the spheroids and disks of \citet{2013MNRAS.430.2622D}
respectively have the
``bottom heavy" and ``top heavy" IMFs seen locally (as defined above), then in very qualitative terms, it would appear
that the IMF evolution should be increasingly dominated by ``bottom heavy" systems at higher redshift, inconsistent
with the allowed evolution from the SFH/SMD constraint. Such a coarse analysis clearly neglects many
effects that need to be investigated in more detail, but this illustration hopefully indicates the scope of opportunities
for continuing to explore and refine our understanding of the IMF.

It might be argued, adopting the traditional approach, that all of the work above may be considered
``consistent with a (poorly specified) universal IMF (with large uncertainties)", given the variety of conflicting results,
counter-claims, and limitations. I hope by this point that the specious nature of this conclusion is clear. There appears
to be clear and growing evidence, albeit with a variety of associated limitations, for some degree of variation in the IMF,
and it is appropriate for the conversation to move on to constraining such variations rather than dismissing them.

On that note, I briefly explore simulation work in \S\,\ref{sims} below, aiming to highlight the need for modelers
to focus not on reproducing a particular IMF behaviour, but on identifying which physical conditions lead to
what kind of IMF behaviour, and under what assumptions. Only by reframing the question to one
that asks how the IMF varies and how do different assumptions or physical conditions impact such variation
can we begin to make self-consistent progress in understanding the IMF itself.

\section{IMF MEASUREMENT APPROACHES: SIMULATIONS AND MODELS}
\label{sims}
\subsection{Simulating star formation}
The physics of star formation is an enormous field, and I do not pretend to provide a thorough review here. The purpose
of the current summary is to highlight the complexity of the field, and the challenge in directly linking fundamental
astrophysical processes to the form of an IMF. For details of work in this area, interested readers are referred to reviews
by \citet{2014PhR...539...49K}, and \citet{2014prpl.conf...53O}, work by \citet{2013MNRAS.430.1653H}, \citet{2014MNRAS.437...77B}, \citet{2015MNRAS.450.4137G}, \citet{2015MNRAS.449.2643B}, \citet{2016ApJ...824...17K},
\citet{2016MNRAS.458..673G}, \citet{2017MNRAS.468.4093G}, and references therein.

For ease of readability I refer below to ``top heavy"
or ``bottom heavy" IMFs in reference to work that uses those terms. These correspond respectively to
$\alpha_h\gapp -2.35$ (usually for $m>0.5\,M_{\odot}$, sometimes $\alpha\gapp -2.35$ for the full mass range) and
$\alpha_l\lapp -2.35$ (often for $m<1\,M{\odot}$, but about as often also $\alpha\lapp -2.35$ for the full mass range).
Since this summary is largely qualitative, this usage should not be too ambiguous.

The recent work by \citet{2016MNRAS.460.3272K} provides a concise introduction to the key elements considered by
most star formation simulations. In brief, the thermal Jeans mass, turbulence, magnetic fields, radiative feedback and
mechanical feedback are all considered by various authors to play more or less significant roles. That analysis extends
work by \citet{2011ApJ...743..110K}, who quantifies how radiative feedback can set the stellar mass scale, in turn building
on earlier work by \citet{2009MNRAS.392.1363B} and \citet{2006ApJ...641L..45K}. He argues that radiative processes are
the dominant mechanism in determining the gas temperature and ultimately the origin of the peak in the IMF.

Early work proposed an IMF characteristic mass determined by the thermal Jeans mass
\citep[e.g.,][]{1998MNRAS.301..569L,2005MNRAS.359..211L}. Being temperature dependent, this would lead naturally to
higher $m_c$ in extreme environments such as super star
clusters or galactic nuclei, or high redshift galaxies. Other potential drivers, such as the role of metallicity, have
subsequently been explored. There are arguments that, while metallicity plays an important role in cooling for the
formation of Population III stars, it is unlikely to have a direct effect on the IMF for later stellar generations
\citep{2014MNRAS.442..285B,2005MNRAS.363..363B},
apart from increasing the lower mass limit for lower metallicity systems.
While \citet{2005MNRAS.363..363B} notes that metallicity may have
an indirect impact because of its role in setting the Jeans mass during cloud fragmentation, to the degree that the Jeans
mass of the cloud may affect the characteristic mass of the IMF, \citet{2014MNRAS.442..285B} find stellar
mass functions that are consistent for metallicities ranging from 1/100 to 3 times solar. Similarly,
using two numerical simulations corresponding to Jeans masses different by a factor of three,
\citet{2005MNRAS.356.1201B} argue that any potential IMF variation appears through
a change in the characteristic mass of the system rather than a change in slope at the high-mass end.

It is clear that there is enormous complexity and interplay of the astrophysical processes involved in star formation.
Given this complexity, it is easy to understand that a ``universal" IMF is an attractive end-state to aim at
achieving with models and simulations, as a form of validation. Introducing IMF variations removes this
touchstone, making the work of the theorists more challenging, but as the observational constraints
become more complex, so too do the models in their efforts at addressing them. This has led in more
recent work to the goal of testing how particular models fare in reproducing the range of
popular published IMF variations.

\citet{2013MNRAS.433..170H} uses the excursion set formalism to calculate mass functions from the density field in a
supersonically turbulent interstellar medium. This analysis predicts that IMF variations are most likely to appear at the
low mass end, with remarkably uniform slope for high masses, for reasonable choices of temperature,
velocity dispersion and gas surface density. A different approach, based on a Press-Schechter formalism,
by \citet{2013ApJ...770..150H}, extends their earlier work by including time dependence and the impact of
magnetic field, and reaches similar conclusions.
\citet{2014ApJ...796...75C} shows how the turbulent Jeans mass leads to the peak of
the IMF shifting toward lower masses, to reproduce ``bottom heavy" IMF shapes.
Subsequently \citet{2016MNRAS.462.4171B} identified conditions in two suites of hydrodynamic
simulations that lead to IMF variations at the high mass end. Recently, though, \citet{2017MNRAS.465..105L} have
argued against supersonic turbulence being the primary driver in the shape of the IMF, based on two sets of simulations
with different turbulent modes, finding statistically indistiguishable differences in the resulting IMFs.

It seems that despite the growing sophistication of our theoretical understanding of star formation, there is still scope
for refinement in identifying the various dominant physical mechanisms in different astrophysical environments.
\citet{2007ASPC..362..269E} notes that ``most detailed theoretical models reproduce the IMF, but because they use
different assumptions and conditions, there is no real convergence of explanations yet." In the subsequent decade,
although the models have become more sophisticated, subtle and complex, so have the observational constraints, and the
outcome remains largely the same.

\subsection{Simulating galaxy evolution}
Moving from the complexity of astrophysical processes in individual star formation to the larger scale of galaxies
requires a different form of modeling and simulations. As above, this is a vast field in its own right, and only briefly and
incompletely summarised here with the aim of identifying some of the developments and challenges.

Galaxy populations are typically modeled through semi-analytic recipes embedded in large cosmological simulations,
and individual or small numbers of galaxies through detailed hydrodynamical simulations with better physical resolution
than the cosmological models. In the absence of confirmed physical drivers underlying the shape of the IMF,
such simulations tend to invoke a range of empirical or phenomenological relations that have some physical motivation.
The outcome is that most observational evidence for IMF variations is able to be reproduced by a suitable choice of
physical dependencies for the IMF, although not all results are consistent with each other, or necessarily self-consistent.

\citet{2005MNRAS.356.1191B} modeled the abundance of Lyman break galaxies and submillimetre galaxies, successfully
reproducing luminosity functions and the optical and infrared properties of local galaxy populations, but found a need for
a ``top-heavy" IMF to reproduce the observed $850\,\mu$m galaxy number counts.
Without such a change to the IMF in the model, the constraint from the global SFR density led
to the predicted number counts being too low. Allowing the IMF to be ``top-heavy" increases the $850\,\mu$m flux for
a given (lower) SFR because of the relative increase in the number of high mass stars, allowing the model to consistently
reproduce both the number counts and the SFR density.
\citet{2012MNRAS.423.3601N} explore the impact of allowing $m_c$ to scale with the Jeans mass in giant molecular clouds,
showing that this simple assumption leads to a reduction in the SMH/SMD discrepancy, as well as reducing the
tensions in several other observational constraints. \citet{2013MNRAS.436.2892N} extend this work to show that such an
assumption leads to galaxies experiencing both ``top heavy" and ``bottom heavy" IMFs at different stages of their evolution,
with the bulk of stars forming in a ``top heavy" phase. \citet{2012MNRAS.422.2246M} use a model of rapid gas expulsion
to produce more ``top heavy" IMFs in systems with increasing density and decreasing metallicity.
\cite{2013MNRAS.436.2254B} show that such density and metallicity dependencies for the IMF can
lead, among other effects, to lower SFRs than with a fixed IMF, and that [Mg/Fe] is higher for a given metallicity.
Similarly, \citet{2013ApJ...765L..22B} are able to reproduce the ``top heavy" IMF results
of \citet{2011MNRAS.415.1647G} by allowing the IMF to depend on local densities and metallicities of the interstellar medium.
In contrast, \citet{2013ApJ...779....9B} show that ``bottom heavy" IMFs can also be reproduced with suitable choices
of metallicity and gas density in the star forming gas clouds.

Taking a different approach, \citet{2014MNRAS.442.3138F} uses a semi-analytic model to test the impact of different IMF
prescriptions, broadly falling into two classes of SFR-dependent ``top heavy" models and stellar mass-dependent
or $\sigma$-dependent ``bottom heavy" models. He finds that the ``bottom heavy" models lead to variations in stellar
mass and SFR functions similar to the uncertainty in the determination of those quantities, while the ``top heavy"
models lead to an underestimate in the high mass end of the galaxy stellar mass function, compared to a fixed
\citet{2001MNRAS.322..231K} IMF.

\citet{2017ApJ...845..136B} also explore the impact of observed IMF variations on models. They implement various
IMF dependencies by tagging stellar particles in their simulation with individual IMFs using observationally derived
dependencies on velocity dispersion, metallicity or star formation rate. They then find that the IMFs recovered in the simulated
$z=0$ galaxies no longer reproduce the imposed relations. This leads them to conclude that even more extreme
physical IMF relations for some stellar populations are required to reproduce the observed level of variation.
\citet{2017MNRAS.465.2397S} explore the evolution of $\alpha_{mm}$ (defined here as the ratio of the true stellar mass
to that inferred assuming a Salpeter IMF) using cosmological $N$-body simulations. They find that dry mergers do not
strongly impact the relation between $\alpha_{mm}$ and $\sigma$. They note, though, that the underlying
dependence of the IMF on stellar mass or $\sigma$ is mixed through the dry merger process, making it
observationally challenging to infer which quantity was originally coupled with the IMF. \citet{2010MNRAS.402.1536S} tested
the impact of a ``top-heavy" IMF at high gas pressures, finding that it reduced the need to invoke self-regulated feedback
from accreting black holes to reproduce the observed decline in the cosmic SFR density at $z<2$.
\citet{2015MNRAS.446.3820G} argue that  a ``top-heavy" IGIMF best reproduces the [$\alpha$/Fe]-stellar mass relation
for elliptical galaxies when there is an SFR-dependence for the IGIMF slope.
\citet{2017MNRAS.464.3812F} also explore the implications of the IGIMF method in their semi-analytic model, finding that
it leads to a more realistic [$\alpha$/Fe]-stellar mass relation than with a fixed IMF.

It is clear that numerical simulations and semi-analytic models can provide valuable insights into the way we
understand the IMF. In particular, they can be used to test how different physical prescriptions for star formation manifest,
and the properties of the observational constraints on IMFs that they produce, as well as what accessible observational
tracers give the most discrimination in
measuring the IMF. It is important that models are used to make predictions for how different IMF prescriptions should
present observationally, defining observational tests to refine or rule out particular forms of physical dependencies or
underlying variation. There is perhaps more value in using the models in this way than merely through tweaking
some underlying dependencies to reproduce a select subset of observational constraints.
Because of the fundamental nature of the IMF it is important that models and simulations are
used to test as broad a suite as possible of observational implications, rather than merely focusing on one
or two in particular. This is to ensure that some observational constraints are not violated in the models while
attempting to assess the impact on others.

Large cosmological hydrodynamical simulations are now available, such as E{\sc agle}
\citep{2015MNRAS.446..521S}, Illustris \citep{2014MNRAS.444.1518V} and Magneticum
\citep{2015IAUGA..2250156D} within which detailed galaxy simulations can be
created, for example. By selecting sub-volumes sampling a broad range of galaxy environments
and re-simulating those subregions at high resolution, it should be possible to identify the impact of different simulated
IMFs on the physical properties of the resulting galaxies.
Simulation outputs should be produced that are directly comparable to observables (e.g., luminosities as well
as stellar masses or SFRs) to avoid the need to reconstruct such derived properties from observational datasets,
and potentially introducing inconsistent assumptions regarding the IMF in doing so.
By incorporating population synthesis approaches
that link directly to the observational metrics being used in inferring IMF properties, there may be the opportunity
to directly assess how underlying IMF dependencies are subsequently quantified observationally.

In summary, there is an opportunity to begin linking the numerical, semi-analytic and population synthesis
model approaches to self-consistently assess whether observational approaches for inferring IMFs are actually
providing the quantitative conclusions expected, or whether other underlying effects may dominate.

\section{A CONSISTENT APPROACH}
\label{consistency}
\subsection{IMF constructs}
The IMF has been used as a tool in a broad range of different contexts, as illustrated above. But if the IMF is
not universal, then the quantity actually being measured in these different contexts is not necessarily the same.
When inferring an IMF based on the integrated light from
a galaxy, the quantity being measured is not the same as when inferring an IMF from star counts or luminosity functions.
Likewise, when using a cosmic census approach such as the SFH/SMD constraint,
the quantity in question is different again.

Such spatial dependence of the IMF has been recognised since the earliest work, with \citet{1955ApJ...121..161S}
describing the IMF as ``the number of stars in [a given] mass range \ldots created in [a given] time interval \ldots per
cubic parsec." The spatial dependence, though, can easily be glossed over, and especially with the idea of a
``universal" IMF guiding the thinking, it is easy to conflate IMFs associated with different spatial volumes
(star clusters, galaxies) and treat them as the same entity when they may well not be. This can lead to artificial or apparent
inconsistencies that may not necessarily be in conflict.

\begin{figure}[ht]
\begin{center}
\includegraphics[width=8.5cm, angle=0]{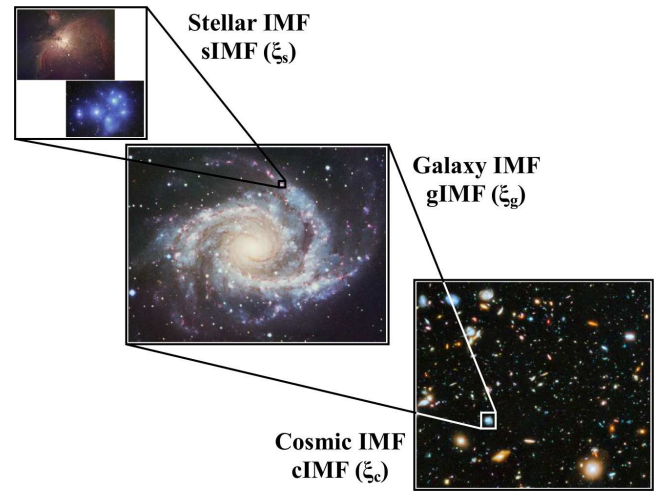}
\caption{A framework to aid in clarifying discussions of the IMF. If the IMF is not universal, then the sIMF, gIMF and cIMF
are not necessarily the same, and all may have a time-dependence. Different measurement techniques and observational
samples probe these different quantities, and what has been referred to uniformly in published work to date as ``the IMF"
conflates these distinct properties. This may well contribute to much of the current tension between different
IMF estimates in different contexts.}\label{schematic}
\end{center}
\end{figure}

The notion that the IMF within a star forming region is potentially a different quantity than the effective IMF for
a galaxy, and different again from the effective IMF for a population of galaxies at a given epoch, is an important
foundational concept. Here I define these three quantities as the ``stellar IMF" or sIMF ($\xi_s$), the ``galaxy IMF"
or gIMF ($\xi_g$), and the ``cosmic IMF" or cIMF ($\xi_c$) as illustrated in Figure~\ref{schematic}. Lower case
prefixes and subscripts are chosen here explicitly with the aim of minimising ambiguity between other
commonly used variants such as IGIMF (for a galaxy-wide IMF), or CIMF (the cluster IMF for stellar
clusters, or ``core" IMF for dense gas cores).

We can generalise the formalism of the IMF by writing the dependence on time and spatial volume explicitly:
\begin{equation}
\xi(m,t,V) = \frac{dN(m,t,V)}{dmdtdV}
\end{equation}
where $dN$ is the number of stars in mass interval $dm$ created in the time interval $dt$ within
the spatial volume $dV$. In this generalisation it is important to note that the time dependence explicit
to $\xi$ allows for the form of the IMF to vary with time. It is different from the time-dependent mass function
scaling that arises from a varying SFR, as defined for example by \citet{1959ApJ...129..243S}.

This approach describes the number of stars of a given mass that have formed up to a given time, for some
spatial volume. Over a (short) finite time period and (small) spatial volume this is identically the stellar mass function
(not accounting for stellar evolution):
\begin{equation}
\int_V \int_t \xi (m,t,V) dt dV = dN(m)/dm
\end{equation}
and corresponds to what might be considered as a ``traditional" IMF.
This approach eliminates the ambiguity between a nominal, or Platonic ideal, IMF from which real star clusters must
be populated, and the actual instantiated mass function, since what is defined here is the real physical quantity of interest,
the number of stars formed as a function of mass, time and location.
The ``universal" IMF scenario can arguably be recovered by asserting that $\xi$ has no temporal or
spatial dependence, $\xi(m) = dN(m)/dm$, but this reopens the issue of accounting for a finite duration for star formation,
and the effects of stellar evolution, since such a $\xi(m)$ is in principle never observable \citep[e.g.,][]{2013pss5.book..115K}.
In contrast, $\xi(m,t,V)$ is directly observable in principle, although in practice doing so may be highly challenging.

Integrating $\xi(m,t,V)$ over different volumes gives the (time dependent) sIMF, gIMF and cIMF:
\begin{equation}
\xi_s(m,t) = \int_{V_s} \xi(m,t,V) dV = \left[\frac{dN(m,t,V)}{dmdt}\right]_{V_s},
\end{equation}
where $V_s$ is a volume characteristic of a star forming region, for example. Relations for $\xi_g$ and $\xi_c$
are analogous.

Clearly there must be a lower limit to the spatial scale or volume over which an IMF is a well-defined quantity,
since it makes little sense to attempt to define an IMF at the scale of individual stars, for example. A natural
minimum spatial scale is that sufficient to encompass a star cluster, and much of the work on the IMF
focuses on the properties of stellar clusters or uses them to probe the IMF \citep[as reviewed by, e.g.,][]{2010ARA&A..48..339B}.
Also, the volumes referred
to here are not specific fixed or comoving volumes in space (although they could be defined as such, in a numerical
simulation for example), but refer instead (for convenience) to a particular spatial scale, and which I illustrate
through these three representative characteristic scales.

Considering the time dimension too is a revealing mental exercise. The process of star
formation is not instantaneous. As stars form within a nascent star cluster there will be
different mass functions extant depending on the time step sampled \citep[e.g.,][]{2013pss5.book..115K}.
There is an extensive literature on the protostellar mass function, explicitly to understand this time dependence
in the way that the IMF is generated. \citet{2010ApJ...716..167M}, for example, present a formalism using models for mass
accretion by protostars to link the IMF to its progenitor protostellar mass function,
extended to the protostellar luminosity function by \citet{2011ApJ...736...53O}. A more recent analytic model
for the mass gained by protostars is presented by \citet{2014ApJ...781...33M}.
\citet{2016ARA&A..54..135H} review accretion onto pre-main-sequence stars, and their Figure~13
highlights the different stellar mass and luminosity functions expected based on different accretion models.

For a large ensemble of clusters, throughout a galaxy say, each at a different stage in its formation
process, the stellar mass function sampled over the full ensemble at a given time step may more
closely resemble the mass function expected from a well-defined physical process, such as gravo-turbulent
models \citep[e.g.,][]{2013ApJ...770..150H,2013MNRAS.433..170H}, although the mass function for any individual region
may well be rather different, due to the complex feedback effects from stellar evolution during the formation event itself,
and the local environment which may be influenced by adjacent regions of star formation or other astrophysical
processes.

In this way of thinking, if there is a ``universal" physical process that drives star formation, then it is likely
to be better sampled on the scales of galaxies (the gIMF) than within individual star forming regions. Extending this
idea, and to account for the possibility of variations between galaxies (due to a range of metallicities,
star formation environments, and so on), any ``universal" physical process might be most accurately sampled
through the effective stellar mass function over an entire galaxy population (the cIMF). This leads to the need
for the physical processes of star formation to be able to explain potential variations in the stellar mass function
from the scale of star clusters to galaxies (which may arise through effects unrelated to the star formation process itself),
ultimately converging on a model prediction when sampled over sufficiently large regions.
This may be written as $\xi (m,t) = \int_V \xi (m,t,V) dV \rightarrow {\rm IMF}$ for $V \rightarrow V_c$ where
$V_c$ is some large volume encompassing one or more galaxies, and ``IMF" here is being
used to describe the stellar mass function expected from a nominal ``universal" physical process.

As an alternative, rather than the sampling of a large spatial volume at a fixed time step, a small volume may be
considered over a long period of time to equally ensure that all phases of the physical process of star formation are
sampled. This might be summarised as $\xi (m,V) = \int_t \xi (m,t,V) dt \rightarrow {\rm IMF}$ for $t \rightarrow t_c$
where $t_c$ is large compared to the duration of a star formation event, perhaps capturing multiple such
events within the volume $V$, and ``IMF" is used as above. Observationally this is not a practical approach,
while the former is, but it may be of value in simulations.

This is another way of considering the arguments posed by \citet{2014MNRAS.439.3239K}, as this concept equally
applies to the gas clouds from which the stars are forming, and the associated ``core" mass functions, or
the mass functions of stellar clusters. Any given gas reservoir may not be representative of the full population
of star forming gas clouds throughout a galaxy, and only by sampling a sufficient number of them will the
statistics of the density distribution be accurately represented.

\subsection{Linking mass functions between different spatial scales}
With differing stellar mass functions on different spatial scales, a natural question arises regarding
how to relate individual mass functions on small scales to those measured
on the larger scales, i.e., how to link the sIMF for multiple star forming regions to the gIMF for the
galaxy comprising those stars. The IGIMF method \citep{2003ApJ...598.1076K,2005ApJ...625..754W,2013pss5.book..115K}
is one approach, which broadly speaking considers a summation of many sIMFs to construct the gIMF.
This method assumes that stars form in self-regulated embedded clusters, which follow a
relationship between the total mass of a stellar cluster and the mass of its highest mass star. Their sIMFs
are therefore empirically constrained by the stellar cluster mass. They can be summed to calculate the gIMF,
or the IMF of a region within a galaxy containing multiple stellar clusters, and can lead to a variable gIMF
\citep{2017A&A...607A.126Y}. This approach allows a gIMF to be calculated given a knowledge of how the sIMF depends
on the physical conditions of star formation.
The early results using this technique favoured galaxy-wide IMF slopes somewhat steeper at the high mass end
than that of the individual star forming regions. Later work incorporating constraints on the variation of the sIMF
from \citet{2012MNRAS.422.2246M} extended this approach to show how flatter IGIMF slopes at the
high mass end could be produced in galaxies with high SFRs \citep{2017A&A...607A.126Y}.

The way that sIMFs themselves arise, or their dependencies on associated astrophysical processes,
may differ from the IGIMF assumptions. Some level of variation would seem likely given that
star formation happens in a complex multiphase medium, with regions of star formation potentially
overlapping, and triggering or suppressing one another in highly non-linear ways, all of which may evolve with time.
So the gIMF may not necessarily comprise a sum over a discrete set of identical, or even simply modeled sIMFs.
If each such star formation event can be characterised by its own sIMF, though (whether or however it is influenced
by, or overlapping with, neighbouring events) we can write
\begin{equation}
\xi_g(m,t) = \int_{V_g} \xi_s (m,t,V) dV
\end{equation}
where $V_g$ is the volume of the galaxy in question. It is important to distinguish this generalisation
from the IGIMF method, since in the current approach each $\xi_s (m,t,V)$ may arise from
different physical dependencies or processes to those assumed in the IGIMF approach.

Likewise, the cIMF may be able to be approximated as a simple sum over gIMFs. Of course, galaxy interactions are
an important channel for galaxy evolution, and they are clearly associated in many cases with
significant levels of star formation. But assuming for the present argument that stars formed in
this mode are a negligible fraction of all stars formed, or alternatively can be accounted for through separate
characterisation with their own $\xi_g (m,t,V)$ we can write
\begin{equation}
\xi_c(m,t) = \int_{V_c} \xi_g (m,t,V) dV
\end{equation}
where $V_c$ is the cosmic volume being probed.

In this formalism there is no analogue to the process of generating a stellar mass function
by ``populating" or ``drawing from" some underlying IMF, since $\xi(m,t,V)$ here is in effect
the stellar mass function itself, incorporating its spatial and temporal variations. Instead, the link to
be highlighted is over what spatial or temporal scales this mass function needs to be sampled in order
to compare with predictions from various physical models of star formation.
In this context questions such as whether the sIMF is drawn from a gIMF, or how to ``populate" an sIMF,
are poorly posed, and not helpful in developing our understanding of star formation.

\subsection{Derived quantities}
The SFR and total stellar mass are directly linked to the IMF \citep[e.g.,][]{1959ApJ...129..243S}, and there is some value
in presenting this explicitly.
The SFR, $S(t,V)$, is the mass of stars formed in a time interval $dt$ and volume $dV$:
\begin{equation}
S(t,V) \equiv \frac{dm}{dt}(t,V) = \int_{m_l}^{m_u} m\xi(m,t,V)dm, \\
\end{equation}
and the total mass in stars ever formed in that volume over some time period $t_1$ to $t_2$ is then:
\begin{eqnarray}
m_{\rm total}(V) & = & \int_{t_1}^{t_2} S(t,V) dt \nonumber \\
 & = & \int_{t_1}^{t_2} \int_{m_l}^{m_u} m\xi(m,t,V)dm dt.
\end{eqnarray}
The mass remaining in stars at a time $\tau$ is $m_\tau=(1-R)m_{\rm total}$, where $R$ is
the mass recycled into the interstellar medium (ISM) due to stellar evolutionary processes, and is dependent
on the stellar mass distribution. Recycling fractions can be calculated for a given mass distribution if the
mass returned to the ISM is known as a function of initial stellar mass \citep[e.g.,][]{1981A&A....94..175R,1995ApJS..101..181W}.
This is also expected to depend on metallicity.

\subsection{Implementation}
The value of a ``traditional" IMF is largely through the ability to use it as a tool to infer the presence of
stellar populations not directly observed. Depending on the techniques being used, observables are often limited
either to the high mass (e.g., through H$\alpha$, UV, or infrared tracers) or the low mass (e.g., direct star
counts, or gravity sensitive spectral features) range of the stellar mass distribution, and accounting for the
stellar populations not directly measured is done by invoking an IMF. With an assumed ``universal" IMF simply
characterised through a well-defined parametric form, such extrapolations are straightforward.

In the general case posed above, a number of simplifying assumptions need to be incorporated in order
to regain the utility of the simple ``universal" model. The value of the general approach is that these
assumptions now become explicit, rather than implicit, defining the form (or absence) of any temporal or spatial variations
(which may reflect other underlying physical dependencies). The same parameterisations (incorporating
physical dependencies if desired) can be applied as always, using the general formalism, but assumptions
about the spatial scale or epoch to which such parameterisations apply become clear. This hopefully
enables distinctions to be drawn between stellar mass functions that should not necessarily be compared
directly, to avoid artificial inconsistencies. It should also facilitate the exposing of internal contradictions within
analyses.

The mass functions as parameterised by, say, \citet{2001MNRAS.322..231K} and \citet{2003PASP..115..763C} are
not inconsistent with this approach, and they can now be explicitly defined as the integrals over some spatial and temporal
scale. So, to the degree that these Milky Way stellar mass functions (${\rm IMF_{MW}}$) correspond to star formation on
the scale of star clusters over a period of several Myr, we may write, for example,
${\rm IMF_{MW}} = \int_t^{t+5\,{\rm Myr}} \int_{V_{s}} \xi (m,t,V) dV dt$, where $V_s$ is the volume sufficient to encompass
the star cluster, and 5\,Myr is nominally taken as a timescale sufficient to allow the full range of masses for all stars to form.
When exploring potential physical dependencies for a stellar mass function, the various observational constraints
can be used in this fashion as boundary conditions.

\subsection{A broader context}
This more general approach can help in setting the context for the broader range of work on the IMF. In particular, by
making explicit the potential spatial and temporal dependencies, which are likely a consequence of any
underlying physical dependencies, the scales over which certain observational constraints apply also
become explicit. It also means that analyses can be clear about the spatial and temporal scales for
the stellar mass functions they are using, or making predictions for. For example, the investigation of
\citet{2017ApJ...845..136B} adopts an observed gIMF, which is then implemented as an sIMF in simulations. They find,
perhaps not surprisingly given the discussion above, that this does not lead to the observed gIMF being
reproduced in the simulated galaxies, concluding that sIMFs need to be more extreme than adopted in
order to replicate the observed gIMF.

The approach presented here provides the potential for self-consistent explorations in models and
simulations, enables a clearer link between what the models predict and what the observations measure,
and avoids conflation between constraints that apply to different physical scales. It provides a
framework in which the subtle biases associated with an implicit tendency toward a ``universal" IMF are
eliminated, allowing for a more critical evaluation of the constraints on potential variations between stellar mass functions,
and their link to the underlying physics of star formation.

With these considerations at hand, I now revisit the variety of observational constraints discussed above
and position them in this self-consistent framework, in order to re-examine the extent to which the IMF may vary.

\section{DISCUSSION}
\label{discussion}
\subsection{What do the observations really tell us?}
With the extensive sets of measurements, inferences and constraints summarised above, it is helpful in
the discussion of the degree of consistency or otherwise to present the results separately for the cIMF ($\xi_c$),
gIMF ($\xi_g$) and sIMF ($\xi_s$), to ensure only comparable quantities are being examined.
For the Milky Way and nearby galaxies, it may be possible to show these in diagrams representing each of
$\xi_g$ and $\xi_s$ depending on whether the full galaxy, or star forming regions within it, are shown.
This consideration also reveals a distinction between ``field star" IMFs and those for stellar clusters, since
the ``field stars" probe, in a sense, the gIMF, while the stellar clusters probe the sIMF.

It quickly becomes clear when doing this that many published IMF constraints are actually not directly comparable,
and that, in fact, there is an extensive range of parameter space to explore in addressing the question of
how to characterise the IMF. In the figures below, representative regions are shown for simplicity and
by way of illustration, rather than attempting to reproduce individual measurements in detail, especially because
in some cases they are not available, although a range of IMF slopes has been given.

In the presentation here I focus on comparisons of IMF shapes as characterised generally by the
low and high mass slopes, $\alpha_l$ and $\alpha_h$, largely because that is the most common approach taken
in the published work. In many cases, though, it may be possible to explain the observed results
by a different approach to modifying the IMF shape, such as an increase in $m_c$
rather than a more positive (flatter) value of $\alpha_h$, or reducing $m_u$ rather than a more negative
(steeper) value of $\alpha_h$, for example. Other measures, too, such as $\alpha_{mm}$, will be important to
include in the development of a suite of diagnostic diagrams for constraining the measurement of the IMF. These points
should be borne in mind when considering the discussion below.

\begin{figure}[h]
\begin{center}
\vspace{3mm}
\includegraphics[width=8cm, angle=0]{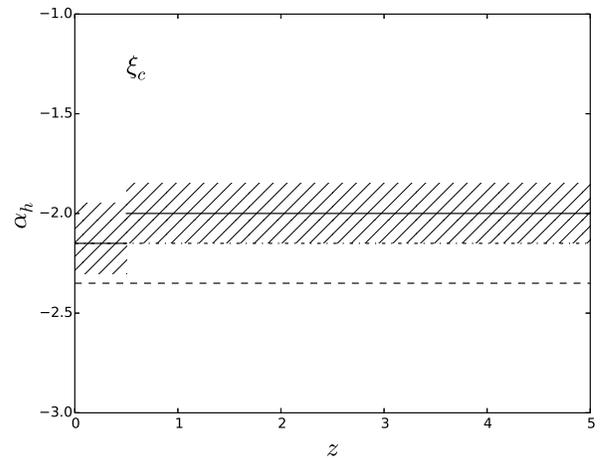}
\caption{The possible variation in $\alpha_h$ for $\xi_c$ from \citet{2008MNRAS.391..363W} (solid lines and hatched
regions). The dashed line is the Salpeter slope ($\alpha_h=-2.35$), and represents the ``universal" IMF from
\citet{2014ARA&A..52..415M}. The dot-dashed line is $\alpha_h=-2.15$ from \citet{2003ApJ...593..258B}.}\label{xi_c}
\end{center}
\end{figure}

There are relatively few observations inferring $\xi_c$, illustrated in Figure~\ref{xi_c}, and only $\alpha_h$ is
typically constrained. A value for $\alpha_l$ as steep as Salpeter is ruled out
\citep[e.g.,][]{2006ApJ...651..142H,2014ARA&A..52..415M}, but most analyses
then assume $\alpha_l=-1.3$ \citep{2008MNRAS.391..363W} or $\alpha_l=-1.5$
\citep{2003ApJ...593..258B,2006ApJ...651..142H} to be consistent with that for the
Milky Way. Figure~\ref{xi_c} shows the Salpeter slope from \citet{2014ARA&A..52..415M}, $\alpha_h=-2.15$ from
\citet{2003ApJ...593..258B}, and the evolving high mass slope of \citet{2008MNRAS.391..363W}. The ``paunchy" IMF from
\citet{2007MNRAS.379..985F} is not shown, as there are multiple $\alpha_h$ values (the slope is different for different
mass ranges above $m>0.5\,M_{\odot}$, see \S\,\ref{census}), but these values bracket those shown in Figure~\ref{xi_c}.
The relatively small variation seen in $\alpha_h$ demonstrates the potential of the cosmic
census approaches in strongly constraining $\xi_c$. Already variations in $\alpha_h$ for $\xi_c$
at the 10\% level are potentially being discriminated between, and future constraints will be even tighter.
It seems fairly clear from this comparison, though, that if the cIMF does evolve, the extent of any evolution
needed to resolve the SFH/SMD constraint is relatively mild, and at the level of $10-20\%$ in $\alpha_h$.

\begin{figure}[h]
\begin{center}
\vspace{3mm}
\includegraphics[width=8cm, angle=0]{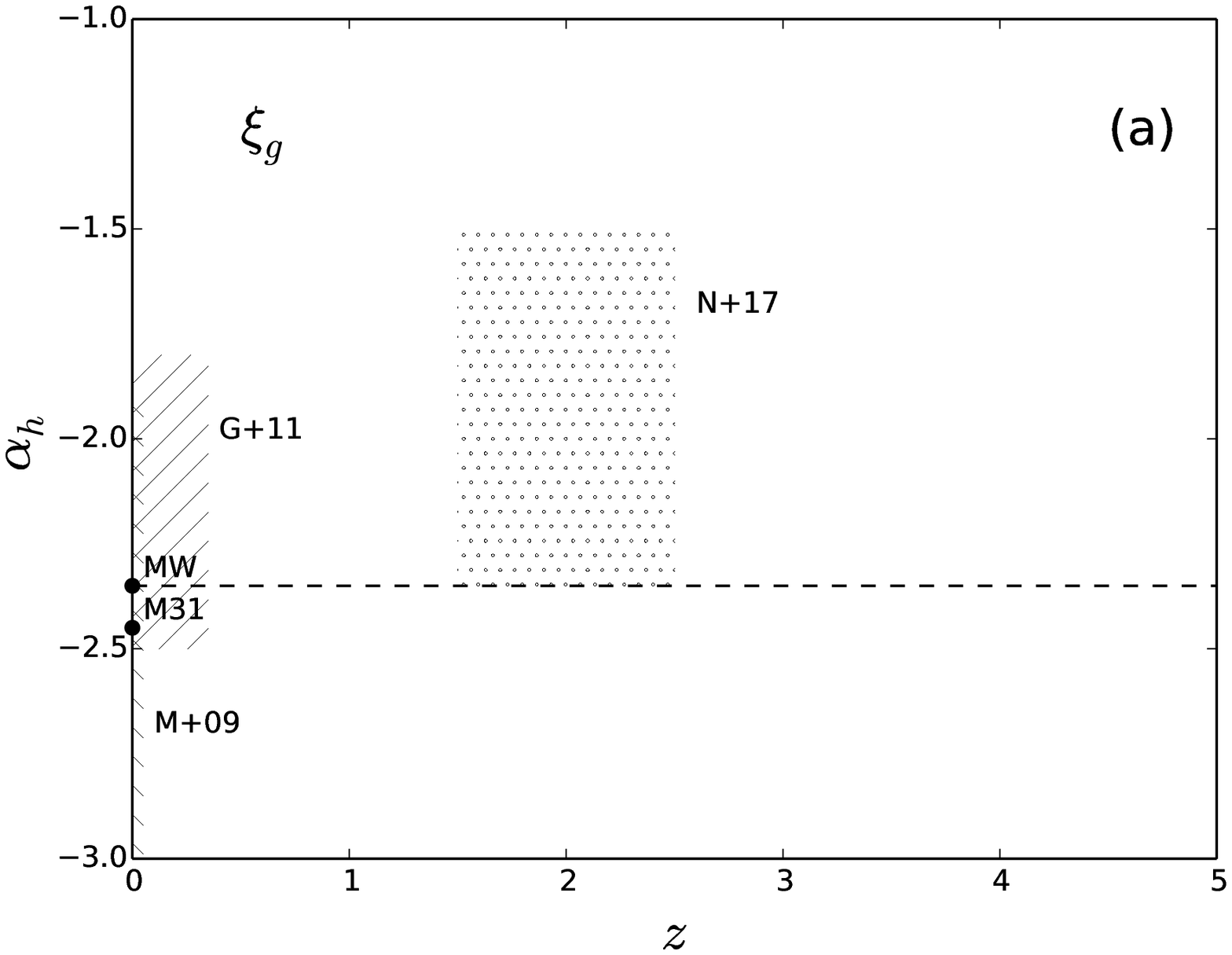}
\includegraphics[width=8cm, angle=0]{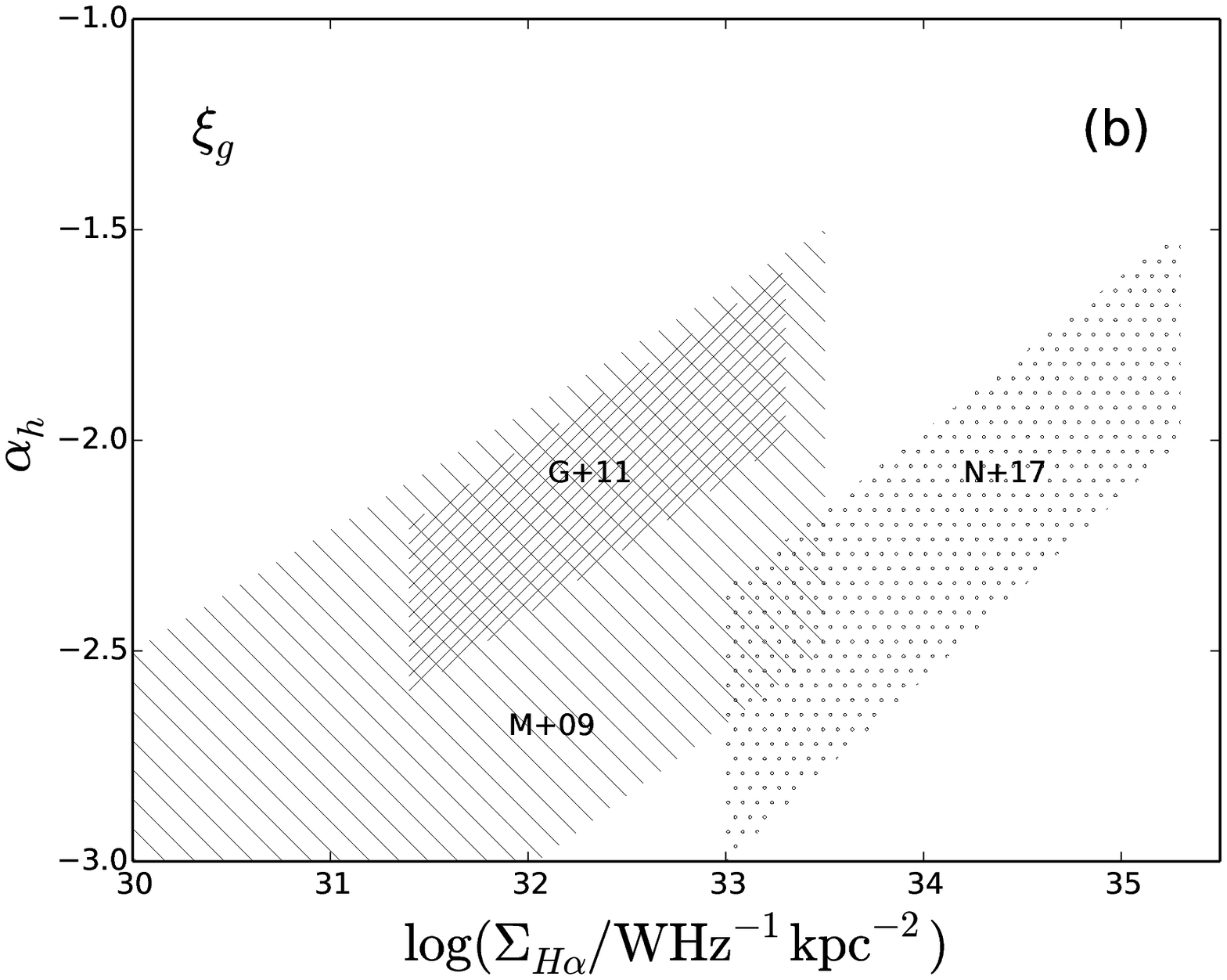}
\caption{(a) The possible variation in $\alpha_h$ for $\xi_g$ from \citet{2009ApJ...695..765M},
\citet{2011MNRAS.415.1647G}, and \citet{2017MNRAS.468.3071N},
shown as hatched and dotted regions. The dashed line is the Salpeter slope ($\alpha_h=-2.35$). Values
for the Milky Way (MW) and M31 \citep{2015ApJ...806..198W} are also shown. Note that the full range of $\alpha_h$ is
indicated, and the dependencies on sSFR or other physical property are not represented here. (b) The
approximate dependence of $\alpha_h$ on H$\alpha$ surface density, inferred from each of
\citet{2011MNRAS.415.1647G}, \citet{2009ApJ...695..765M} and \citet{2017MNRAS.468.3071N}.}\label{xi_g}
\end{center}
\end{figure}

The gIMF has constraints on $\alpha_h$ for the star forming population, and on $\alpha_l$ for the passive
population, but there is no real overlap between the two. The studies constraining $\alpha_h$ are not sensitive to
$\alpha_l$ and vice versa. Figure~\ref{xi_g}a shows, as an illustration, the range of values for $\alpha_h$ from
\citet{2009ApJ...695..765M}, \citet{2011MNRAS.415.1647G} and \citet{2017MNRAS.468.3071N} as a function of redshift,
compared with that for the Milky Way \citep[e.g.,][]{2001MNRAS.322..231K} and M31 \citep{2015ApJ...806..198W}. The
results of \citet{2017MNRAS.468.3071N} at high redshift
may well extend to include steeper values, $\alpha_h<-2.35$, although the focus in their discussion is on the possibility
of flatter slopes for the extremely high H$\alpha$ equivalent width systems measured. The dependence of $\alpha_h$
on galaxy properties is illustrated in Figure~\ref{xi_g}b, shown as a function of $\Sigma_{H\alpha}$, as inferred from
\citet{2011MNRAS.415.1647G}, \citet{2009ApJ...695..765M} and \citet{2017MNRAS.468.3071N}.
The relation of $\alpha_h$ with $\Sigma_{\rm SFR}$ from Figure~13 of \citet{2011MNRAS.415.1647G}
has been converted to one with $\Sigma_{H\alpha}$ using their SFR conversion factor (their Equation~5).
Combining information from Figures~3 and 10b of \citet{2009ApJ...695..765M}, we can infer the range of $\alpha_h$
as a function of $\Sigma_{H\alpha}$ to compare with the results of \citet{2011MNRAS.415.1647G} in Figure~\ref{xi_g}b.
The H$\alpha$ SFR from Figure~21 from \citet{2017MNRAS.468.3071N} can be used, with an assumed galaxy
size of approximately $3-4\,$kpc \citep{2017ApJ...834L..11A}, and the range of $\alpha_h$ inferred from their earlier
figures to reconstruct a rough estimate of how their data may populate this relation.

The point made by \citet{2011MNRAS.415.1647G} is worth reiterating. They state that if an IMF-dependent SFR
calibration were used this would have the effect of reducing the range in SFR probed, but would not change
their conclusion of an SFR-dependence for $\alpha_h$, since the variation is monotonic and the ordering
of the SFRs would not be affected. There is some scope for future work here to develop a self-consistent
constraint on $\alpha_h$ with SFR-related parameters.

\begin{figure}[h]
\begin{center}
\vspace{3mm}
\includegraphics[width=8cm, angle=0]{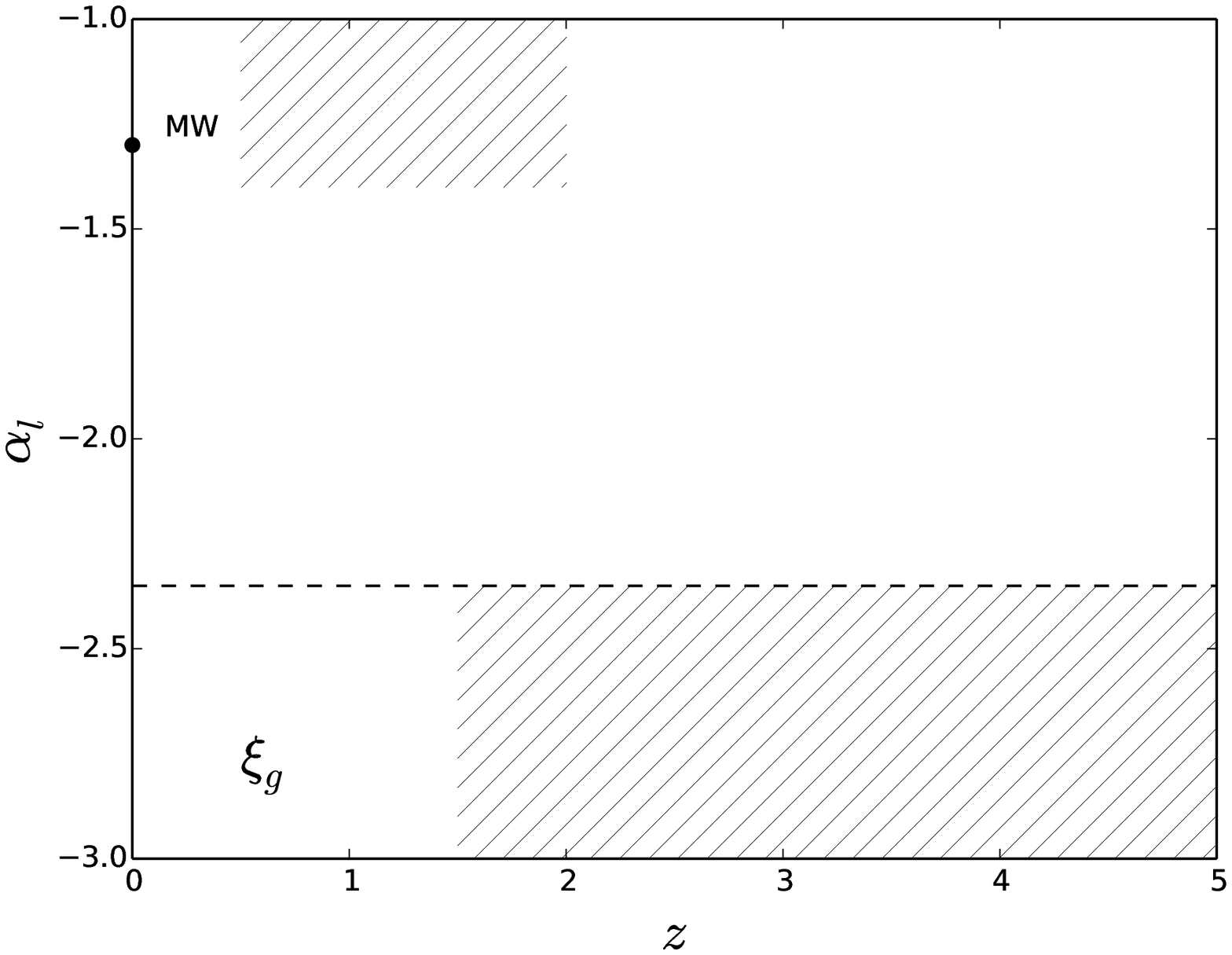}
\caption{The low mass slope of the IMF ($\alpha_l$) as a function of redshift, representative of the formation
time of the stars involved. The \citet{2001MNRAS.322..231K} value for the Milky Way is shown as the data point, and
the range of values for $\alpha_l$ from \cite{2010ARA&A..48..339B} for Milky Way stars is shown as the upper hatched region,
corresponding broadly to formation (lookback) times spanning $5\lapp t/{\rm Gyr} \lapp 10$. The lower hatched region
shows the steep low mass IMF slopes, at formation times approximately $9.5\lapp t/{\rm Gyr} \lapp 12.3$, for the
passive galaxies discussed in \S\,\ref{passive} above. Note that the broad range of $\alpha_l$ for these galaxies
is indicated, and potential dependencies on $\sigma$ or [M/H] are not represented.}\label{xi_g3}
\end{center}
\end{figure}

The low mass slope ($\alpha_l$) for $\xi_g$ is illustrated as a function of redshift in Figure~\ref{xi_g3}. While the galaxies
or stars observed in order to infer these measurements are all at very low redshift, the results are shown at illustrative
formation times for the stars being analysed, making the coarse assumptions that Milky Way field stars formed
$5-10$\,Gyr ago, while the ages for the passive galaxy stellar populations are taken as approximately
$9.5-12.3$\,Gyr. This approach is taken in order to begin unifying our picture for $\xi_g$. So, for example, at $z\approx 2$
we have from \citet{2017MNRAS.468.3071N} a high mass slope for at least some galaxies up to $\alpha_h\approx -1.5$
(Figure~\ref{xi_g}a), with a low mass slope for at least some galaxies of either $\alpha_l \approx -1.3$ or
$\alpha_l \lapp -2.35$. It is interesting to note that there do not seem to be any measurements having intermediate values for
the low mass slope of $\xi_g$. Observations seem to favour either $\alpha_l \gapp -1.5$ or $\alpha_l \lapp -2.35$. This
may echo the bimodality in low mass sIMF shapes implied from the M/L ratios for globular cluster systems from
\citet{2014ApJ...796...71Z}. They argue that these are consistent with either $\alpha =-2.35$ for $0.1<m/M_{\odot}<100$ for
the high M/L systems \citep[similar to the results for passive galaxies from, e.g.,][]{2012ApJ...760...71C}
and a Milky Way style \citep{1993MNRAS.262..545K} IMF (with $\alpha_l\approx-1.3$) for the low M/L systems.

The lensing, kinematic and dynamical constraints on the overall mass normalisation of the gIMF are harder to
capture in these kinds of diagrams, as they do not give an explicit constraint on the gIMF shape, and the same
mass normalisation can be achieved through various combinations of $\alpha_l$ and $\alpha_h$ (\S\,\ref{passive}).
In particular, some analyses for passive galaxies use a single power law gIMF to achieve a given mass normalisation, while
others constrain $\alpha_l=-1.3$ and achieve the same mass normalisation through inferring a steeper $\alpha_h$.
In the latter case, values as steep as $\alpha_h\approx -4$ \citep[e.g.,][]{2015ApJ...798L...4M} are seen for galaxies
observed at $z\approx 1$ that have formation redshifts around $z\approx 2$ (contrast with results shown in
Figure~\ref{xi_g}a). As noted by \citet{2013ARA&A..51..393C}, to retain a given gIMF mass normalisation, more
positive (flatter) values for $\alpha_h$ \citep[such as those of][]{2017MNRAS.468.3071N} would imply a need also
for more positive (flatter) values of $\alpha_l$, inconsistent
with the more negative (steeper) values inferred for passive galaxies from the dwarf-to-giant ratio approach. In other
words, the progenitors of low redshift passive galaxies must have had a gIMF different from that observed in situ
in high redshift star forming galaxies. This may be a further argument for a bimodality in the gIMF, possibly
linked to the stellar M/L ratios for galaxies, that discriminates between spheroid and disk galaxy progenitors.

For the sIMF, $\xi_s$, it would be of interest to show $\alpha_h$ and $\alpha_l$ as a function of age, although
here the effects of dynamical as well as stellar evolution would first need to be accounted for \citep{2010ApJ...718..105D}.
It would also be valuable to explore the sIMF parameters as a function of spatial scale, or total cluster mass,
as well as stellar M/L ratio, to assess the potential for a link to $\xi_g$. With new telescopes such as the JWST and the GMT
enabling the opportunity to explore $\xi_s$ for more resolved stellar systems within nearby galaxies, it will
be valuable to begin quantifying the range of physical parameters currently probed for existing stellar clusters
and associations, in order that larger samples spanning a broader range of environments can be put in a common
context. In particular, as the sample sizes grow for measuring $\xi_s$, there is an opportunity to begin to reduce
the sampling errors and discriminate physical effects from sampling effects, in order to establish whether
apparent differences between super star clusters, field stars or low SFR regions, and globular
clusters are confirmed, and can be attributed to one or more specific physical processes.

The hope is that by discriminating explicitly between $\xi_s$, $\xi_g$ and $\xi_c$, and beginning to explore
each self-consistently through diagnostic diagrams such as those in Figures~\ref{xi_c}, \ref{xi_g} and \ref{xi_g3},
and others that should easily be apparent, clear constraints on any IMF variations should be able to
start being quantified.
 
\subsection{Unifying our understanding of the IMF} 
 
It is possible to bring this complex suite of observational constraint and inference together by recasting
astrophysical questions in this self-consistent approach. \citet{1998MNRAS.301..569L} presented a selection of evidence
that argued for a ``top-heavy" IMF at early times. These included the G-dwarf problem,
perhaps now updated as the CEMP star problem, and the
evolution of the cosmic luminosity density. The first can now be cast as a constraint on the gIMF,
and the second as a constraint on the cIMF (discussed in detail in \S\,\ref{census} above).
Importantly, such questions can now be addressed in a quantitative fashion, in order to establish whether
any given gIMF, for example, can resolve both the CEMP star constraint and the Kennicutt diagnostic
results from \citet{2008ApJ...675..163H}, \citet{2011MNRAS.415.1647G} and \citet{2017MNRAS.468.3071N}. Similarly,
such gIMFs can easily be assessed to see if they are consistent with the mass normalisation required for the low redshift
passive galaxy population. The link between the gIMF and its contributing sIMFs must also be tested,
to ensure consistency with the constraints from stellar clusters.

By bringing together the observational constraints in a tractable way, the issue can evolve
from individual analyses identifying which IMF form best fits their data, to a set of constraints that
must all be met by any successful IMF. This may or may not involve IMF variations, but a strong
set of quantitative boundary conditions ensures that any limits on variations can be well measured.

The use of stellar M/L ratios as a unifying property is worth exploring given that it is a quantity that can
be applied across a broad range of spatial scales. The similarity shown by \citet{2014ApJ...796...71Z} for the
two populations of globular clusters compared to the passive and disk galaxies is tantalising (Figure~\ref{Z14fig9}).
There may still be a disconnect related to age, though, since the high $\Upsilon_{*}$ globular clusters tend to
be those that are young, while the similar $\Upsilon_{*}$ passive galaxies have much older stellar populations.
The situation is reversed for the low $\Upsilon_{*}$ globular clusters, which tend to be those with older stellar
populations compared to the younger stellar populations with similarly low $\Upsilon_{*}$ in actively star forming
disk galaxies. Regardless, there is clearly scope to explore this link further.

It is perhaps worth postulating two broad scenarios at this stage, following from the suggestion above that
there may be evidence for a bimodality in forms taken by the gIMF. Scenario 1 is the ``bottom heavy"
mode, characterised by $\alpha = -2.35$ over $0.1<m/M_{\odot}<100$, and possibly with even steeper $\alpha_l<-2.35$.
This mode is that which seems to characterise passive galaxies and their progenitors, high $\Upsilon_{*}$ globular
clusters, and possibly low surface brightness or low SFR dSph galaxies or star forming regions.
Scenario 2 is the ``top heavy" mode, characterised by $\alpha_h>-2.35$ (with $\alpha_l \approx -1.3$),
that appears to be required for high SFR galaxies and star forming regions, and possibly also
at high redshift by the cosmic census constraints. It may not be the first time such a model has been proposed, but
linking these broad scenarios directly and explicitly with the different sIMF, gIMF, and cIMF constraints will
hopefully aid work on the underlying physics of star formation to help clarify which observations (how and in what conditions)
the modeled or simulated instantiated mass functions need to be reproduced.
A variety of models and simulations already exist that can reproduce such behaviour (see \S\,\ref{sims}) in at least
some circumstances, and the degree to which they self-consistently also reproduce other constraints needs to be tested.

\subsection{Is the IMF universal?}

I am hopeful that at this point we can dispense with this question as either misleading or poorly posed. The
more relevant question is whether there is a ``universal" physical process for star formation. The IMF as a concept
is perhaps better presented directly as the evolving and spatially varying stellar mass function explicitly, $\xi(m,t,V)$
(\S\,\ref{consistency}). Clearly the observed stellar mass functions may vary dramatically between different
stellar clusters, associations and galaxies, as a consequence of dynamical and stellar evolution and
physical conditions. The stellar mass distribution on the scale of galaxies is not necessarily expected to be
the same as that for a star cluster, nor that for a population of galaxies as a whole. There are numerous
lines of evidence, summarised above, that the gIMF in particular may show two qualitatively different shapes.
In order to assess this further, models and simulations should consider explicitly distinguishing between
the IMF on different scales, testing and comparing against observations on appropriate scales. In particular,
if there is a ``universal" physical process of star formation, that process must lead to the full range of
IMF variations seen on the different scales and in the different contexts presented above.

\subsection{Future work}
There are many areas where work on understanding the IMF can be developed further, through simulations
and observations, some of which are briefly touched on here.
These opportunities are qualitatively different for the sIMF, gIMF and cIMF
although in all cases, presenting results in terms of IMF independent observables (such as luminosity) as well as
derived (IMF dependent) quantities (such as masses and SFRs) is an important aid to clarity.

For the sIMF, especially where stellar systems need to be resolved, new telescopes such as JWST or the GMT will enable
significant new breakthroughs. For the cIMF, there is an opportunity to update the review of \citet{2014ARA&A..52..415M}
by conducting a critical assessment of the many published SFH and SMD measurements, and other
cosmic census constraints like the extragalactic background radiation density and supernova rates.
This is needed to develop a ``gold standard" reference set of observations to serve as the boundary
conditions for any cosmic census approach.

As chemical abundance measurements are highly sensitive to the IMF, precision abundance measurements of
a large population of stars can be used to improve such constraints. The GALAH survey \citep{2015MNRAS.449.2604D,2017MNRAS.465.3203M} is one such project, to deliver precision chemical abundances for a million Milky Way stars,
with currently about 0.5 million spectra in hand. Using the technique of ``chemical tagging"
\citep{2002ARA&A..40..487F,2004PASA...21..110B,2007AJ....133..694D,2010ApJ...713..166B},
the preserved chemical compositions of stars allow the reconstruction of original star-formation events
that have long dispersed into the Galaxy background, and possibly even the residual signatures of the first stars in the
early universe \citep{2010ApJ...721..582B}.
In consequence, the large numbers of elemental abundances and large sample size measured by GALAH may enable
the most robust measurement yet using this technique of the historical IMF of the Milky Way
(G. De Silva, personal communication).

Measurements of the gIMF will continue to be able to draw on a wealth of observational data from existing and
upcoming large survey programs, including the large integral field survey SAMI
\citep{2012MNRAS.421..872C,2018MNRAS.475..716G} and the Taipan galaxy survey \citep{2017PASA...34...47D}
that aims to obtain spectra and redshifts for around 2 million galaxies over the Southern hemisphere.
The way the gIMF is measured relies heavily on SPS models, but at present different models are used
in different contexts (passive galaxies are analysed one way, star forming galaxies another). There is an
opportunity to develop SPS models that can provide the information used in multiple IMF metrics
self-consistently. This would allow, for example, the stellar absorption features used in the dwarf-to-giant
ratio approach for an old stellar population to be linked directly to the Kennicutt diagnostics for that same
stellar population at a younger age, in order to self-consistently assess a passive galaxy population and
a star forming galaxy population within a common framework.

Such advances in SPS modeling also need to incorporate the effects of stellar rotation and binarity or
multiplicity in stellar systems. There is also an opportunity through new stellar surveys, such as GALAH
\citep{2015MNRAS.449.2604D}, to extend the range of metallicity and abundances used in stellar evolutionary libraries,
on which the SPS models rely. More comprehensively spanning the observed range of stellar properties in this way would
reduce the potential that inferred IMF properties might be a consequence of model systematics. This comes,
though, at the cost of a greater range of model parameters which may not be observationally well-constrained,
and will likely extend these areas of research in their own right.

There is scope to further explore mass-to-light ratio constraints self-consistently with the SPS approaches,
by applying both uniformly to a well-defined set of stellar clusters and galaxies. Using both together can
break the degeneracy in IMF shape arising from the mass-to-light ratio constraint alone. This approach
could potentially lead to a method that links or even unites star cluster constraints with galaxy constraints.

With a field as broad and far-reaching as that of the IMF, there are clearly many more areas in which
to pursue improvements in our understanding. Those listed here are just a few that may be valuable
to address, directly arising from the discussion and considerations above.

\section{CONCLUSIONS}
\label{conc}
The stellar initial mass function is a critical property influencing almost all aspects of star and galaxy evolution,
and it has been the focus of a prodigious wealth of research spanning more than 60 years. This review
has adopted a particular emphasis on the growing range of observational approaches to inferring or constraining the IMF,
and exploring their strengths, weaknesses and apparent conflicts.

I have pedantically recommended that unambiguous terminology be adopted, echoing similar pleas dating
back twenty years, and argued for the use of a standard
nomenclature convention to minimise ambiguity (\S\,\ref{nomenclature} and \ref{conv}).

I have explored the issue of a ``universal" IMF, and raised concerns about the Occam's Razor default toward
``universality." This is accompanied by a recommendation that the most general scenario, that the IMF is not
``universal," rather than the simplest, be taken as the baseline assumption (\S\,\ref{confusion}). This baseline approach
should lead to a clearer presentation in the literature around degrees of uncertainty and
the physical parameter ranges being probed, to aid in defining the extent of any possible IMF variation.

Relying heavily on previous reviews where available, I have summarised results from a selection of
studies that infer the IMF, spanning the scale of stellar clusters to
galaxies and galaxy populations, along with simulations that model it (\S\S\,\ref{stellar}-\ref{sims}).
This was followed by the introduction of a general and self-consistent approach (\S\,\ref{consistency}).
This approach makes the temporal and spatial dependencies of the stellar mass
function explicit, $\xi(m,t,V)$, leading to clear distinctions between $\xi_s$, $\xi_g$, and $\xi_c$ representative
of the spatial scale of stellar clusters, galaxies and cosmic census probes of galaxy populations. These quantities
should not in general be expected to be the same and should not be conflated or compared.

Using this self-consistent approach, a selection of new diagnostic diagrams were introduced to explore the IMF shapes
from a selection of published results (\S\,\ref{discussion}), complementing and extending the ``alpha plot"
\citep{1998ASPC..142..201S,2002Sci...295...82K,2010ARA&A..48..339B}. These diagnostics were used to assess the
degree to which published IMF properties
are consistent or not. If the cIMF evolves, the degree of evolution only needs to be mild in order to resolve the
SFH/SMD constraint. The gIMF, in contrast, does seem to show some evidence for a bimodality in the IMF
shape for star forming galaxies and the progenitors of low redshift passive galaxies. There is scope now to
begin presenting the many sIMF measurements in diagnostic diagrams similar to these in order to further
quantitatively explore possible dependencies on a range of physical conditions.

This review has, inevitably, been limited in many ways. I have endeavoured to capture the current state of
the field through summarising representative work, in order to tease out where tensions actually exist.
I have presented a general approach that may be of value in supporting simulations and models to
more precisely compare against the most relevant observational constraints. It is my hope that this
review provides a unifying perspective for this fundamental aspect of the formation and evolution of stars and galaxies.

\begin{acknowledgements}
I am deeply grateful to Jill Rathborne for extensive discussion and invaluable support through the development and
writing of this review. I thank the two anonymous reviewers for detailed and positive feedback that helped improve
and refine the discussion presented here. My thanks also to Dennis Zaritsky, Madusha Gunawardhana,
Pieter van Dokkum, and Michele Cappellari for their willingness to share the original versions of the figures from their
papers that are reproduced here. I thank Pavel Kroupa for his careful reading of this lengthy work, identification of
otherwise overlooked references, and his input in refining the descriptions of the IGIMF method. I greatly appreciate
numerous conversations with Madusha Gunawardhana that have helped inform this review, and I thank her also for
providing Figure~\ref{tracks}. I thank Matthew Colless for input on this project, and his advice, mentoring, and guidance
on many others over the past decade. I thank Warrick Couch for his valuable support in encouraging me to pursue this
project and for enabling me to manage my time while working on this review.
I gratefully acknowledge input as well from the following colleagues in helping to refine various
different elements of this review: Joss Bland-Hawthorn, Elisabete da Cunha,
Gayandhi De Silva, Chris Evans, and Ned Taylor.

I thank the Editorial Board of PASA for giving me the opportunity to contribute this work to the Dawes Review series.
I have found it incredibly rewarding to explore the many areas and subtleties of this fascinating field, and it has been a
privilege to share my perspective on this fundamental topic.
\end{acknowledgements}

\bibliography{IMFDawes}
\bibliographystyle{pasa-mnras}


\end{document}